%% file: paper.tex
\documentclass[preprint,times,3p]{elsarticle}
\usepackage[usenames,dvipsnames]{xcolor}
\usepackage{amsthm}
\usepackage{amsmath}
\usepackage{amssymb}
\usepackage{mathtools}
\usepackage{bm}
\usepackage{stanli}
\usepackage{soul}
\usepackage[normalem]{ulem}
\usepackage{subcaption}
\usepackage[normalem]{ulem}
\usepackage{csquotes}
\usepackage{algpseudocode}
\usepackage[ruled]{algorithm}
\usetikzlibrary{arrows, arrows.meta}
\usepackage[colorlinks]{hyperref}

\makeatletter
\providecommand{\doi}[1]{%
	\begingroup
	\let\bibinfo\@secondoftwo
	\urlstyle{rm}%
	\href{http://dx.doi.org/#1}{%
		doi:\discretionary{}{}{}%
		\nolinkurl{#1}%
	}%
	\endgroup
}
\makeatother

\title{Modular-topology optimization of structures and mechanisms with free material design and clustering}
\author[1]{Marek Tyburec\corref{cor1}}
\author[1]{Martin Do\v{s}k\'a\v{r}}
\author[1]{Jan Zeman}
\author[2]{Martin Kru\v{z}\'ik}
\address[1]{Department of Mechanics, Faculty of Civil Engineering, Czech Technical University in Prague, Th\'{a}kurova 7, 166 29 Prague 6, Czech Republic}
\address[2]{Department of Physics, Faculty of Civil Engineering, Czech Technical University in Prague, Th\'{a}kurova 7, 166 29 Prague 6, Czech Republic}
\cortext[cor1]{Corresponding author}
\date{}

\def\imagetop#1{\raisebox{-\height+\baselineskip}{#1}}

\newcommand{\pd}[2]{\ensuremath{\frac{\partial #1}{\partial #2}}}
\newcommand{\WangTileS}[2]{%
		\begin{scope}[shift={(#1,#2)}]
		\draw[fill=black!10,black!10] (1,1) node{}
		-- (0,1) node{}
		-- (0.5,0.5) node{}
		-- cycle;%
		\draw (0,1) -- (0.5,0.5) -- (1,1);
		\draw[white] (0,1) node{}
		-- (0,0) node{}
		-- (0.5,0.5) node{}
		-- cycle;%
		\draw (0,1) -- (0.5,0.5) -- (0,0);
		\draw[fill=black!10,black!10] (0,0) node{}
		-- (1,0) node{}
		-- (0.5,0.5) node{}
		-- cycle;%
		\draw (0,0) -- (0.5,0.5) -- (1,0);
		\draw[white] (1,0) node{}
		-- (1,1) node{}
		-- (0.5,0.5) node{}
		-- cycle;
		\draw (1,0) -- (0.5,0.5) -- (1,1);
		\end{scope}
}

\newcommand{\bmath}[1]{\ensuremath{\bm{#1}}}

\mathchardef\mhyphen="2D

\newcommand{\atx}{\ensuremath{(\tens{x})}}

\newcommand{\domain}{\Omega}

\newcommand{\norm}[1]{\left\lVert#1\right\rVert}

\newcommand{\scal}[1]{\mathnormal{#1}}
\newcommand{\tens}[1]{\bm{#1}}				            
\newcommand{\tenss}[1]{\mathbf{#1}} 					
\newcommand{\tensf}[1]{\bmath{\mathbf{#1}}} 			

\newcommand{\x}{\tens{x}} 								
\newcommand{\trn}{^\mathrm{T}}

\newcommand{\semtrx}[1]{\ifcat\noexpand#1\relax\bm{#1}\else\mathsf{#1}\fi} 					
\newcommand{\sevek}[1]{\ifcat\noexpand#1\relax\bm{#1}\else\mathsf{#1}\fi} 					

\newcommand{\Fref}[1]{Fig.~\ref{#1}}
	
\newcommand{\Eref}[1]{Eq.~(\ref{#1})}

\newcommand{\Sref}[1]{{Section~\ref{#1}}}

\newcommand{\replace}[2]{#2}

\begin{document}

\begin{abstract}
\replace{}{Topology optimization of modular structures and mechanisms enables balancing the performance of automatically-generated individualized designs, as required by Industry 4.0, with enhanced sustainability by means of component reuse.} \replace{In the optimal design of modular structures and mechanisms}{For optimal modular design}, two key questions must be answered: (i) what should the topology of individual modules be like and (ii) how should modules be arranged at the product scale? We address these challenges by proposing a bi-level sequential strategy that combines free material design, clustering techniques, and topology optimization. First, using free material optimization enhanced with checkerboard suppression, we determine the distribution of elasticity tensors at the product scale. To extract the sought-after modular arrangement, we partition the obtained elasticity tensors with a novel deterministic clustering algorithm and interpret its outputs within Wang tiling formalism. Finally, we design interiors of individual modules by solving a single-scale topology optimization problem with the design space reduced by modular mapping, conveniently starting from an initial guess provided by free material optimization. We illustrate these developments with three benchmarks first, covering compliance minimization of modular structures, and, for the first time, the design of \replace{}{non-periodic} compliant modular mechanisms. Furthermore, we design a set of modules reusable in an inverter and in gripper mechanisms, which ultimately pave the way towards the rational design of modular architectured (meta)materials.
\end{abstract}


\maketitle

\section{Introduction}
\label{sec:introduction}

\replace{}{This contribution addresses challenges emerging in the design of structures, mechanisms, and architectured materials (archimats) related to two contemporary research topics: (i) automation and individualization of the design and manufacturing processes (aligned with Industry 4.0 precepts), and (ii) environmental sustainability. In relation to the former challenge, topology optimization coupled with additive manufacturing seems to provide the definite answer \citep{zhu2021,csaldi2021}. The sustainability aspect, however, requires not only efficient material usage, but also structural reusability after the end of the structure lifetime \citep{liu2016}, repairability \citep{tajsZieliska2021} and reconfigurability. In general, these aspects have seldom been considered in topology optimization.}

\replace{}{Aligned with these goals, we believe that modularity is an ingredient that would allow topology optimization to become more sustainable by forging a pathway, via modules, towards more circular element design, because m}odular architectured materials, (self-)assembled \citep{jilek2021} from a number of patterned components called modules, boost stru\-ctu\-ral sustainability with possibly (re)configurable, multi-functional structures and mechanisms \citep{yasuda2021,ion2017,nezerka2018,tugilimana2019,wu2021b} while enabling efficient mass manufacturing through prefabrication, e.g., \citep{franchetti2016,Jenett2020}.

\begin{figure}[!t]
	\centering
	\includegraphics[width=0.35\linewidth]{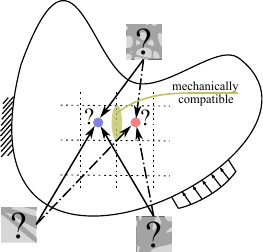}
	\caption{\replace{}{Schematical illustration of the modular-topology optimization problem.}}
	\label{fig:scheme}
\end{figure}

\replace{}{However, o}ptimal design of these modular materials is still in its infancy. In principle, the design problem involves solving a~bilevel \textit{modular-topology} optimization problem incorporating (a) the \replace{}{topological} design of \replace{the}{a small number of} module\replace{ topologie}{}s, and (b) the spatial arrangement of the modules\replace{}{, see Fig.~\ref{fig:scheme}}, both of which drive the resulting performance and versatility of the final structure. While the lower level \replace{problem }{(a)} alone introduces only a slight modification to standard topology optimization (TO) procedures \cite{tyburec2020,tugilimana2019,wu2021}, the upper-level assembly plan design \replace{}{(b)} adds another combinatorial dimension to the structure of the problem \citep{tyburec2020}. For the case of continuum topology optimization, the module design problem \replace{}{(a)} lacks convexity and appears to be strongly dependent on initial guesses, which makes a rigorous assessment of assembly plan performance challenging.

Following a brief introduction to microstructural optimization in Section~\ref{sec:intromicro}, we focus on modular designs and divide the contemporary methods for solving the modular-topology optimization problem into \replace{three}{four} groups: \replace{}{intuition-based, }multi-material-based and meta-heuristic methods \replace{(both }{}covered in Section~\ref{sec:intromodular}\replace{)}{}, and clustering-based procedures in Section~\ref{sec:introclustering}. In \replace{any of }{general, }these paradigms enable either a \textit{concurrent} design scheme, in which the two levels are solved simultaneously and converged designs are (locally) optimal with respect to both the assembly plan and the module topologies, or a \textit{sequential} scheme, in which the assembly plan design precedes the design of modules\replace{}{, or vice versa}. 

In this contribution, we develop a novel sequential clustering method for designing modular structures and mechanisms based on Wang tiling formalism, Section~\ref{sec:wangtiles}. \replace{M}{Finally, the m}ain challenges \replace{of the methods }{}and our contributions are summarized in Section~\ref{sec:noveltyFMO}.

\subsection{Microstructural optimization}\label{sec:intromicro}

Microstructural topology optimization has been a very vivid research area \citep{wu2021}, particularly due to the link to numerical homogenization \citep{bourgat1979}. However, conversely to \replace{}{the} homogenization, which evaluates \replace{}{the} effective properties of periodic media, the \textit{inverse} homogenization approach proposed by \citet{sigmund1994} has established a new framework for designing periodic materials with prescribed constitutive properties. The local periodicity requirement for homogenization-based methods relies on an infinite scale separation in order to maintain material connectedness of non-uniform microstructures and theoretical, superior performance. Because non-uniform distributions are optimal in general \citep{rodrigues2002,tyburec2020}, but since only finite-scale separations can be practically considered, the disconnectedness issue of neighboring microstructural cells has challenged researchers for years \citep{schury2012,garner2019}.

Except for periodic-unit-cell (PUC) designs, which are compatible by definition but usually provide poor structural performance \citep{liu2008,tugilimana2016,wu2021b}, several conceptual ideas have been proposed to resolve this problem for practical finite-scale separations during the last decade: prescribing connectors of neighboring, often parametric microstructures \citep{zhang2018a}, design post-processing \citep{groen2017,allaire2019,schury2012}, and maintaining natural \textit{mechanical} connectivity via single-scale optimization of the whole domain \citep{tyburec2020,tugilimana2019} or subdomains \citep{garner2019}.

Among these methods, the de-homogenization approach \citep{pantz2008,allaire2019,groen2020,geoffroydonders2020} has proven to be a~superior tool for single-load-case compliance optimization problems. With it, optimal microstructures are known to be rank-2 or rank-3 laminates whose stiffness is aligned with principal stress directions. Consequently, the optimal structures are found by projecting and smoothing optimal parametric microstructures. In more general cases such as multiple load\replace{}{ing scenarios}, modular structures, or compliant mechanism designs, the topology of optimal microstructures is not known\replace{, and hence the method cannot be applied}{. Although well-performing microstructures can still be obtained for multiple loading scenarios and compliant mechanisms using suitable microstructural parametrizations, such as fiber-reinforced composites \cite{Jung2022}, these approaches do not extend to modular structures.}

\subsection{\replace{The optimum d}{D}esign of modular \replace{}{(meta)}materials}\label{sec:intromodular}

Unlike the theoretically-optimal non-uniform distributions of microstructures, modularity limits itself to a finite small number of pattern types. As a consequence, it only approximates the optimal distribution, unless the optimal distribution is periodic. There are two extreme cases of modular designs: PUC microstructures, which repeat a single module in the design domain; and the optimal, infinite number of module types. As a result, modularity drastically reduces the number of design variables compared to the theoretically-optimal solution, but also provides a better structural response than PUC designs \citep{tyburec2020}. Furthermore, modularity may also improve the susceptibility of structures to local damage or imperfections \citep{wu2021b}.

As we have already outlined, the \replace{main}{grand} challenge in the design of modular structures and mechanisms remains a coupled design of module topologies and their connectable spatial\replace{}{ly-varying} arrangements. Hence, early methods for modular design relied on a predefined distribution of modules \replace{}{\citep{sivapuram2016,ferrer2018,alexandersen2015,alexandersen2015a}}. Although \replace{these}{the optimized} designs performed better than PUCs in general, the\replace{y were unable to handle}{ir non-optimality with respect to} the macro-scale module-assembly design problem\replace{, which}{} substantially influenced their performance.

\replace{}{
	An inverse approach has often been adopted in modular metamaterial design. The particular topologies of modules are then usually designed by intuition, combining a few design patterns, and the parametric module properties with their spatial distribution drives the exotic material response \cite{Jenett2020}, such as textured deformation \cite{coulais2016}, extreme stiffness-to-density ratio \cite{Jenett2020}, tunable Poisson ratio \cite{nezerka2018}, tunable band-gap properites \cite{wu2020}, or on-demand acoustic wave filter, guide, lens, or cloak devices \cite{yang2020}. While some of these modular metamaterials rely on lattices \cite{Jenett2017,Jenett2020,cheung2013,gregg2018,yang2018}, other are based on origami or kirigami principles \cite{silverberg2014,yang2017} possibly involving compatible mechanisms \cite{ma2022}. Benefiting from a library of metamaterials already presented in the literature, \citet{wu2020} proposed a structural-evolution-based method to generate meta-atoms of metamaterials with specified properties.
}

\replace{S}{Substituting the intuition-based design with topology optimization, s}everal of the early methods for modular-topology optimization relied on meta-heuristics coupled with numerical optimization. Among these, \citet{tugilimana2016} optimized \replace{}{first} the spatially-varying rotations of a truss module, and, ultimately, both rotations and placements of multiple module types \citep{tugilimana2019}. In the latter work, a simulated annealing was used for selecting an appropriate assembly plan, and a non-convex mathematical program for designing module topologies. Another approach was developed by \citet{tyburec2020}, who optimized truss modules and their assemblies in the form of Wang tilings. In this method, a genetic algorithm was combined with convex mathematical programming, which provided a reliable way of assessing \replace{}{and comparing} the performance of individual assembly plans.

Another class of approaches to modular design relies on concurrent multi-material-based methods. Among these, \citet{zhang2018a} developed an approach based on ordered SIMP interpolation. Its main idea encompassed a repeated solution to two subproblems: solving a multi-material distribution problem for fixed material properties, and finding the material properties for a given material distribution. Notice that the former subproblem is not considered in the above meta-heuristic approaches. A similar setting to \citep{zhang2018a} can also be found in \citep{lu2021}, where a rotation of a single parameterized unit cell was optimized for compliant mechanism problems.

\subsection{Clustering-based methods}\label{sec:introclustering}

The third class of methods relies on clustering algorithms. Because these methods are particularly relevant to us, we review relevant developments in a dedicated section. For the readers' convenience, we categorize the \replace{}{clustering} methods according to the quantities that are clustered: densities, the stress or strain tensors, and material stiffness tensors.

\paragraph{Density-based clustering} Density-based clustering determines candidate modular arrangements \replace{as}{from} clustered pseudo-densities of the variable thickness sheet problem. Since these densities usually interpolate isotropic materials, the isotropic homogenized properties of the modules are implicitly assumed. Due to the continuity of the density field, an infinite number of isotropic microstructures is needed in order to describe a general macro-domain. To effectively limit this number, \replace{}{\citet{zhang2006},} \citet{li2018} and \citet{gao2019a,gao2019b} partitioned \replace{their}{the pseudo-density field} based on predefined \replace{}{range} thresholds \replace{}{or unique values}. Each of these clustered densities was subsequently associated with a unique microstructure found\replace{}{, e.g.,} by an inverse-homogenization\replace{based on the level set }{-based} method.

A related concept was adopted in Yan~et~al.'s \citep{yan2020} concurrent method, in which the strain energy of thermo-elastic lattice structures was minimized and the discreteness of the macro-scale structure assembly was secured by the \texttt{kmeans} algorithm which clustered similar strain energies of the cells without the need to specify fixed thresholds a priori. \replace{}{In another work, \citet{Nikravesh2022} partitioned the strain energies by a~data-driven approach, with the final designs recovered using a mapping of isotropic microstructures onto the partition elements.}

\paragraph{Stress- and strain-based clustering} \replace{Because the idea of}{The previous} density-based clustering rel\replace{ies}{ying} on the assumption of iso\-tro\-pic microstructures\replace{, it }{} is usually too restrictive, since it reduces all microstructural information into a scalar field. \replace{A}{Consequently, a} more general viewpoint incorporat\replace{es}{ing} the directional information, i.e., the data provided by the stress and strain tensors\replace{}{, has emerged}.

In one of the earliest works leading towards stress-based clustering, \citet{xu2018} investigated the concurrent setting of minimum-compliant multiscale topology optimization. In their work, the requirement of multiple unique microstructures was enforced by using so-called misplaced material volume constraints, which penalized the assignments of microstructures out of their assumed range of principal stress directions. Individual microstructures were designed by the inverse homogenization approach, but lacked mutual connectivity. Another recent concurrent method from \citet{qiu2020} proposed using dynamic clustering to cluster vectors containing the direction of the principal stresses as well as their ratios. The clustering task was solved by the \texttt{kmeans} algorithm. To ensure similar local optima of the clustering algorithm among the iterations of topology optimization, the cluster centers of \texttt{kmeans} were initialized based on the previous run.

In contrast to density-based clustering, stress- or strain-based clustering usually operates without microstructural volume fraction information. This drawback has been circumvented by the combined clustering proposed by \citet{kumar2019}, who clustered the vectors containing densities and the ratios of principal strains.

\paragraph{Material stiffness clustering} However, even \replace{this}{the latter} perspective lacks full generality: it \replace{}{designs} each microstructure based on the principal directions only, i.e., rank-$3$ laminates in $3$D. Such settings thus cannot handle cases with different optimum microstructures properly. These cases include multiple loading scenarios in minimum-compliance optimization or compliant mechanism design.

The most structurally-efficient designs can only be found when exploiting all available information\replace{. F}{, i.e., f}or microstructural design, \replace{this}{}information follow\replace{s}{ing} from the fourth-order material stiffness tensor. Considering isotropic materials only, \citet{yang2019} optimized the Poisson ratios of isotropic materials by introducing free isotropic material optimization. Their work combined a hierarchical clustering algorithm with a second-order conic formulation in the transformed polyhedral $P$-$Q$ space, optimizing the structural complementary strain energy. A generalization, developed by \citet{hu2020}, explored the entire design freedom of positive definite material tensors by free material optimization (FMO), e.g.~\citep{zowe1997}. The optimal spatially-varying material tensors were clustered by a hierarchical clustering and individual micro-structures were obtained using the inverse homogenization method with embedded physics-based connectivity control.

\subsection{Wang tilings}\label{sec:wangtiles}

\begin{figure}[!htbp]
	\centering
	\begin{subfigure}{2.5cm}
		\centering
		\includegraphics[height=4cm]{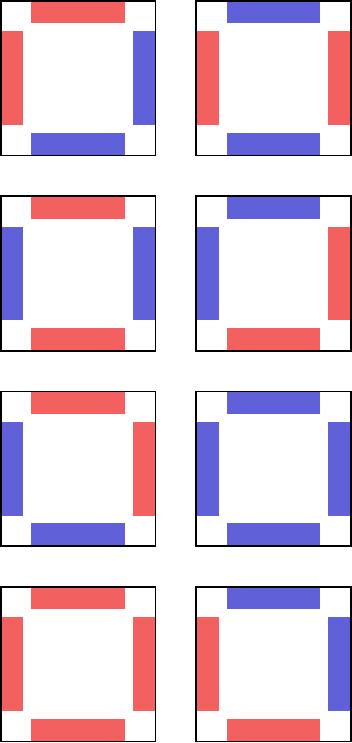}
		\caption{}
	\end{subfigure}
	\begin{subfigure}{5.5cm}
		\centering
		\includegraphics[height=4cm]{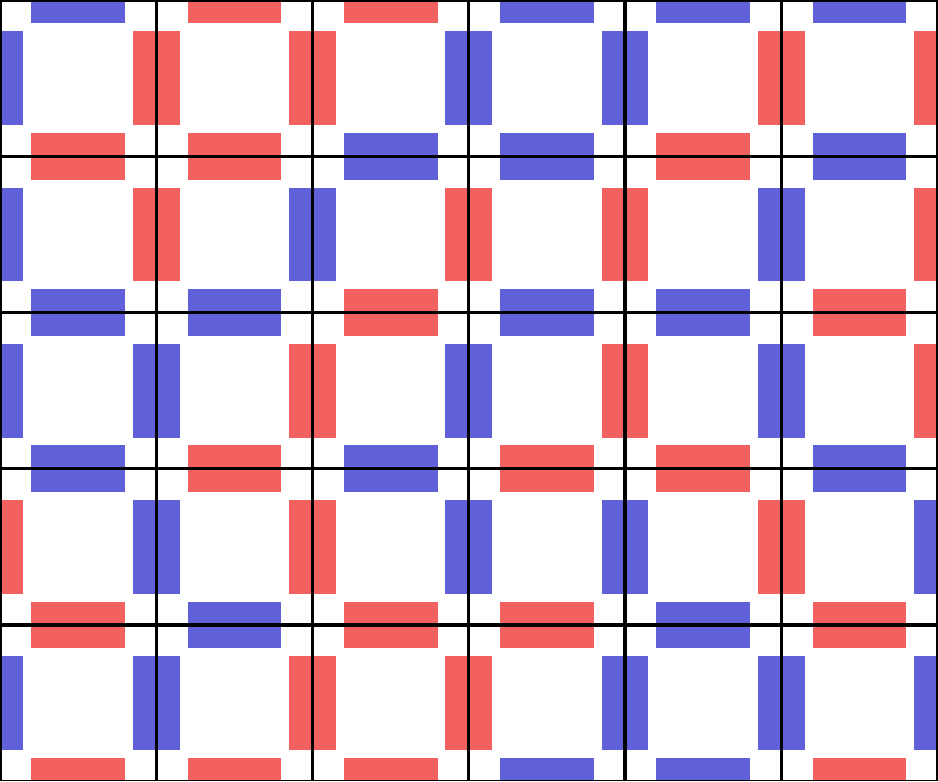}
		\caption{}
	\end{subfigure}
	\caption{Illustration of Wang tiles: (a) a tileset of $8$ tiles over two colors, and (b) a random $6 \times 5$ valid tiling satisfying the domino rule.}
	\label{fig:wt}
\end{figure}

In this work, we adopt the concept of Wang tiles---a natural generalization of the PUC---as a convenient formalism for describing modular assemblies. Visualized as unit squares with colored edges and a~fixed orientation, the concept of Wang tiles was developed by, and is named in honor of, Hao Wang to simulate the $\forall \exists \forall$ decision problem of predicate calculus in mathematical logic. In an equivalent domino problem, \citet{wang1961} assumed an arbitrary but fixed finite set of unitary tiles or dominoes and asked whether one can determine that their copies can cover an infinite plane while satisfying the domino rule of matching colors at shared edges of adjacent pieces; see \Fref{fig:wt} for an illustration.

An answer to the domino problem flourished in the work of Wang's student \citet{berger1966}, who proved that, in addition to tilesets that can cover the infinite plane by repeating periodic patterns, \textit{aperiodic} tilesets also exist which satisfy the tiling objective yet forbid any periodic arrangement. Hence, the domino problem is not decidable. Since then, numerous aperiodic tilesets have emerged; see \citep[Chapter 11]{grunbaum1987} for a thorough introduction.

Berger's original proof relied on a reduction of the Turing machine halting problem into the domino problem. This reduction implies that Wang tiles are Turing-complete, and hence their (self)-assembly can simulate desired Turing computations. This property has motivated applications, e.g., at the DNA level \citep{winfree1998,winfree2000}, and, recently, by \citet{jilek2021} on the centimeter-scale.

Visually appealing, nonperiodic yet compressed arrangements have impelled the adaptation of Wang tiles in computer graphics to generate naturally-looking textures \cite{cohen2003,zhang2008} and Poisson disk distributions \cite{cohen2003,hiller2001}. Sharing objectives, these applications have inspired the use of Wang tiles in modeling materials microstructures, both for geometrical representation~\cite{novak2012,doskar2014,doskar2020} and calculations~\cite{novak2013,doskar2016,doskar2018,doskar2021}. Because the cardinality of the Wang tileset balances the computation efficiency of the PUC with precise (micro-)structural description, Wang tilings also appear to be useful in securing automatic connectivity in the topology optimization of modular truss structures \cite{tyburec2020}.

\subsection{Aims and novelty}\label{sec:noveltyFMO}

As follows from the literature survey, the design of modular architectured materials has only recently become an active research domain in the topology optimization field. Although several significant results have already appeared, there still remain fundamental questions to be answered.

Because the overall response of modular structures follows from two constituents, with the smooth macro-domain response driven by the effective constitutive parameters that are impaired by the presence of heterogeneities \citep{doskar2021} in the form of module topologies, finding an appropriate assembly plan is the major challenge that, when successfully addressed, would allow attaining the desired structural effectiveness.

\replace{}{In our recent work \cite{tyburec2020}, we introduced a concurrent modular-topology optimization method for designing minimum-compliant modular truss structures encoded with the corner Wang tiling formalism. The method relied on a bilevel optimization involving convex conic optimization subproblems (leading to the optimal compliance for a fixed assembly plan), and a genetic algorithm to optimize the module arrangement. The convexity of the module design subproblem is indeed an essential assumption; otherwise, the assembly plan performance cannot be assessed reliably. Unfortunately, for the case of continuum topology optimization, the module design problem lacks convexity in general, so the method in \cite{tyburec2020} lacks applicability.
}

\replace{I}{Thus, i}n this contribution, we rely on the \replace{}{sequential} clustering approach, recall \Sref{sec:introclustering}, operating over material stiffnesses, and a standard single-scale topology optimization \citep{bendsoe2003} extended to structural modularity. Each of our modules is defined as a quadruple of edge labels, thus defining a Wang tile, recall \Fref{fig:wt}. In this dual point of view, we ask \textit{how do we place at most $m$ different interface types to make the resulting modular structure behave efficiently?} The answer is indeed in the form of a valid Wang tiling over at most $m$ colors.

To make our setting fully general, our clustering dataset originates from a solution to an FMO problem, which we extend to handle both minimum-compliance and compliant mechanism design objectives, \Sref{sec:formulation}. To the best of our knowledge, the procedures for designing modular assembly plans in linear elasticity problems have mostly been limited to minimum-compliance optimization. The only exception we are aware of is \citep{lu2021}, optimizing rotations of a single unit cell for compliant mechanisms via multi-material topology optimization. However, a more general setting has already been requested \citep{wu2021b}.

Surprisingly, as we will show in \Sref{sec:checkerboard}, the optimal elastic properties obtained from the FMO problem often contain checkerboard patterns. Nevertheless, this artifact has not yet been investigated with connection to the FMO problems, although it has appeared in published results, e.g., \citep{hu2020}. \replace{Since we only utilize the elastic properties for a heuristic guidance, we adopt a simple post-processing step and postpone the rigorous treatment of the checkerboard patterns in FMO to future works.}{Here, we mitigate the issue by enhancing the kinematic response though a refined design-element mesh.}

The objective of the clustering algorithms is to partition a given dataset of points into a specific number of clusters. However, standard clustering algorithms have several disadvantages: they are usually stochastic and, therefore, require multiple evaluations to generate quality outputs. Moreover, they also generally fail to maintain structural symmetries. To overcome these shortcomings, we have developed a novel greedy agglomerative clustering algorithm that is deterministic while preserving structural symmetry of the elasticity field, when present (\Sref{sec:clustering}). Using this clustering algorithm on a module interface-based discretization produces a valid Wang tiling.

For appropriate assembly plans obtained via the clustering procedure, the module topologies must also be optimized. Unfortunately, the emergent optimization problems appear to be strongly dependent on the provided starting point. To make the initial choice problem-specific yet well-defined, we propose to use a solution to a macro-scale FMO problem with modularity to construct initial guesses based on the traces of the optimal stiffness tensors obtained from FMO (\Sref{sec:modularityconstr}). A single-scale topology optimization extended to structural modularity is finally used for a computation of the optimized modular topologies in \Sref{sec:topopt}.

We illustrate the developed theory on four selected examples, including minimum-compliance optimization of the MBB beam problem, \Sref{sec:mbb}, and three compliant mechanisms: inverter (\Sref{sec:inverter}), gripper (\Sref{sec:gripper}), and a module set reusable among both the inverter and gripper mechanisms (\Sref{sec:reusable}). We conclude that our approach to designing modular structures provides fairly efficient and well-behaving designs.

In this contribution, we adopt the following notation: scalar variables are written in plain, lower-case italic letters ($a$). First- and second-order tensors appear in lower-case, bold italic ($\tens{a}$) and bold regular ($\tenss{a}$) letters, respectively, and the fourth-order tensors are denoted as bold capital regular letters ($\tensf{A}$). Finally, column vectors and matrices appear in sans-serif lower and upper-case bold characters ($\sevek{a}$, $\semtrx{A}$).

\section{Methods}\label{sec:methods}

This section is devoted to establishing a sequential method that provides a heuristic solution to the modular-topology optimization problem with the modular assembly plan parameterized as valid Wang tilings, recall \Sref{sec:wangtiles}. Our approach to designing modular structures and mechanisms encompasses \replace{five}{four} sequential steps: (i) optimum free material design performed on a module interface-based mesh, \Sref{sec:formulation}, \replace{}{including a proper handling of emerging checkerboard patterns (Section \ref{sec:checkerboard}),} (ii) \replace{removal of potential modules whose interfaces lack any significant stiffness, and (iii)}{and} clustering of \replace{}{the optimal} elasticity matrices by a newly-developed agglomerative clustering algorithm to output valid Wang tilings, \Sref{sec:clustering}. For the resulting assembly plan, we (i\replace{v}{ii}) optimize the material properties of individual modules via an FMO extended to modularity, \Sref{sec:modularityconstr}, which provides an appropriate initial guess for (\replace{}{i}v) the standard continuum topology optimization extended to modularity in \Sref{sec:topopt}.

\subsection{Assembly plan design}\label{sec:assemblyplan}

Based on the inherent bilevel nature of the modular-topology optimization problem, we split the above \replace{six}{four} steps into two parts---a heuristic procedure for assembly plan design, and then topology optimization of the modules. In this section, we aim to answer the question of \textit{how should we design efficient assembly plans}? To this goal, we introduce free material optimization in \Sref{sec:formulation}, \replace{}{handling of checkerboard patterns in \Sref{sec:checkerboard}}, and a novel clustering procedure in \Sref{sec:clustering}.

\subsubsection{Free material optimization}\label{sec:formulation}

Let us consider a two-dimensional continuum body $\Omega$ under appropriate static and kinematic boundary conditions. In this domain, the free material optimization (FMO) method seeks the spatially-varying material elasticity tensors $\tensf{E}$ that make the structural response optimal with respect to a specified performance functional $a(\tens{u})$ \citep{zowe1997} of a displacement field $\tens{u}$.

In this manuscript, we assume a linear strain-displacement relation
\begin{equation}\label{eq:strain}
\tenss{\varepsilon}_{ij} \atx = \frac{1}{2} \left( \frac{\partial u_i\atx}{\partial x_j} + \frac{\partial u_j\atx}{\partial x_i} \right)
\,,\quad
\forall \x \in \domain\,,
\end{equation}
where $\tenss{\varepsilon}$ denotes a strain tensor. The sought elastic material properties are represented by the effective elasticity tensor $\tensf{E}$ that satisfies linear Hooke's law written in indicial notation as
\begin{equation}\label{eq:Hooke_tens}
\scal{\sigma}_{ij}\atx = \scal{E}_{ijkl}\atx \, \scal{\varepsilon}_{kl}\atx,
\quad
\forall \x \in \domain\,,
\end{equation}
in which $\tenss{\sigma}$ is a stress tensor.

In what follows, we adopt the (engineering) Voigt notation that allows us to rewrite the tensors $\tenss{\varepsilon}$, $\tenss{\sigma}$, and $\tensf{E}$ as column and square matrices, respectively. To this goal, let 
\begin{equation}
\sevek{\varepsilon}\atx = \begin{pmatrix}
\varepsilon_{11}\atx & \varepsilon_{22}\atx & 2 \varepsilon_{12}\atx
\end{pmatrix}^\mathrm{T} = \begin{pmatrix}
\varepsilon_{11}\atx & \varepsilon_{22}\atx & \gamma_{12}\atx
\end{pmatrix}^\mathrm{T}
\end{equation}
be the strain column matrix, with $\gamma_{12}$ being the engineering shear strain, and let
\begin{equation}
\tenss{\sigma}\atx = \begin{pmatrix}
\sigma_{11}\atx & \sigma_{22}\atx & \sigma_{12}\atx
\end{pmatrix}^\mathrm{T} = \begin{pmatrix}
\sigma_{11}\atx & \sigma_{22}\atx & \tau_{12}\atx
\end{pmatrix}^\mathrm{T}
\end{equation}
denote the stress matrix. Then, the matrix form of the Hooke law \eqref{eq:Hooke_tens} reads as
\begin{equation}
\sevek{\sigma}\atx = \semtrx{E}\atx \, \sevek{\varepsilon}\atx
\,,\quad
\forall \x \in \domain\,,
\end{equation}
with the material stiffness matrix
\begin{equation}
\semtrx{E}\atx = \begin{pmatrix}
\scal{E}_{1111}\atx & \scal{E}_{1122}\atx & \scal{E}_{1112}\atx\\
\scal{E}_{1122}\atx & \scal{E}_{2222}\atx & \scal{E}_{2212}\atx\\
\scal{E}_{1112}\atx & \scal{E}_{2212}\atx & \scal{E}_{1212}\atx
\end{pmatrix}
\end{equation}
constructed from the (to-be-optimized) components $\scal{E}_{ijkl}$ of the forth-order elasticity tensor $\tensf{E}$. To simplify the further developments, we introduce a vector form of \replace{$\semtrx{E}$}{the fourth-order tensor $\tensf{E}$} as
\begin{equation}
\sevek{e}\atx = \begin{pmatrix}
\scal{E}_{1111}\atx & \sqrt{2} \scal{E}_{1122}\atx & \scal{E}_{2222}\atx & \replace{\sqrt{2}}{2} \scal{E}_{1112}\atx & \replace{\sqrt{2}}{2} \scal{E}_{2212}\atx & \replace{}{2}\scal{E}_{1212}\atx
\end{pmatrix}^\mathrm{T}
\end{equation}
such that
\begin{equation}\label{eq:normeq}
\left\lVert \replace{\semtrx{E}}{\tensf{E}}\atx \right\rVert_\text{F} = \lVert \sevek{e}\atx \rVert_2 \,,
\end{equation}
where $\lVert \bullet \rVert_2$ denotes the standard Euclidean norm of the vector $\bullet$, and $\lVert \bullet \rVert_\text{F}$ is the Frobenius norm \replace{}{of the forth-order tensor}.

Our approach to a numerical solution of the FMO optimization problems relies on approximate solutions to the governing equation obtained with the finite element method: we search for the kinematically admissible displacement field $\sevek{u} \in \mathbb{R}^{n_\mathrm{dof}}$ satisfying the equilibrium equations
\begin{equation}\label{eq:equilibrium}
\semtrx{K}\left(\semtrx{E}^{(1)},\dots,\semtrx{E}^{(n_\mathrm{\ell})}\right) \sevek{u} = \sevek{f},
\end{equation}
where $\sevek{f} \in \mathbb{R}^{n_\mathrm{dof}}$ denotes the column matrix of equivalent nodal forces, $n_\mathrm{dof}$ stands for the number of degrees of freedom, the super-script $\bullet^{(\ell)}$ expresses an assignment of $\bullet$ to the element $\ell$, and $\semtrx{K}\left(\semtrx{E}^{(1)},\dots,\semtrx{E}^{(n_\mathrm{\ell})}\right) \in \mathbb{S}_{\succ 0}^{n_\mathrm{dof}}$ is the symmetric, positive definite structural stiffness matrix---a function of the discretized field of elasticity matrices $\semtrx{E}^{(\ell)} \in \mathbb{S}^{3}_{\succ 0}$ of $n_\mathrm{\ell}$ finite elements. In this notation, $\mathbb{S}^{\bullet}$ refers to a space of symmetric square matrices of the size $\bullet \times \bullet$, and $\succ$ denotes positive definiteness. In addition, we define $\semtrx{X} \succeq \scal{\epsilon}$ to express that the eigenvalues of $\semtrx
{X}$ are greater than or equal to $\scal{\epsilon}$.

The structural stiffness matrix $\semtrx{K}$ follows from an assembly of element stiffness matrices $\semtrx{K}^{(\ell)}$ based on the element gather matrices $\semtrx{L}^{(\ell)}$ and a design-independent term $\semtrx{K}_{0} \in \mathbb{S}^{n_\mathrm{dof}}_{\succeq 0}$,
\begin{equation}\label{eq:assembly}
\semtrx{K}\left(\semtrx{E}^{(1)},\dots,\semtrx{E}^{(n_\mathrm{\ell})}\right) = \semtrx{K}_0 + \sum_{\ell=1}^{n_\mathrm{\ell}} \semtrx{L}^{(\ell)\mathrm{T}} \semtrx{K}^{(\ell)}(\semtrx{E}^{(\ell)}) \semtrx{L}^{(\ell)},
\end{equation}
where the element contributions $\semtrx{K}^{(\ell)} \left(\semtrx{E}^{(\ell)}\right) \in \mathbb{S}_{\succeq 0}$ evaluate as
\begin{equation}\label{eq:ke}
\semtrx{K}^{(\ell)} \left(\semtrx{E}^{(\ell)}\right) = \int_{\Omega_\ell} \semtrx{B}^{(\ell)\mathrm{T}}\atx \, \semtrx{E}^{(\ell)}\atx \, \semtrx{B}^{(\ell)}\atx \mathrm{d} \Omega.
\end{equation}
In \eqref{eq:ke}, $\semtrx{B}^{(\ell)}$ denotes the strain-displacement matrix arising from the adopted displacement approximation, and $\Omega_\ell$ constitutes the element domain. From now on, we assume $\semtrx{E}^{(\ell)}$ to be constant over each element, so that $\semtrx{K}^{(\ell)}$ is a linear function of $\semtrx{E}^{(\ell)}$.

In the spirit of standard topology optimization (TO) \cite{bendsoe2003}, FMO seeks the most efficient distribution of a limited amount of material, constrained by Eqs. \eqref{eq:fmo_volume}--\eqref{eq:fmo_eig}, that satisfies the (elastic) equilibrium, Eq. \eqref{eq:fmo_mutual}. However, compared to TO, the FMO setting opens up a broader design freedom, because the full material stiffness matrix is the subject of optimization. Here, we formalize the FMO problems of our interest as
\begin{subequations}\label{eq:fmo}
	\begin{alignat}{4}
	\min_{s, \semtrx{E}^{(1)}, \dots, \semtrx{E}^{(n_\mathrm{\ell})}}\; && s\hspace{40mm}\label{eq:fmo_obj}\\
	\text{s.t.}\; && s-\sevek{a}^\mathrm{T} \left[\semtrx{K}\left(\semtrx{E}^{(1)},\dots,\semtrx{E}^{(n_{\mathrm{\ell}})}\right)\right]^{-1} \sevek{f} &&\;\ge\;& 0,\label{eq:fmo_mutual}\\
	&& \overline{V}\overline{\epsilon} \sum_{\ell=1}^{n_\mathrm{\ell}} v^{(\ell)} - \sum_{\ell=1}^{n_\mathrm{\ell}} \left[v^{(\ell)}\mathrm{Tr} \left(\semtrx{E}^{(\ell)}\right)\right] &&\;\ge\;& 0,\label{eq:fmo_volume}\\
	&& \overline{\epsilon} - \mathrm{Tr}\left(\semtrx{E}^{(\mathrm{\ell})}\right) &&\;\ge\;& 0, &\;\;\forall \ell \in \{1,\dots,n_\mathrm{\ell}\},\label{eq:fmo_eltrace}\\
	&& \semtrx{E}^{(\ell)} &&\;\succeq\;& \underline{\epsilon}, &\;\;\forall \ell \in \{1,\dots,n_\mathrm{\ell}\}.\label{eq:fmo_eig}
	\end{alignat}
\end{subequations}
In \eqref{eq:fmo}, $s \in \mathbb{R}$ is a slack variable and $\sevek{a} \in \mathbb{R}^{n_\mathrm{dof}}$ stands for a fixed vector representing the discretized performance of the functional $a$. Furthermore, $\underline{\epsilon} \in \mathbb{R}_{>0}$ and $\overline{\epsilon} \in \mathbb{R}_{>0}$ are lower- and upper-bounds on the (sum of the) eigenvalues of $\semtrx{E}^{(\ell)}$, $\forall \ell \in \{
1,\dots, n_\mathrm{\ell}\}$, $\overline{V} \in \langle 0,1 \rangle$ stands for the volume fraction, $v^{(\ell)} \in \mathbb{R}_{>0}$ represents the volume of the $\ell$-th element, and $\mathrm{Tr}$ is the trace operator, amounting to the sum of its argument's eigenvalues. Notice that in contrast to established formulations \cite{zowe1997,kocvara2008,hu2020}, we use an element volume-based weighted average in \eqref{eq:fmo_volume} to facilitate optimization with non-structured spatial discretizations.

When setting $\sevek{a} = \sevek{f}$, the slack variable $s$ becomes an upper bound on the compliance functional, \eqref{eq:fmo_mutual}, rendering \eqref{eq:fmo} to be a compliance minimization problem. Such optimization problem is convex: the functions in \eqref{eq:fmo_obj} and \eqref{eq:fmo_volume}-\eqref{eq:fmo_eig} are linear (matrix) functions, and \eqref{eq:fmo_mutual} possesses a positive definite Hessian, see Appendix \ref{app:hessian}, and it is therefore also convex. For the case of $\sevek{a} \neq \sevek{f}$, the second term in function \eqref{eq:fmo_mutual} amounts to $\sevek{a}^\mathrm{T} \sevek{u}$. Consequently, formulation \eqref{eq:fmo} facilitates optimization of compliant mechanisms by minimizing weighted sums of the displacement vector entries.

Considering Eqs.~\eqref{eq:fmo_volume} and \eqref{eq:fmo_eig}, a natural question arises of how to choose $\underline{\epsilon}$ and $\overline{\epsilon}$. Clearly, the lower bound for eigenvalues must be sufficiently small to mimic void materials but should be large enough to avoid problems with a possibly non-existing inverse in \eqref{eq:fmo_mutual}. In our case, we fix $\underline{\epsilon}= 10^{-3} \overline{\epsilon}$. To determine the upper-bound on the sum of eigenvalues, $\overline{\epsilon}$, we link the FMO problem \eqref{eq:fmo} with our ultimate goal of designing microstructures made of a mixture of an isotropic material and void, optimized with standard continuum topology optimization. Assuming a family of isotropic materials under plane stress that are parameterized by the Young modulus $E^{(\ell)} \in \langle 0, 1 \rangle$ and a fixed Poisson ratio $\nu$, the trace of the isotropic material stiffness matrix $\semtrx{E}_{\mathrm{iso}}^{(\ell)}$ amounts to
\begin{equation}
\mathrm{Tr}(\semtrx{E}_{\mathrm{iso}}^{(\ell)}) = \frac{5-\nu}{2 (1-\nu^2)}E^{(\ell)}.
\end{equation}
If the entire design domain is filled with the solid material, $\overline{V}=1$, the trace constraint \eqref{eq:fmo_volume} implies that
\begin{equation}
\overline{\epsilon} = \frac{5-\nu}{2 (1-\nu^2)}.
\label{eq:trace_upper_bound}
\end{equation}
With the theoretical bounds on the Poisson ratio $\nu \in ( -1, 0.5 \rangle$, Eq.~\eqref{eq:trace_upper_bound} implies that for all solid isotropic materials under plane stress $\overline{\epsilon} \in \langle \frac{5}{4} + \frac{\sqrt{6}}{2}, \infty)$.

The optimization problem \eqref{eq:fmo} can readily be solved by nonlinear semidefinite programming solvers such as \textsc{Penlab} \citep{fiala2013} or \textsc{Pennon} \citep{kocvara2003}. However, for the special case of $\sevek{a}=\sevek{f}$, one can also adopt any of the more widely available (convex) linear semidefinite programming optimizers. In this contribution, we rely on the \textsc{Pennon} solver kindly provided by its authors. For compliance minimization problems, we use both the first- and second-order derivatives, see Appendix~\ref{app:sensitivities}. For compliant mechanism problems, it appeared to be more efficient to use an identity matrix instead of the exact Hessian term from Eq.~\eqref{eq:hessian}. In addition, practical computations revealed that it is more beneficial to solve a scaled version of \eqref{eq:fmo}, we refer the reader to Appendix~\ref{app:scaling} for details.

\subsubsection{Checkerboard pattern}\label{sec:checkerboard}

\begin{figure}[!t]
	\begin{minipage}{0.965\linewidth}
		\begin{subfigure}{0.5\linewidth}
			\centering
			\begin{tikzpicture}
			\node[inner sep=0pt] (im) at (0,0)
			{\includegraphics[width=0.825\linewidth]{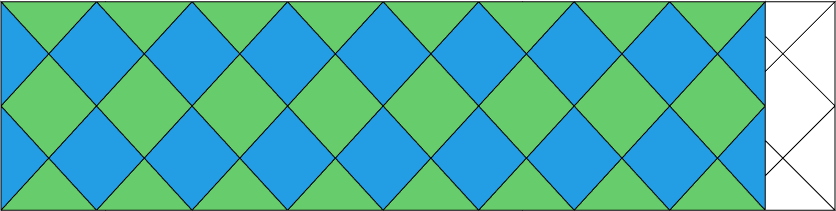}};
			\begin{scope}[color=gray]
			\coordinate [left=0.3 of im.north east] (l1);
			\coordinate [left=0.3 of im.south east] (l2);
			\coordinate [below=0.3 of im.west] (sb);
			\coordinate [above=0.3 of im.west] (sa);
			\lineload{1}{l1}{l2}[0.3][0.3][0.2];
			\support{3}{sa}[270];
			\support{3}{sb}[270];
			\end{scope}
			\end{tikzpicture}
			\caption{}
		\end{subfigure}%
		\begin{subfigure}{0.5\linewidth}
			\centering
			\begin{tikzpicture}
			\node[inner sep=0pt] (im) at (0,0)
			{\includegraphics[width=0.825\linewidth]{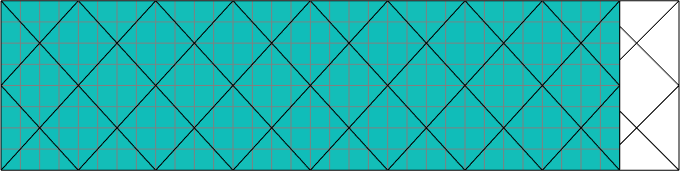}};
			\begin{scope}[color=gray]
			\coordinate [left=0.3 of im.north east] (l1);
			\coordinate [left=0.3 of im.south east] (l2);
			\coordinate [below=0.3 of im.west] (sb);
			\coordinate [above=0.3 of im.west] (sa);
			\lineload{1}{l1}{l2}[0.3][0.3][0.2];
			\support{3}{sa}[270];
			\support{3}{sb}[270];
			\end{scope}
			\end{tikzpicture}
			\caption{}
		\end{subfigure}\\
		\begin{subfigure}{0.5\linewidth}
			\centering
			\begin{tikzpicture}
			\node[inner sep=0pt] (im) at (0,0)
			{\includegraphics[width=0.825\linewidth]{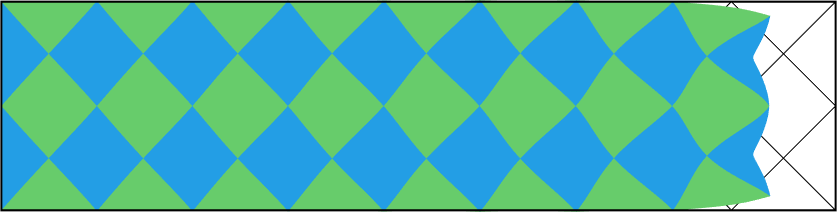}};
			\begin{scope}[color=gray]
			\coordinate [left=0.3 of im.north east] (l1);
			\coordinate [left=0.3 of im.south east] (l2);
			\coordinate [below=0.3 of im.west] (sb);
			\coordinate [above=0.3 of im.west] (sa);
			\lineload{1}{l1}{l2}[0.3][0.3][0.2];
			\support{3}{sa}[270];
			\support{3}{sb}[270];
			\end{scope}
			\end{tikzpicture}
			\caption{}
		\end{subfigure}%
		\begin{subfigure}{0.5\linewidth}
			\centering
			\begin{tikzpicture}
			\node[inner sep=0pt] (im) at (0,0)
			{\includegraphics[width=0.825\linewidth]{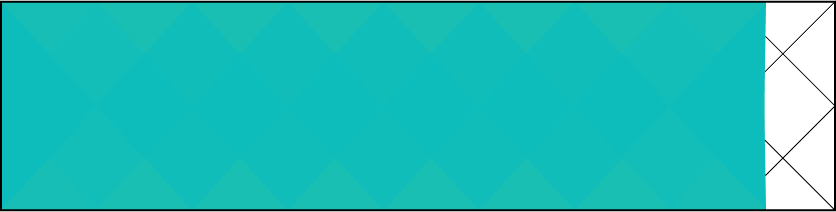}};
			\begin{scope}[color=gray]
			\coordinate [left=0.3 of im.north east] (l1);
			\coordinate [left=0.3 of im.south east] (l2);
			\coordinate [below=0.3 of im.west] (sb);
			\coordinate [above=0.3 of im.west] (sa);
			\lineload{1}{l1}{l2}[0.3][0.3][0.2];
			\support{3}{sa}[270];
			\support{3}{sb}[270];
			\end{scope}
			\end{tikzpicture}
			\caption{}
		\end{subfigure}
	\end{minipage}%
	\begin{minipage}{0.035\linewidth}
		\includegraphics[width=\linewidth]{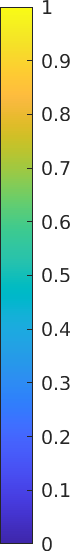}\vspace{4mm}
	\end{minipage}
	\caption{A bar of length $4$ and height $1$ under uniaxial compression, designed for $\overline{V} = 0.5$ and $\overline{\epsilon} = 1$ with visualized traces of elasticity matrices and deformed shape. (a) Optimum design obtained by solving an FMO problem \replace{}{using design element discretization} ($s_{\mathrm{FMO}}=32.13$) and (c) its projection on a \replace{}{$100\times$ re}fine\replace{r}{d} mesh ($s_\mathrm{FMO}^{\mathrm{refined}}=34.33$), and \replace{}{(b) optimum FMO} design \replace{after post-processing}{obtained for a refined design element mesh} ($s_\mathrm{FMO,f}=32.13$) and (d) \replace{projected}{its projection} onto a \replace{}{$100\times$} refined mesh ($s_\mathrm{FMO,f}^{\mathrm{refined}}=32.13$).}
	\label{fig:bar}
\end{figure}
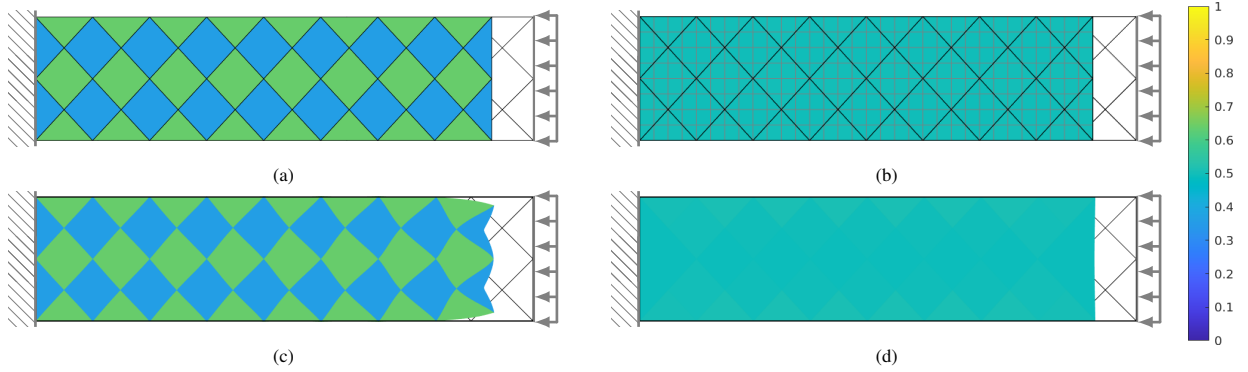

The FMO generates (locally-)optimal elasticity matrices $\semtrx{E}^{(\ell)}$ that are, however, prone to the checkerboard pattern phenomenon \citep{daz1995}, especially in the case of the bilinear quadrilateral finite elements, which are the most widely used elements in the TO community \citep{andreassen2010}. In Fig.~\ref{fig:bar}, we show this emerging issue on traces of the optimal material stiffnesses in compliance optimization of a bar under uniaxial compression. Please note that while this issue is apparent at the element stiffness traces in this case, it may be difficult to recognize in general because FMO operates directly over stiffness components and, hence, it has no scalar unknown field as in standard density-based TO, and the checkerboard may be lost as a result of visualization.

There are probably multiple reasons why the checkerboard issue occurs. First, we optimize over element design variables, whereas the finite element system \eqref{eq:equilibrium} operates over a nodal-based displacement field. Therefore, the output of \eqref{eq:fmo} shall be interpreted as optimal discretized (nodal) stiffnesses in the form of the stiffness matrix $\semtrx{K}\left(\semtrx{E}^{(1)},\dots,\semtrx{E}^{(n_\mathrm{\ell})}\right)$, which seems to be checkerboard-free. However, discretized nodal stiffnesses result from a (weighted) sum of the stiffnesses of connected elements, recall \eqref{eq:assembly}. Clearly, the element stiffnesses do not play any role unless they change a value in their assembled sum $\semtrx{K}\left(\semtrx{E}^{(1)},\dots,\semtrx{E}^{(n_\mathrm{\ell})}\right)$. Hence, the element stiffnesses may indeed form a checkerboard pattern, which is often falsely recognized in finite element analyses as structurally more efficient \citep{daz1995}. We note here that associating the design variables with the nodes does not resolve the problem, because the element stiffnesses \eqref{eq:ke} are usually evaluated at the Gauss points inside the finite element interior, not at the nodal points. Furthermore, the bilinear quadrilateral elements used in this manuscript are known to be prone to locking, and, therefore, they tend to overestimate the shear effects \cite{prathap1985,zienkiewicz2013}. Again, this may result in a load transfer that is nonphysical.

To our knowledge, the checkerboard issue has not been investigated with connection to the FMO problem \eqref{eq:fmo} yet, probably because only regular orthogonal mesh discretizations have been adopted, and, in that case, the issue is \replace{usually}{often} not visually significant but indeed occurs, see e.g., \citep[Figs. 4b and 4e]{hu2020}. For another discretizations, such as the rotated one used in this manuscript, Fig.~\ref{fig:bar}, the checkerboard is more apparent.

Traditionally, two approaches have been adopted in the density-based topology optimization to mitigate the checkerboard patterns: the sensitivity and density filters \citep{bendsoe2003}. These filters suppress the checkerboard pattern by a convolution of the first-order derivatives, or of the pseudo-density field. They can also be applied to the FMO problem directly by filtering the entries of the elasticity matrices independently because the space of positive semi-definite matrices is convex. \replace{However, we adopt a different, simpler approach. Because we use FMO as a heuristic tool only, we perform a single post-processing step. In addition to being computationally more efficient, our post-processing maintains near-optimality in the structural performance for sufficiently fine discretizations.
	
	With $n_\mathrm{n}$ denoting the number of nodes, let $\semtrx{F} \in \mathbb{R}^{n_\mathrm{\ell} \times n_{\mathrm{n}}}_{\ge 0}$ be a \textit{filtering} matrix, whose $\ell$-th row is associated with the element $\ell$ and $j$-th column belongs to the node $j$. Components of $\semtrx{F}$ follow from 
	\begin{equation}
	\scal{F}_{\ell j} = \int_{\Omega_\ell} N_j \atx {\mathrm{d}\Omega}\,,
	\end{equation}
	where $N_j$ is the shape function associated with the $j$-th node. Operating coefficient-wise, we define a vector $\sevek{e}_k = \left(e_{k}^{(1)}, \dots, e_{k}^{(n_\mathrm{\ell})}\right)^\mathrm{T}$ collecting $k$-th component across all elements $\ell$ and perform post-processing according to
	\begin{equation}\label{eq:filter}
	\widehat{\sevek{e}}_{k} = 
	\left[\mathrm{diag}\left(\semtrx{F} \sevek{1}\right)^{-1} \semtrx{F} \right] 
	\left[\mathrm{diag}\left(\semtrx{F}^\mathrm{T} \sevek{1}\right)^{-1} \semtrx{F}^\mathrm{T} \right] 
	\sevek{e}_{k},
	\end{equation}
	with $\semtrx{1}$ being an all-one column vector of appropriate dimensions and the $\mathrm{diag}()$ operator constructing a square matrix with its argument vector placed at the major diagonal. The procedure in \eqref{eq:filter} consists of two projections: interpolating element material stiffnesses onto nodal material stiffnesses by $\left[\mathrm{diag}\left(\semtrx{F}^\mathrm{T} \sevek{1}\right)^{-1} \semtrx{F}^\mathrm{T} \right] $, and a backwards averaging of nodal material stiffnesses onto the element ones, $\left[\mathrm{diag}\left(\semtrx{F} \sevek{1}\right)^{-1} \semtrx{F} \right]$. Because both these projections are convex, such post-processing preserves the trace value in the resource constraint \eqref{eq:fmo_volume} and the convex eigenvalue constraints \eqref{eq:fmo_eig} and \eqref{eq:fmo_eltrace}.
	
	Nevertheless, the post-processing step may impair the quality of the objective function value \eqref{eq:fmo_obj}. Our numerical experiments with projecting the originally element-wise constant material properties onto a refined mesh, e.g. $800\times200$ elements in Fig.~\ref{fig:bar}c, leads to significant discrepancy in the predicted objective when the post-processing is not used. On the other hand, with post-processed densities being projected, the detoriation of the performance is less pronounced, and more importantly the objective is lower compared to a refined design based on the non-filtered design. In addition, the increase caused by the post-processing is mostly aligned with the fineness of the finite element discretization. Because optimal material designs are locally periodic in the limit \citep{bendsoe1988}, we have $\widehat{\sevek{e}}_{k} \rightarrow \sevek{e}_{k}$. In fact, a zero increase occurs for any periodic problem, even with a coarse discretization, recall Fig.~\ref{fig:bar}. Therefore, there is a zero increase in the objective function value for infinitely-fine discretizations.}{}

\replace{}{Another option is to enhance the kinematic response by adopting higher-order finite elements, or by a refined design-element discretization, where the original mesh is used for the material design variables and a refined variant for the evaluation of structural response \citep{Mukherjee2021}. Here, we follow the latter approach and adopt the refined discretization shown in Fig.~\ref{fig:bar}b.
}

\subsubsection{Clustering for modular Wang tile design}\label{sec:clustering}

\replace{Having smoothed the material stiffness field}{Using the optimal, checkerboard-free material stiffness field}, our next goal is to determine structurally-efficient modular assemblies in the form of valid Wang tilings. In this work, we follow the direction of clustering the material stiffnesses because it \replace{is}{represents} the most general option: it generalizes both the density clustering for isotropic materials with a fixed Poisson ratio and the strain-based clustering for problems possessing optimal design in the form of rank-3 laminates. Our approach thus extends to the remaining linear elasticity problems including multi-load designs and compliant mechanisms. To this goal, this section develops a greedy agglomerative clustering algorithm that maintains structural symmetries when present.

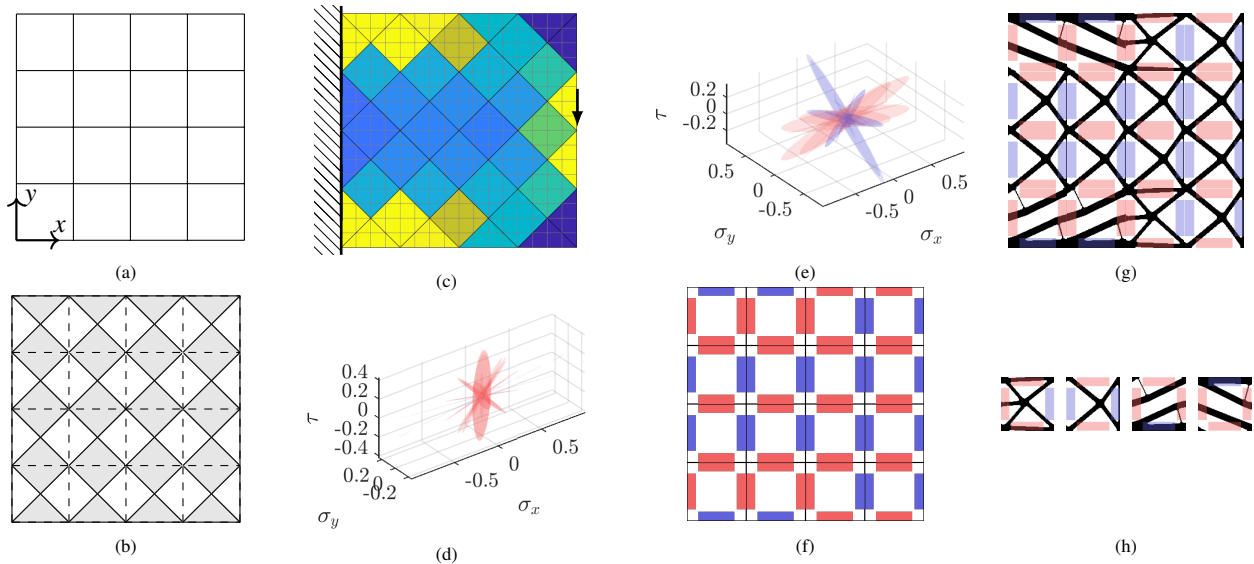
\begin{figure}[!t]
	\begin{minipage}{0.2\linewidth}
		\begin{subfigure}{\linewidth}
			\centering
			\vbox to 31mm {\vfil%
				\begin{tikzpicture}[scale=0.75]
				\draw[step=1.0,black,thin] (0.0,0.0) grid (4.0,4.0);
				\draw[->,thick] (0.0,0.0)--(0.75,0.0);
				\draw[->,thick] (0.0,0.0)--(0.0,0.75);
				\node at (0.75,0.25) {$x$};
				\node at (0.25,0.75) {$y$};
				\end{tikzpicture}%
				\vfil}
			\caption{}
		\end{subfigure}\\
		\begin{subfigure}{\linewidth}
			\centering
			\vbox to 31mm {\vfil%
				\begin{tikzpicture}[scale=0.75]
				\WangTileS{0}{0};\WangTileS{1}{0}; \WangTileS{2}{0}; \WangTileS{3}{0};
				\WangTileS{0}{1};\WangTileS{1}{1}; \WangTileS{2}{1}; \WangTileS{3}{1};
				\WangTileS{0}{2};\WangTileS{1}{2}; \WangTileS{2}{2}; \WangTileS{3}{2};
				\WangTileS{0}{3};\WangTileS{1}{3}; \WangTileS{2}{3}; \WangTileS{3}{3};
				\draw[step=1.0,black,thin,dashed] (0.0,0.0) grid (4.0,4.0);
				\draw (0,0) rectangle (4,4);
				\end{tikzpicture}%
				\vfil}
			\caption{}
		\end{subfigure}%
	\end{minipage}%
	\hfill\begin{minipage}{0.25\linewidth}
		\begin{subfigure}{\linewidth}
			\centering
			\begin{tikzpicture}
			\node[inner sep=0pt] (im) at (0,0)
			{\includegraphics[height=3.1cm]{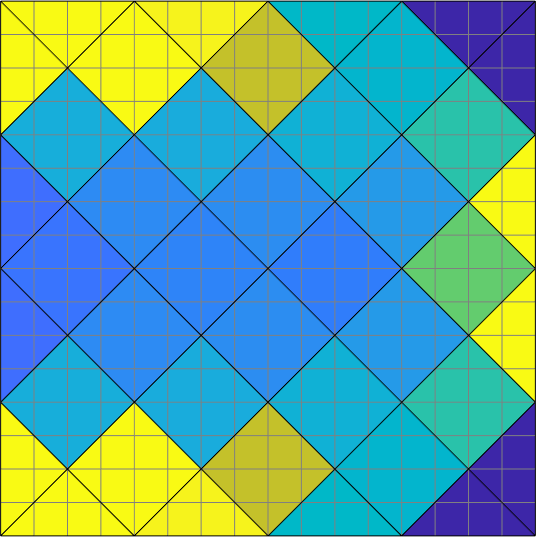}};
			\coordinate [below=0.1 of im.east] (l);
			\coordinate [below=0.6 of im.west] (sb);
			\coordinate [above=0.6 of im.west] (sa);
			\coordinate [above=1.05 of im.west] (sc);
			\coordinate [below=1.05 of im.west] (sd);
			\load{1}{l}[90][0.5][0];
			\support{3}{sa}[270];
			\support{3}{sb}[270];
			\support{3}{sc}[270];
			\support{3}{sd}[270];
			\end{tikzpicture}
			\caption{}
		\end{subfigure}\\
		\begin{subfigure}{\linewidth}
			\centering
			\vbox to 31mm {\vfil%
				\includegraphics[height=3.1cm]{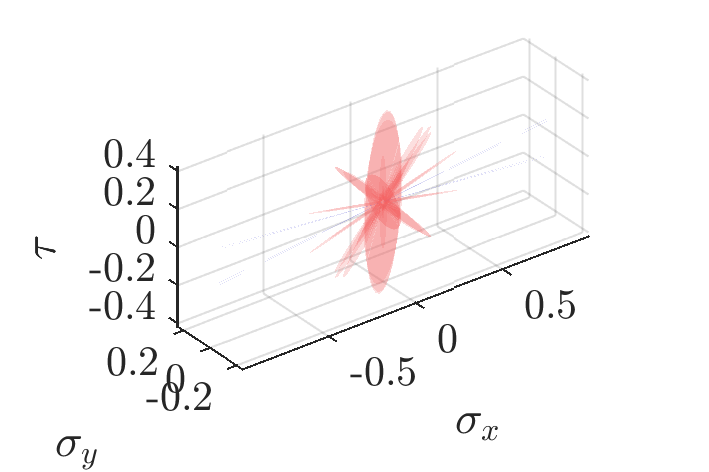}%
				\vfil}%
			\caption{}
		\end{subfigure}%
	\end{minipage}
	\hfill\begin{minipage}{0.25\linewidth}
		\begin{subfigure}{\linewidth}
			\centering
			\vbox to 31mm {\vfil%
				\includegraphics[height=3.1cm]{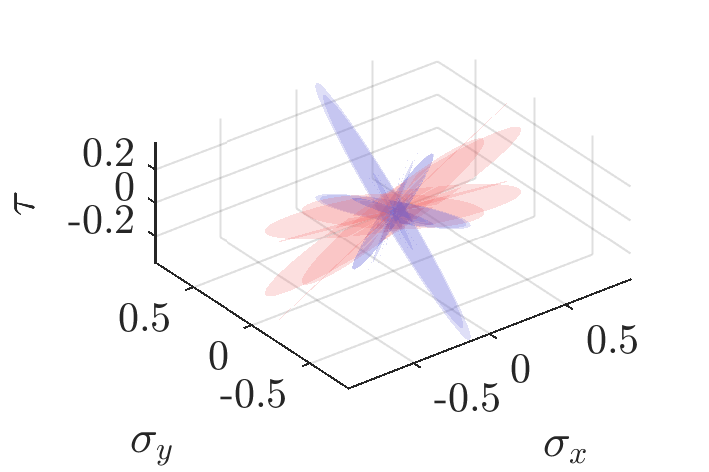}%
				\vfil}%
			\caption{}
		\end{subfigure}\\
		\begin{subfigure}{\linewidth}
			\centering
			\vbox to 31mm {\vfil%
				\includegraphics[height=3.1cm]{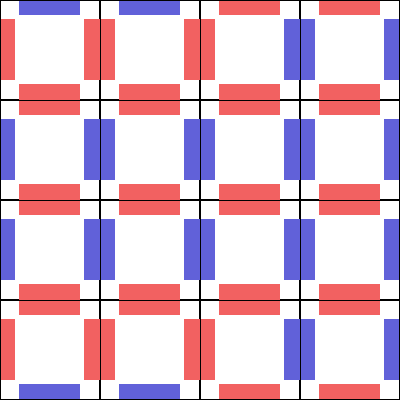}%
				\vfil}%
			\caption{}
		\end{subfigure}
	\end{minipage}%
	\hfill\begin{minipage}{0.2\linewidth}
		\begin{subfigure}{\linewidth}
			\centering
			\vbox to 31mm {\vfil%
				\includegraphics[height=3.1cm]{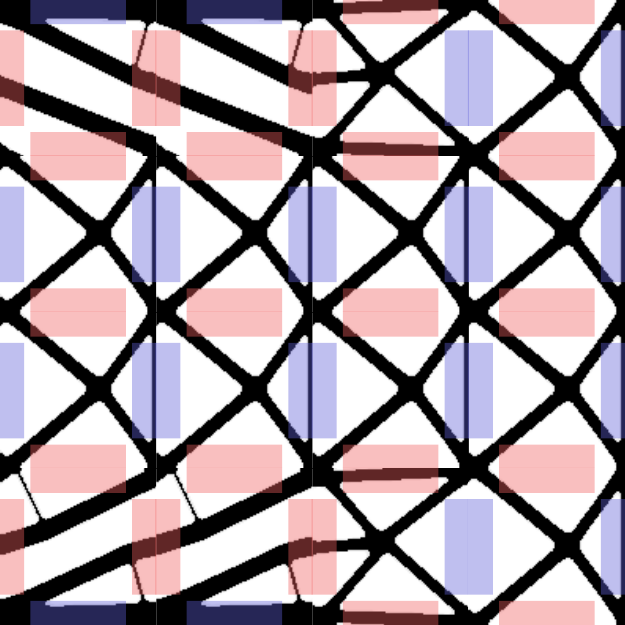}%
				\vfil}%
			\caption{}
		\end{subfigure}\\
		\begin{subfigure}{\linewidth}
			\centering
			\vbox to 31mm {\vfil%
				\includegraphics[width=\linewidth]{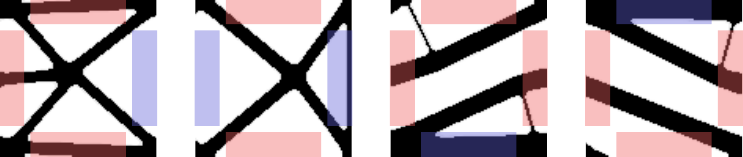}%
				\vfil}%
			\caption{}
		\end{subfigure}
	\end{minipage}
	\caption{Illustration of the proposed heuristic approach to generate efficient valid Wang tilings: First, we (a) discretize a design domain into modules, and (b) associate a finite element with each edge: elements belonging to horizontal and vertical edges are shown in gray and white, respectively. For \replace{this mesh}{a refined interface-based discretization}, we (c) solve the free material optimization problem \eqref{eq:fmo}, represented in this figure by the traces of material stiffnesses\replace{, and (d) post-process them. Finally,}{. Subsequently,} the material stiffnesses visualized in the \replace{(e)}{(d)} horizontal and \replace{(f)}{(e)} vertical edges as envelopes of the stress components $\sigma_x$, $\sigma_y$ and $\tau$ \replace{induc}{generat}ed by arbitrary unit strain vectors, are clustered independently based on the Frobenius norm, establishing \replace{(g)}{(f)} a Wang tiling assembly plan. \replace{}{The assembly plan forms the input to (g) topology optimization of the structure} \replace{and  the}{consisting of} (h) \replace{}{a set of modules}.}
	\label{fig:clustering}
\end{figure}

As discussed already in Section~\ref{sec:wangtiles}, valid Wang tilings are determined uniquely by the edge color codes. Therefore, instead of the modules, Fig. \ref{fig:clustering}a, we must cluster material stiffnesses corresponding to the horizontal and vertical edges. These stiffnesses are obtained by solving the free material optimization problem \eqref{eq:fmo} on a~rotated, module interface-based finite element mesh, see Fig.~\ref{fig:clustering}b. When comparing with Fig.~\ref{fig:clustering}a, it clearly follows that each of the modules is defined by a quadruple of (edge-related) elements in Fig.~\ref{fig:clustering}b. Hence, after clustering, the edge labels automatically define Wang tiles and a valid tiling, see Figs.~\ref{fig:clustering}f--h. We note here that a similar approach relying on rotated mesh has been applied to generate Wang tile-based textures in computer graphics \cite{cohen2003} and microstructure modeling~\cite{doskar2014}.
\replace{Using the post-processed FMO outputs, it may happen that there would exist modules without any stiffness in their edges, i.e., the maximum eigenvalue of the edge-related stiffness tensors is small. Such modules can conversely be interpreted as voids, and we remove them prior to the clustering procedure and leave their locations empty.}{}

The common aim of the clustering algorithms is to partition a set of points $\mathcal{P} \coloneqq \left\{\sevek{e}^{(1)},\dots,\sevek{e}^{(n_\mathrm{\ell})}\right\}$ into (at most) $m$ groups, such that a specified distance-based criterion is minimized \citep{gan2020}. Among these algorithms, \texttt{kmeans} is probably the most widely used. It starts with a random assignment of points to clusters, computes a centroid of each cluster based on a specified distance function, and re-assigns each point to the cluster with the nearest centroid. This procedure repeats until convergence, but the resulting clustering strongly depends on the initial partitioning.

The second group of clustering algorithms, spectral clustering, exploits a representation of the points $\mathcal{P}$ in a graph form. For this representation, one computes the graph Laplacian matrix and finds a partitioning based on its eigenvectors. Based on our testing (not included), neither \texttt{kmeans}, nor spectral clustering led to convincing designs.

In this manuscript, we rely on the third class of algorithms, the hierarchical clustering, because this class may avoid randomization and can be very simply tailored to satisfy additional constraints imposed on the clustering data. In particular, the latter feature appears to be very difficult to achieve with the two previous classes of algorithms. Hierarchical clustering algorithms can be categorized into \textit{divisive} clustering---starting with all points belonging to a single cluster, partitioned recursively---and \textit{agglomerative} clustering---where each point initially defines a cluster, and these are joined repetitively based on a linkage criterion.

In our implementation, we follow the agglomerative clustering approach and adopt the squared Frobenius norm to measure the distances of the material stiffness \replace{matrices}{tensors}, which is equivalent to the Euclidean norm \replace{among}{of the difference in} the stiffness vectors $\sevek{e}^{(\ell)}$, recall Section~\ref{sec:formulation}. To compute a distance between clusters, we represent each cluster $\mathcal{C}_i$ by its centroid with components
\begin{equation}\label{eq:centroid}
\sevek{t}_{i,j} = \frac{\sum_{\ell \in \mathcal{C}_i} w_\ell e_j^{(\ell)}}{\lvert \mathcal{C}_i \rvert},
\end{equation}
where $\sevek{w} \in \mathbb{R}^{n_\mathrm{\ell}}_{>0}$ are weights of the input data and $\lvert \mathcal{C}_i \rvert$ denotes the cardinality of points in $\mathcal{C}_i$, and by the cluster cost
\begin{equation}\label{eq:cost}
c_i = \sum_{\ell \in \mathcal{C}_i} w_\ell \lVert \sevek{t}_i - \sevek{e}^{(\ell)} \rVert^2_2.
\end{equation}
To determine which clusters shall be merged, we compute for all pairs of clusters $(i,j)$ the deterioration cost
\begin{equation}\label{eq:deterioration}
d_{i,j} = \left\{ \begin{matrix*}[l]
c_{i,j} - c_i - c_j & \text{if $i$ and $j$ are allowed to merge},\\
\infty & \text{otherwise},
\end{matrix*}\right.
\end{equation}
where $c_{i,j}$ stands for the cost of the cluster $\mathcal{C}_i \cup \mathcal{C}_j$. We accept the merger that least deteriorates the objective function. Notice that this linkage criterion is not equivalent to merging the nearest clusters, as more distant clusters containing fewer points may be cheaper to merge than the nearer ones with more data points. In this sense, we believe that this clustering procedure provides a more balanced number of points inside clusters.

Assume now that the material stiffness distribution exhibits symmetry with respect to a vertical or horizontal axis. For such a~setting, we find it reasonable that the resulting \replace{clusterings}{clusters} preserve this symmetry pattern. In our implementation, this can be guaranteed by restricting which clusters cannot join, i.e., the deterioration cost \eqref{eq:deterioration} is set to $\infty$ for some non-admissible pairs $(i,j)$.

To this goal, we split the material stiffnesses into three groups $\bm{\alpha}$, $\bm{\alpha}'$ and $\bm{\beta}$. $\bm{\beta}$ contains material stiffnesses of the elements that are either located on the axis of symmetry or do not have symmetric counterparts, and groups $\bm{\alpha}$ and $\bm{\alpha}'$ consists of the symmetric complements. More precisely, let us assume a fixed $i \in \{1,\dots,\lvert \mathcal{C}\rvert\}$ and $j \in \{1,\dots,\lvert \mathcal{C}\rvert\} \setminus i$ such that the entries of these centroids satisfy $t_{i,1} = t_{j,1}$, $t_{i,2} = t_{j,2}$, $t_{i,3} = t_{j,3}$, $t_{i,4} = -t_{j,4}$, $t_{i,5} = -t_{j,5}$, and $t_{i,6} = t_{j,6}$, which certifies symmetry of the corresponding material stiffness centroids with respect to the vertical or horizontal axis, cf. Appendix~\ref{app:symmetry} for details. If no such $j$ exists for a fixed $i$ or when $t_{i,4}=t_{i,5}=0$, we set $\sevek{t}_i \in \bm{\beta}$. If the corresponding pair exists, we set $\sevek{t}_i \in \bm{\alpha}$ if $t_{i,4}>0$ or $t_{i,4}=0 \land t_{i,5}>0$, and $\sevek{t}_i \in \bm{\alpha}'$ otherwise.

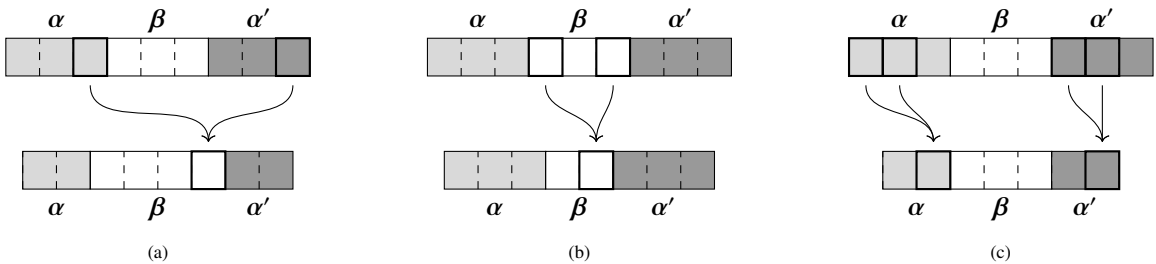
\begin{figure}[!b]
	\begin{subfigure}{0.333\linewidth}
		\centering
		\begin{tikzpicture}
		\draw[fill=black!15] (0,0) rectangle (1.333,0.5);
		\draw[] (1.333,0) rectangle (2.666,0.5);
		\draw[fill=black!40] (2.666,0) rectangle (4,0.5);
		\draw[dashed] (0.444, 0)--(0.444, 0.5); \draw[dashed] (0.888, 0)--(0.888, 0.5);
		\draw[dashed] (1.777, 0)--(1.777, 0.5); \draw[dashed] (2.222, 0)--(2.222, 0.5);
		\draw[dashed] (3.111, 0)--(3.111, 0.5); \draw[dashed] (3.555, 0)--(3.555, 0.5);
		\draw[thick] (0.888,0) rectangle (1.333,0.5);
		\draw[thick] (3.555,0) rectangle (4,0.5);
		\node[align=center,text height=1.5ex,text depth=.25ex] at (0.666, 0.75) {$\bm{\alpha}$};
		\node[align=center,text height=1.5ex,text depth=.25ex] at (2, 0.75) {$\bm{\beta}$};
		\node[align=center,text height=1.5ex,text depth=.25ex] at (3.333, 0.75) {$\bm{\alpha}'$};
		\node[] (a01) at (1.111,0) {};
		\node[] (a02) at (3.777,0) {};
		\begin{scope}[shift={(-0.222,-1.5)}]
		\draw[fill=black!15] (0.444,0) rectangle (1.333,0.5);
		\draw[] (1.333,0) rectangle (3.111,0.5);
		\draw[fill=black!40] (3.111,0) rectangle (4,0.5);
		\draw[dashed] (0.444, 0)--(0.444, 0.5); \draw[dashed] (0.888, 0)--(0.888, 0.5);
		\draw[dashed] (1.777, 0)--(1.777, 0.5); \draw[dashed] (2.222, 0)--(2.222, 0.5);
		\draw[dashed] (2.666, 0)--(2.666, 0.5); \draw[dashed] (3.555, 0)--(3.555, 0.5);
		\draw[thick] (2.666,0) rectangle (3.111,0.5);
		\node[align=center,text height=1.5ex,text depth=.25ex] at (0.888, -0.25) {$\bm{\alpha}$};
		\node[align=center,text height=1.5ex,text depth=.25ex] at (2.222, -0.25) {$\bm{\beta}$};
		\node[align=center,text height=1.5ex,text depth=.25ex] at (3.555, -0.25) {$\bm{\alpha}'$};
		\node[] (b01) at (2.888,0.5) {};
		\end{scope}
		\draw[->] (a01) to [out=270,in=90] (b01);
		\draw[->] (a02) to [out=270,in=90] (b01);
		\end{tikzpicture}
		\caption{}
	\end{subfigure}
	\begin{subfigure}{0.333\linewidth}
		\centering
		\begin{tikzpicture}
		\draw[fill=black!15] (0,0) rectangle (1.333,0.5);
		\draw[] (1.333,0) rectangle (2.666,0.5);
		\draw[fill=black!40] (2.666,0) rectangle (4,0.5);
		\draw[dashed] (0.444, 0)--(0.444, 0.5); \draw[dashed] (0.888, 0)--(0.888, 0.5);
		\draw[dashed] (1.777, 0)--(1.777, 0.5); \draw[dashed] (2.222, 0)--(2.222, 0.5);
		\draw[dashed] (3.111, 0)--(3.111, 0.5); \draw[dashed] (3.555, 0)--(3.555, 0.5);
		\draw[thick] (1.333,0) rectangle (1.777,0.5);
		\draw[thick] (2.222,0) rectangle (2.666,0.5);
		\node[align=center,text height=1.5ex,text depth=.25ex] at (0.666, 0.75) {$\bm{\alpha}$};
		\node[align=center,text height=1.5ex,text depth=.25ex] at (2, 0.75) {$\bm{\beta}$};
		\node[align=center,text height=1.5ex,text depth=.25ex] at (3.333, 0.75) {$\bm{\alpha}'$};
		\node[] (a01) at (1.555,0) {};
		\node[] (a02) at (2.444,0) {};
		\begin{scope}[shift={(0.222,-1.5)}]
		\draw[fill=black!15] (0,0) rectangle (1.333,0.5);
		\draw[] (1.333,0) rectangle (2.222,0.5);
		\draw[fill=black!40] (2.222,0) rectangle (3.555,0.5);
		\draw[dashed] (0.444, 0)--(0.444, 0.5); \draw[dashed] (0.888, 0)--(0.888, 0.5);
		\draw[dashed] (1.777, 0)--(1.777, 0.5); \draw[dashed] (2.222, 0)--(2.222, 0.5);
		\draw[dashed] (2.666, 0)--(2.666, 0.5); \draw[dashed] (3.111, 0)--(3.111, 0.5);
		\draw[thick] (1.777,0) rectangle (2.222,0.5);
		\node[align=center,text height=1.5ex,text depth=.25ex] at (0.666, -0.25) {$\bm{\alpha}$};
		\node[align=center,text height=1.5ex,text depth=.25ex] at (1.777, -0.25) {$\bm{\beta}$};
		\node[align=center,text height=1.5ex,text depth=.25ex] at (2.888, -0.25) {$\bm{\alpha}'$};
		\node[] (b01) at (1.999,0.5) {};
		\end{scope}
		\draw[->] (a01) to [out=270,in=90] (b01);
		\draw[->] (a02) to [out=270,in=90] (b01);
		\end{tikzpicture}
		\caption{}
	\end{subfigure}
	\begin{subfigure}{0.333\linewidth}
		\centering
		\begin{tikzpicture}
		\draw[fill=black!15] (0,0) rectangle (1.333,0.5);
		\draw[] (1.333,0) rectangle (2.666,0.5);
		\draw[fill=black!40] (2.666,0) rectangle (4,0.5);
		\draw[dashed] (0.444, 0)--(0.444, 0.5); \draw[dashed] (0.888, 0)--(0.888, 0.5);
		\draw[dashed] (1.777, 0)--(1.777, 0.5); \draw[dashed] (2.222, 0)--(2.222, 0.5);
		\draw[dashed] (3.111, 0)--(3.111, 0.5); \draw[dashed] (3.555, 0)--(3.555, 0.5);
		\draw[thick] (0.444,0) rectangle (0.888,0.5);
		\draw[thick] (0,0) rectangle (0.444,0.5);
		\draw[thick] (2.666,0) rectangle (3.111,0.5);
		\draw[thick] (3.111,0) rectangle (3.555,0.5);
		\node[align=center,text height=1.5ex,text depth=.25ex] at (0.666, 0.75) {$\bm{\alpha}$};
		\node[align=center,text height=1.5ex,text depth=.25ex] at (2, 0.75) {$\bm{\beta}$};
		\node[align=center,text height=1.5ex,text depth=.25ex] at (3.333, 0.75) {$\bm{\alpha}'$};
		\node[] (a01) at (0.666,0) {};
		\node[] (a02) at (0.222,0) {};
		\node[] (a03) at (2.888,0) {};
		\node[] (a04) at (3.333,0) {};
		\begin{scope}[shift={(0,-1.5)}]
		\draw[fill=black!15] (0.444,0) rectangle (1.333,0.5);
		\draw[] (1.333,0) rectangle (2.666,0.5);
		\draw[fill=black!40] (2.666,0) rectangle (3.555,0.5);
		\draw[dashed] (0.444, 0)--(0.444, 0.5); \draw[dashed] (0.888, 0)--(0.888, 0.5);
		\draw[dashed] (1.777, 0)--(1.777, 0.5); \draw[dashed] (2.222, 0)--(2.222, 0.5);
		\draw[dashed] (2.666, 0)--(2.666, 0.5); \draw[dashed] (3.555, 0)--(3.555, 0.5);
		\draw[thick] (0.888,0) rectangle (1.333,0.5);
		\draw[thick] (3.111,0) rectangle (3.555,0.5);
		\node[align=center,text height=1.5ex,text depth=.25ex] at (0.888, -0.25) {$\bm{\alpha}$};
		\node[align=center,text height=1.5ex,text depth=.25ex] at (2, -0.25) {$\bm{\beta}$};
		\node[align=center,text height=1.5ex,text depth=.25ex] at (3.111, -0.25) {$\bm{\alpha}'$};
		\node[] (b01) at (1.111,0.5) {};
		\node[] (b02) at (3.333,0.5) {};
		\end{scope}
		\draw[->] (a01) to [out=270,in=90] (b01);
		\draw[->] (a02) to [out=270,in=90] (b01);
		\draw[->] (a03) to [out=270,in=90] (b02);
		\draw[->] (a04) to [out=270,in=90] (b02);
		\end{tikzpicture}
		\caption{}
	\end{subfigure}
	\caption{Illustration of all allowed cluster aggregations to generate Wang tilings preserving the symmetric distribution of material stiffness matrices when present. The set $\bm{\beta}$ consists of material stiffnesses of the elements located on the axis of symmetry and those that do not have symmetric counterparts, and $\bm{\alpha}$ with $\bm{\alpha}'$ contain\replace{}{ing} the symmetric complements.}
	\label{fig:aggregations}
\end{figure}

Using these groups, a finite deterioration cost is associated only with (i) a merger of the corresponding pair $i \in \bm{\alpha}$ and $j \in \bm{\alpha}'$, producing a cluster belonging to $\bm{\beta}$, (ii) joining clusters $\left\{i,j: i,j \in \bm{\beta}, i\neq j\right\}$, with the result also belonging to $\bm{\beta}$, and (iii) a merger of $\left\{i,j: i,j \in \bm{\alpha}, i \neq j \right\}$, which requires \replace{to merge concurrently}{concurrently merging} the corresponding pair $\left\{i,j: i,j \in \bm{\alpha}', i \neq j \right\}$. In this last case, the number of clusters is decreased by two, with the merged outputs belonging to $\bm{\alpha}$ and $\bm{\alpha}'$, respectively. All the possible aggregations are illustrated in Fig.~\ref{fig:aggregations}.

In each iteration, the aggregation procedure reduces $\lvert\mathcal{C}\rvert$ by $1$ at least, building an aggregation tree called dendrogram. When $\lvert \mathcal{C} \rvert \leq m$, the clustering algorithm terminates, see Algorithm~\ref{alg:clustering} for the pseudo-code. As an output, we obtain an assignment of the input data (horizontal or vertical edges) into clusters, Fig.~\ref{fig:clustering}e--f, and this partitioning is uniquely representable by a valid Wang tiling over (at most) $m$ colors, Fig.~\ref{fig:clustering}g--h.

\begin{algorithm}[!htbp]
	\begin{algorithmic}[1]
		\State $\bm{\alpha}, \bm{\alpha}', \bm{\beta}, \mathcal{C} \gets$ \texttt{initialize}($\sevek{e}$)
		\While{$\lvert \mathcal{C}\rvert>m$}
		\State pair$_1$, cost$_1$ $\gets$ \texttt{findCheapestMerge}($\bm{\alpha}$, $\bm{\alpha}'$)
		\State pair$_2$, cost$_2$ $\gets$ \texttt{findCheapestMerge}($\bm{\beta}$, $\bm{\beta}$)
		\State pairs$_3$, cost$_3$ $\gets$ \texttt{findCheapestMerge}($\bm{\alpha}$, $\bm{\alpha}$, $\bm{\alpha}'$, $\bm{\alpha}'$)
		\If{$\lvert\mathcal{C}\rvert -m > 1$}
		\State $\bm{\alpha}, \bm{\alpha}', \bm{\beta}, \mathcal{C} \gets$ \texttt{mergeCheapest}(pair$_1$, cost$_1$, pair$_2$, cost$_2$, pairs$_3$, cost$_3/2$)
		\Else
		\State $\bm{\alpha}, \bm{\alpha}', \bm{\beta}, \mathcal{C} \gets$ \texttt{mergeCheapest}(pair$_1$, cost$_1$, pair$_2$, cost$_2$, pairs$_3$, cost$_3$)
		\EndIf
		\EndWhile
		\State \Return $\mathcal{C}$
	\end{algorithmic}
	\caption{Pseudo-code of the proposed greedy agglomerative clustering algorithm.}
	\label{alg:clustering}
\end{algorithm}

\subsection{Design of topologies of modules}\label{sec:moduletop}

Once an assembly plan---in the form of valid Wang tiling---is obtained, the dimension of design space in TO usually reduces \replace{drastically}{significantly} because $n_\mathrm{m}<n_\mathrm{\ell}$, with $n_\mathrm{m}$ being the number of Wang tiles or modules and $n_\mathrm{\ell}$ denoting the number of module positions in Fig.~\ref{fig:clustering}a. However, it seems that---at least for compliant mechanism design problems---the landscape of the objective function used in TO is very complex and optimized topologies depend strongly on a supplied initial guess \replace{}{(see Section \ref{sec:reusableguess} for an example)}. This issue becomes even more relevant when dealing with the design of modules reusable among multiple structures (Section~\ref{sec:reusable}), since an inappropriate initial guess may prevent the mechanisms \replace{to work}{from functioning} at all.

In this section, we first recall a variable linking procedure \citep{tyburec2020} to describe modular structures mathematically, and apply it to the FMO problem \eqref{eq:fmo}\replace{, which}{. The modular FMO} allows for an inexpensive computation of the initial guesses for TO based on the element traces, see Section~\ref{sec:modularityconstr}, and also \replace{}{for} a simple comparison of the efficiency of different assembly plans. Description of the adopted TO method is given in Section~\ref{sec:topopt}. 

\subsubsection{Modular free material optimization}\label{sec:modularityconstr}

Let $n_\mathrm{m}$ denote the number of modules and $\{\sevek{g} \in \mathbb{N}^{n_\mathrm{\ell}}, \forall \ell \in \{1, \dots, n_\mathrm{\ell}\}:
g_\ell \le n_\mathrm{m}\}$ be a group vector assigning a module number $m \in \{1, \dots, n_\mathrm{m}\}$ to each element $\ell \in \{1, \dots, n_\mathrm{\ell}\}$ of the module scale, recall Fig.~\ref{fig:clustering}a. In addition, let us define a group matrix $\semtrx{G} \in \mathbb{B}^{n_\mathrm{\ell} \times n_\mathrm{m}}$ as
\begin{equation}\label{eq:group_matrix}
G_{\ell,j} = 
\begin{dcases}
0\quad\text{if } j \neq g_\ell,\\
1\quad\text{if } j  = g_\ell,
\end{dcases} \quad \forall \ell \in \{ 1, \dots, n_\mathrm{\ell} \}, \forall j \in \{ 1, \dots, n_\mathrm{m} \}.
\end{equation}
Using this definition, we can construct the full-domain elasticity properties $\sevek{e}_k$, $k \in \{1, \dots, 6\}$, by mapping the \textit{grouped} modular elastic properties $\sevek{e}_{\mathrm{g},k} = \left( e_{k}^{\langle1\rangle}, \dots, e_{k}^{\langle n_\mathrm{m} \rangle} \right)^\mathrm{T}$ using the group matrix $\semtrx{G}$,
\begin{equation}\label{eq:fmogroup}
\sevek{e}_k = \semtrx{G} \sevek{e}_{\mathrm{g},k}, \quad \forall k \in \{1,\dots, 6\},
\end{equation}
with the notation $\bullet^{\langle m \rangle}$ expressing an association of $\bullet$ with the module $m$. Notice that an association with an element $\ell$ remains denoted by $\bullet^{(\ell)}$.

Then, the optimization problem \eqref{eq:fmo} extended to structural modularity reads as
\begin{subequations}\label{eq:mfmo}
	\begin{alignat}{4}
	\min_{s, \semtrx{E}^{\langle 1\rangle}, \dots, \semtrx{E}^{\langle n_\mathrm{m}\rangle}}\; && s\hspace{41.5mm}\label{eq:mfmo_obj}\\
	\text{s.t.}\; && s-\sevek{a}^\mathrm{T} \left[\semtrx{K}\left(\semtrx{E}^{\langle 1 \rangle},\dots,\semtrx{E}^{\langle n_{\mathrm{m}}\rangle}\right)\right]^{-1} \sevek{f} &&\;\ge\;& 0,\label{eq:mfmo_mutual}\\
	&& \overline{V}\overline{\epsilon} \sum_{\ell=1}^{n_\mathrm{\ell}} v^{(\ell)} - \sum_{\ell=1}^{n_\mathrm{\ell}} \left[v^{(\ell)}\mathrm{Tr} \left(\semtrx{E}^{\langle g_\ell \rangle}\right)\right] &&\;\ge\;& 0,\label{eq:mfmo_volume}\\
	&& \overline{\epsilon} - \mathrm{Tr}\left(\semtrx{E}^{\langle m\rangle}\right) &&\;\ge\;& 0, &\;\;\forall m \in \{1,\dots, n_\mathrm{m}\},\label{eq:mfmo_eltrace}\\
	&& \semtrx{E}^{\langle m\rangle} &&\;\succeq\;& \underline{\epsilon}, &\;\;\forall m \in \{1,\dots, n_\mathrm{m}\},\label{eq:mfmo_eig}
	\end{alignat}
\end{subequations}
with the stiffness matrix assembled as
\begin{equation}
\semtrx{K}\left(\semtrx{E}^{\langle 1\rangle},\dots,\semtrx{E}^{\langle n_{\mathrm{m}}\rangle}\right) = 
\semtrx{K}_0 + \sum_{\ell=1}^{n_\mathrm{\ell}} \semtrx{L}^{(\ell)\mathrm{T}} \semtrx{K}^{(\ell)}(\semtrx{E}^{\langle g_\ell \rangle}) \semtrx{L}^{(\ell)}.
\end{equation}
Because of $n_\mathrm{m} < n_\mathrm{\ell}$, the optimization problem \eqref{eq:mfmo} is usually solved much faster than \eqref{eq:fmo}. After its solution, we obtain the optimal elasticity matrices of individual modules, and their traces are used to build an initial guess for TO as constant pseudo-density fields of individual modules according to
\begin{equation}
\rho_{\mathrm{init}}^{\langle m \rangle} \coloneqq \mathrm{Tr}(\semtrx{E}^{\langle m \rangle})/\overline{\epsilon}.
\label{eq:initial_guess}
\end{equation}

\subsubsection{Topology optimization}\label{sec:topopt}

The above-described procedures based on the free material optimization yield an efficient assembly plan and related optimal material stiffnesses that are constant over individual modules. Ultimately, though, we want to obtain explicit design of modules manufactured from a single material.
As mentioned already in~\Sref{sec:introduction}, one option is to perform inverse homogenization for each module separately, optionally with enforcing geometrical or mechanical continuity across their boundaries~\citep{garner2019,hu2020}. Here, we opted for a different approach in which the functional continuity is resolved automatically by performing the single-scale topology optimization with a modularity projection following from the optimized assembly plan.

In particular, we adopted the widely-used Solid Isotropic Material with Penalization (SIMP) approach~\cite{bendsoe1999,bendsoe2003}. The design is thus parameterized by a pseudo-density field $0 \leq \rho\atx \leq 1$ that, in turn, dictates the distribution of the material stiffness according to
\begin{equation}
\tensf{E}\atx = \tensf{E}_{\text{min}} + \rho^{p}\atx \left( \tensf{E}_{0} - \tensf{E}_{\text{min}} \right) \,,
\label{eq:SIMPstiffness}
\end{equation}
where $\tensf{E}_{0}\succ0$ and $\tensf{E}_\text{min}\succ0$ are the stiffness tensors of the bulk material and void space (which attain small but positive eigenvalues in order to avoid a singular Hessian), respectively, and the penalization parameter $p = 3$ is introduced to promote ``black-white'' designs~\cite{bendsoe2003}.
Throughout all the simulations, we assumed an isotropic material under plane stress conditions with fixed Poisson ratio $\nu = 0.3$; the pseudo-density $\rho$ thus effectively scales only the Young modulus.

Similarly to the free material optimization problem~(\ref{eq:fmo}), the formulation of the topology optimization builds on the finite element discretization of the governing equation of linear elasticity with unknown vector $\sevek{\rho}$ containing element-wise constant densities $\rho^{(\mathrm{\ell})}$.
For notation simplicity, we assume that each of $n_\text{m}$ modules is discretized with $n_{\text{m}\ell}$ finite elements. The modularity is then introduced in the formulation via a boolean group matrix $\semtrx{G}_{\text{TO}} \in \left\{0,1\right\}^{n_{\ell} \times \left( n_\text{m} \cdot n_{\text{m}\ell} \right)}$ which maps the module-related densities collected in vector $\sevek{\rho}_\text{g} = \left( \sevek{\rho}^{\langle 1 \rangle},\dots,\sevek{\rho}^{\langle n_\text{m}\rangle} \right)\trn$ onto full-domain densities $\sevek{\rho}$,
\begin{equation}
\sevek{\rho} = \semtrx{G}_{\text{TO}} \, \sevek{\rho}_\text{g} \,,
\label{eq:mto_modularity}
\end{equation}	
recall Eq.~\eqref{eq:fmogroup}.
Keeping the same structure as in~\Eref{eq:fmo}, the modular topology optimization is eventually formalized as
\begin{subequations}
	\label{eq:mto}
	\begin{alignat}{4}
	\min_{s_{\textrm{TO}},\,\sevek{\rho}=\semtrx{G}_{\text{TO}} \sevek{\rho}_\text{g}}\; && s_{\textrm{TO}}  \hspace{26mm}\label{eq:mto_obj}\\
	\text{s.t.}\; &&  s_{\textrm{TO}} - \sevek{a}^\mathrm{T} \left[\semtrx{K}\left(\sevek{\rho}\right)\right]^{-1} \sevek{f} &&\;\ge\;& 0, \label{eq:mto_mutual}\\
	&& \overline{V} \sum_{\ell=1}^{n_\mathrm{\ell}} v^{(\ell)} - \sum_{\ell=1}^{n_\mathrm{\ell}} \rho^{(\ell)} v^{(\ell)} &&\;\ge\;& 0.\label{eq:mto_volume} \\
	&& 1 \;\ge\; \rho^{(\ell)} &&\;\ge\;& 0, &\;\;\forall \ell \in \{1,\dots, n_\mathrm{\ell}\} .\label{eq:mto_bounds}
	\end{alignat}
\end{subequations}
The structural stiffness matrix $\semtrx{K}\left(\sevek{\rho}\right)$ follows from~\Eref{eq:assembly} (including the design-independent part $\semtrx{K}_{0}$) with $\semtrx{E}^{(\ell)}$ depending on $\rho^{(\ell)}$ through~\Eref{eq:SIMPstiffness}.
The slack variable $s_{\textrm{TO}}$ is introduced here only formally to denote the objective.
Contrary to the formulation \eqref{eq:fmo}, the volume constraint (\ref{eq:mto_volume}) is here formulated directly in element densities $\rho^{(\ell)}$ and no analogue to the eigenvalue constraints (\ref{eq:fmo_eltrace}) and (\ref{eq:fmo_eig}) is needed because they are already satisfied due to the interpolation scheme~(\ref{eq:SIMPstiffness}). On the other hand, this comes at the price of a restricted design space compared to the FMO in which the whole material stiffness matrix is unknown.

Having effectively only volume and bound constraints, we solve the problem (\ref{eq:mto}) with the first-order Optimality Criteria (OC) method~\cite{bendsoe2003} due to its simplicity and practical performance. Moreover, we modify the standard OC scheme with the gray-scale suppression proposed by \citet{groenwold2009} to further promote faster convergence towards designs with clearly distinguished phases.

The sensitivity of the objective function $\frac{\partial s_\text{TO}}{\partial \sevek{\rho}_\text{g}}$ under satisfied equality equation~(\ref{eq:mto_mutual}) follows from the chain rule
\begin{equation}
\frac{\partial s_\text{TO}}{\partial \sevek{\rho}_\text{g}} = \frac{\partial s_\text{TO}}{\partial \sevek{\rho}} \frac{\partial \sevek{\rho}}{\partial \sevek{\rho}_\text{g}} \,,
\end{equation}
where $\frac{\partial s_\text{TO}}{\partial \sevek{\rho}}$ is obtained similarly to FMO, see Appendix~\ref{sec:general_sensitivity}, and $\frac{\partial \sevek{\rho}}{\partial \sevek{\rho}_\text{g}} = \semtrx{G}_{\text{TO}}$, recall \Eref{eq:mto_modularity}. The same chain rule applies also to the sensitivity of the volume constraint (\ref{eq:mto_mutual}), denoted as $\frac{\partial V}{\partial \sevek{\rho}_\text{g}}$ further on.
In order to avoid mesh-dependency and checkerboard patterns discussed already in~\Sref{sec:checkerboard} for FMO, we opt for the sensitivity filter~\cite{sigmund2007} that modifies the objective sensitivity by weight averaging. In the discrete setup, where each $\rho^{(\ell)}$ belongs to the center of gravity $\tens{x}^{(\ell)}$ of the corresponding element $\ell$, the filtered non-modular sensitivity $\widehat{\frac{\partial s_\text{TO}}{\partial \sevek{\rho}}}$ is defined as
\begin{equation}\label{eq:filterto}
\widehat{\frac{\partial s_\text{TO}}{\partial \rho^{(i)}}} 
= 
\frac{1}{\rho^{(i)} \sum_{j=1}^{n_{\ell}} w(\tens{x}^{(i)},\tens{x}^{(j)})} 
\sum_{j=1}^{n_{\ell}} \rho^{(j)} \frac{\partial s_\text{TO}}{\partial \rho^{(i)}} \, w(\tens{x}^{(i)},\tens{x}^{(j)}) \,, 
\quad
\forall \ell \in \{1,\dots, n_\mathrm{\ell}\} \,,
\end{equation}
which leads to a linear relation
\begin{equation}
\widehat{\frac{\partial s_\text{TO}}{\partial \sevek{\rho}}} = \semtrx{F}_\text{w}(\sevek{\rho}) \frac{\partial s_\text{TO}}{\partial \sevek{\rho}} \,,
\end{equation} 
with the filtering matrix $\semtrx{F}_\text{w}$ following from \Eref{eq:filterto}. Here, we assume a linear weight function
\begin{equation}
w(\tens{x},\tens{y}) = \max(r - \norm{\tens{x}-\tens{y}}, 0)
\end{equation} 
with the radius $r$.

Denoting the sensitivity vectors $\sevek{g}^{s} \coloneqq \widehat{\frac{\partial s_\text{TO}}{\partial \sevek{\rho}_\text{g}}} = \semtrx{G}_{\text{TO}}\trn \, \semtrx{F}_{w}(\sevek{\rho}) \, \frac{\partial s_\text{TO}}{\partial \sevek{\rho}}$ and $\sevek{g}^{V} \coloneqq \frac{\partial V}{\partial \sevek{\rho}_\text{g}} = \semtrx{G}_{\text{TO}}\trn \, \frac{\partial V}{\partial \sevek{\rho}}$ for terseness, the update of the design variables $\sevek{\rho}_\text{g}$ within the modified OC scheme takes the form
\begin{equation}
\rho^{\text{new}}_{\text{g},i}
=
\begin{cases}
\,\underline{\rho}_{i} &\text{if} \quad \left(\rho_{\text{g},i} \left(B_{i}\right)^{\eta}\right)^{q} \leq \underline{\rho}_{i} \,,\\[4pt]
\,\overline{\rho}_{i} &\text{if} \quad \overline{\rho}_{i} \geq \left(\rho_{\text{g},i} \left(B_{i}\right)^{\eta}\right)^{q} \,,\\[3pt]
\,\left(\rho_{\text{g},i} \left(B_{i}\right)^{\eta}\right)^{q} &\text{otherwise,}
\end{cases}
\end{equation}
where $\underline{\rho}_{i} = \max(\rho_{\text{g},i} - \delta, 0.0)$, $\overline{\rho}_{i} = \min (\rho_{\text{g},i} + \delta, 1.0)$, and
$
B_{i} 
= 
\max( 
-\frac{{g}^{s}_{i}}{\Lambda {g}^{v}_{i}} 
,
\mu  
)\,.
$
The lower index $\bullet_{,i}$ denotes the $i$th component of the appropriate vector $\bullet$, ranging over the modular unknowns, i.e. $i \in \{ 1, \dots, ( n_\text{m} \cdot n_{\text{m}\ell} ) \}$.
Coefficients $\eta$ and $\delta$ are the damping factor and the move limit of the original OC method, respectively, $q$ is the power coefficient pertinent to the power-law gray-scale suppression, and $\mu$ facilitates the inconsistent modification~\cite{bendsoe2003} such that $\mu$ is set to $-\infty$ for the compliance minimization and $10^{-10}$ for mechanisms. Finally, $\Lambda$ denotes the Lagrange multiplier related to the volume constraint (\ref{eq:mto_mutual}), whose optimal value is found with a bisection algorithm; see e.g.~\cite{bendsoe2003} for more details.

The entire implementation and \replace{}{related} dataset are available at~\cite{tyburec2021}. While the free-material optimization and clustering procedures described above were implemented in \textsc{Matlab}, relying on the \textsc{Pennon} optimizer \cite{kocvara2003}, the topology optimization part was implemented in an in-house, object-oriented C++ code, which was linked against Intel\textsuperscript{\textregistered} oneAPI Math Kernel Library and utilized \textsc{Pardiso} direct sparse solver.

\section{Results}\label{sec:resultsFMO}

We illustrate the presented theoretical developments with three selected classical problems in topology optimization. In particular, we show the results for the minimum-compliance optimization of the Messerschmitt-B\"olkow-Blohm (MBB) beam in Section~\ref{sec:mbb}, Sections~\ref{sec:inverter} and \ref{sec:gripper} \replace{}{then} cover the design of compliant mechanisms (an inverter and a gripper, respectively), and a novel reusable modular gripper-inverter design is demonstrated in Section~\ref{sec:reusable}.

For all these problems, we limited the relative volume with $\overline{V}=0.4$. \replace{The empty modules identified in the FMO part were not explicitly discretized in TO; consequently, the volume fraction limit considered in TO was modified to match the volume fraction relative to the original domain size.}{} Modules were discretized with a regular grid of $100\times100$ bilinear fully integrated quadrilateral finite elements. We assumed isotropic, linear-elastic materials under the plane stress with $E=10^{-9}$ and $\nu = 0.3$ for the voids and $E=1$ and $\nu = 0.3$ for the bulk material, which in turn lead to $0 \le \mathrm{Tr}\left( \semtrx{E}^{(\ell)}\right) \le \overline{\epsilon}$ with the upper-bound on the sum of eigenvalues $\overline{\epsilon}=235/91$, recall \Eref{eq:trace_upper_bound}.
In the topology optimization part, radius $r$ of the sensitivity filter was set to $r = 3.5 h_{\ell}$ with $h_{\ell}$ being the length of an element edge. Optimality Criteria's parameters were as follows: move limit $\delta = 0.1$, damping coefficient $\eta = 0.5$ for compliance and $\eta = 0.3$ for mechanisms. 
We used a continuation strategy for the power coefficient $q$ of the gray-scale suppression; starting with the $20^{\text{th}}$ iteration, the initial value $q = 1.0$ was iteratively updated as $q_{\text{new}} = \min(1.01q, 2.0)$.
Within all examples, the maximal number of iterations was set to 150. The convergence criteria were based on the change in the objective, $\lvert\delta s_{\text{TO}}\rvert < 10^{-12}$, and the change in the design variables, $\norm{\delta\sevek{\rho}_\text{g}}_{\infty} < 10^{-2}$.
To avoid premature convergence with significant gray areas, quantified with the measure~\cite{sigmund2007}
\begin{equation}
M_\text{nd} = \frac{\sum_{\ell = 1}^{n_{\ell}} 4 \rho^{(\ell)} (1 - \rho^{(\ell)} ) }{n_{\ell}} \,,
\end{equation}
we added a convergence requirement $M_\text{nd} < 10^{-3}$. All topology optimization problems were initialized using the module-wise constant guess provided by FMO, as specified in~\Eref{eq:initial_guess}.

\subsection{Messerschmitt-B\"olkow-Blohm beam}\label{sec:mbb}

\begin{figure}[!b]
	\centering
	\begin{subfigure}[b]{0.38\linewidth}
		\resizebox{\linewidth}{!}{
			\begin{tikzpicture}
			\scaling{0.75}
			\point{a}{0}{0};
			\point{b}{4}{0};
			\point{c}{8}{0};
			\point{d}{8}{3};
			\point{e}{4}{3};
			\point{f}{0}{3};
			\draw[gray, dashdotted, xstep=0.1875cm, ystep=0.1875cm] (0,0) grid (6,2.3333);
			\beam{2}{a}{b};\beam{2}{b}{c};\beam{2}{c}{d};\beam{2}{d}{e};\beam{2}{e}{f};\beam{2}{f}{a};
			\support{1}{a};\support{2}{c};
			\dimensioning{1}{a}{b}{-1.25}[$0.5$];
			\dimensioning{1}{b}{c}{-1.25}[$0.5$];
			\dimensioning{2}{c}{d}{-0.5}[$0.375$];
			\load{1}{e}[90][0.5][0];
			\notation{1}{e}{$1$}[above right=1mm];
			\end{tikzpicture}}
		\caption{}
		\label{fig:mbb_bc}
	\end{subfigure}\hfill%
	\begin{subfigure}[b]{0.3\linewidth}
		\centering
		\includegraphics[width=\linewidth]{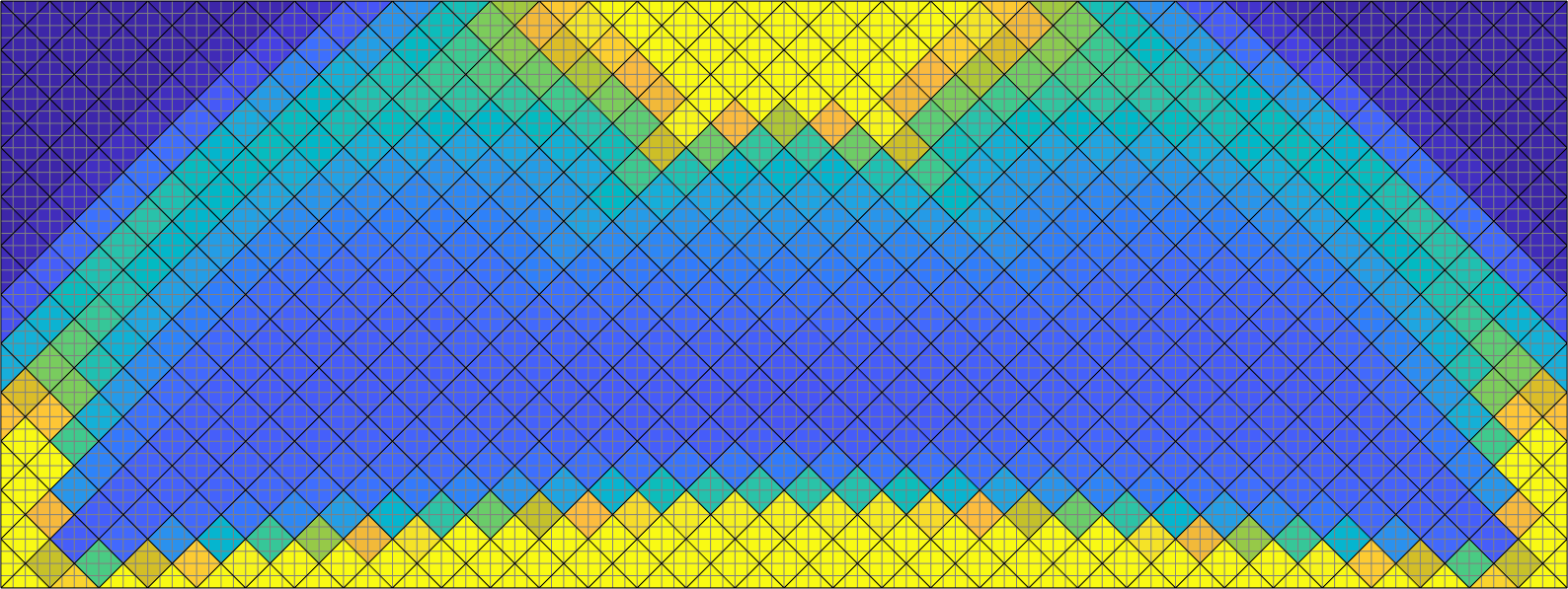}
		\vspace{7.5mm}
		\caption{}
		\label{fig:mbbf}
	\end{subfigure}\hfill%
	\begin{subfigure}[b]{0.3\linewidth}
		\centering
		\includegraphics[width=\linewidth]{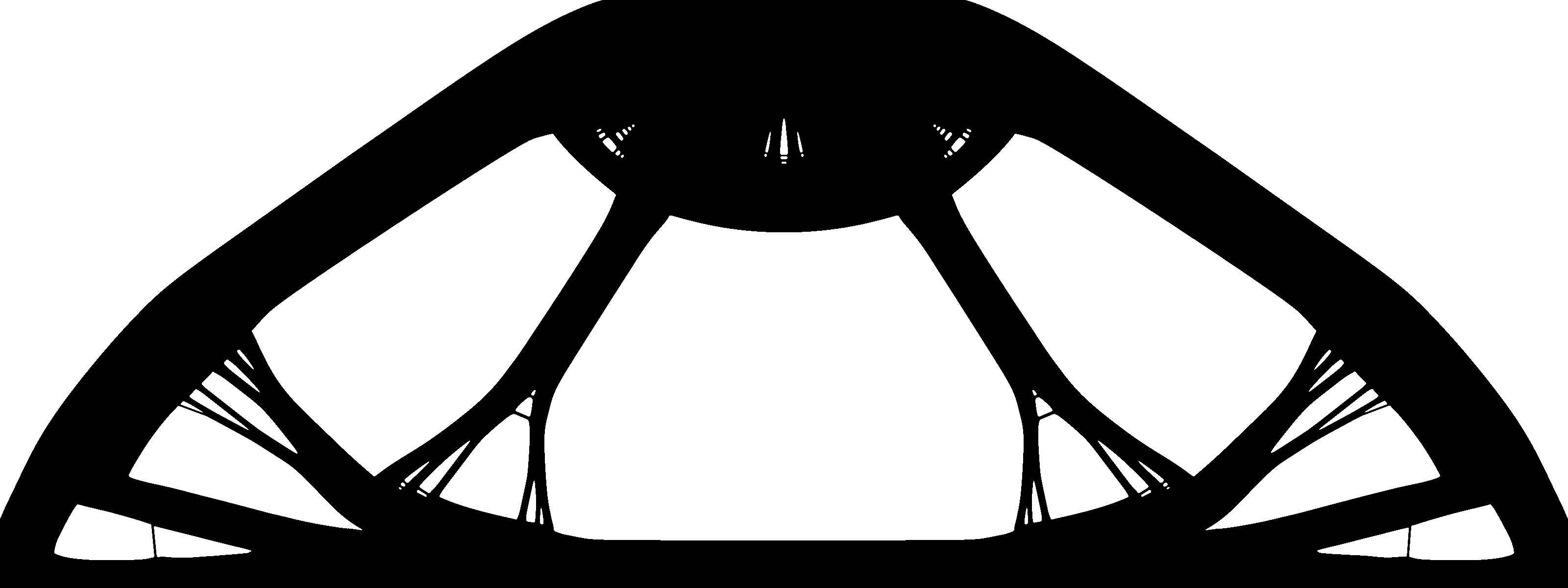}
		\vspace{7.5mm}
		\caption{}
		\label{fig:mbb0}
	\end{subfigure}
	\caption{(a) Boundary conditions and considered modular discretization of the MBB beam and lower-bound designs: (b) optimal traces of material stiffness matrices \replace{susceptible to the checkerboard issue ($s_\mathrm{FMO}=5.60$), (b) post-processed material stiffness distribution ($s_\mathrm{FMO,f}=6.92$), and (c)}{ ($s_\mathrm{FMO}=6.11$), and (c)} non-modular design obtained by TO \replace{($s_\mathrm{TO,\infty}=29.08$)}{($s_{\mathrm{TO},\infty}=20.69$).}}
\end{figure}

First, we examine the performance of our scheme for the Messerschmitt-B\"olkow-Blohm (MBB) beam problem\replace{, i.e., a simply-supported beam with a vertical unit force at its mid-span}{}. The beam is $0.375$ in height and $1.0$ in width, and discretized by $32\times12$ equisized square modules as shown in Fig.~\ref{fig:mbb_bc}. \replace{}{The beam is supported along $1/128$-wide segments at the bottom corners of the beam and loaded by a distributed load with the intensity $64$ acting at a $1/64$-wide segment at the top.}

\begin{figure}[!b]
	\centering
	\setlength{\tabcolsep}{2pt}
	\begin{tabular}{cccc}
		Horizontal clustering & Vertical clustering & Optimized structure & Wang module set\\
		\imagetop{\includegraphics[width=3.5cm]{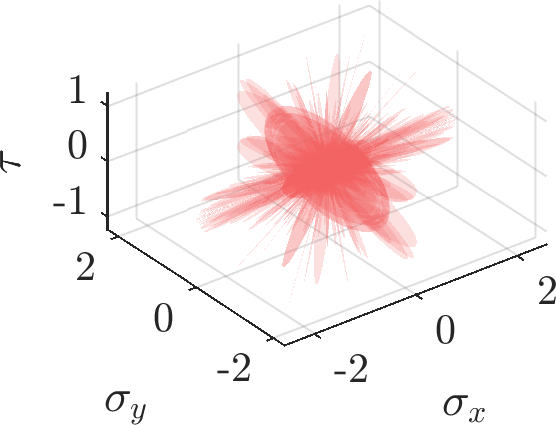}} & \imagetop{\includegraphics[width=3.5cm]{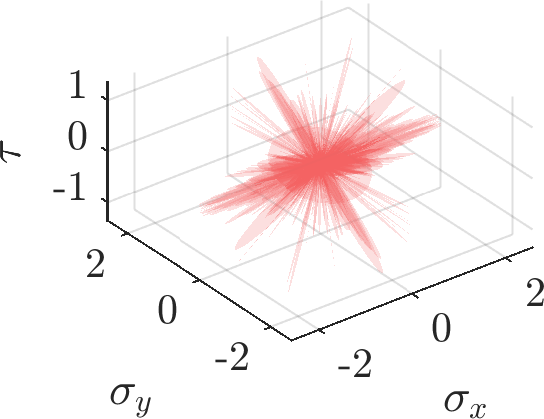}} & \imagetop{\includegraphics[height=0.135\linewidth]{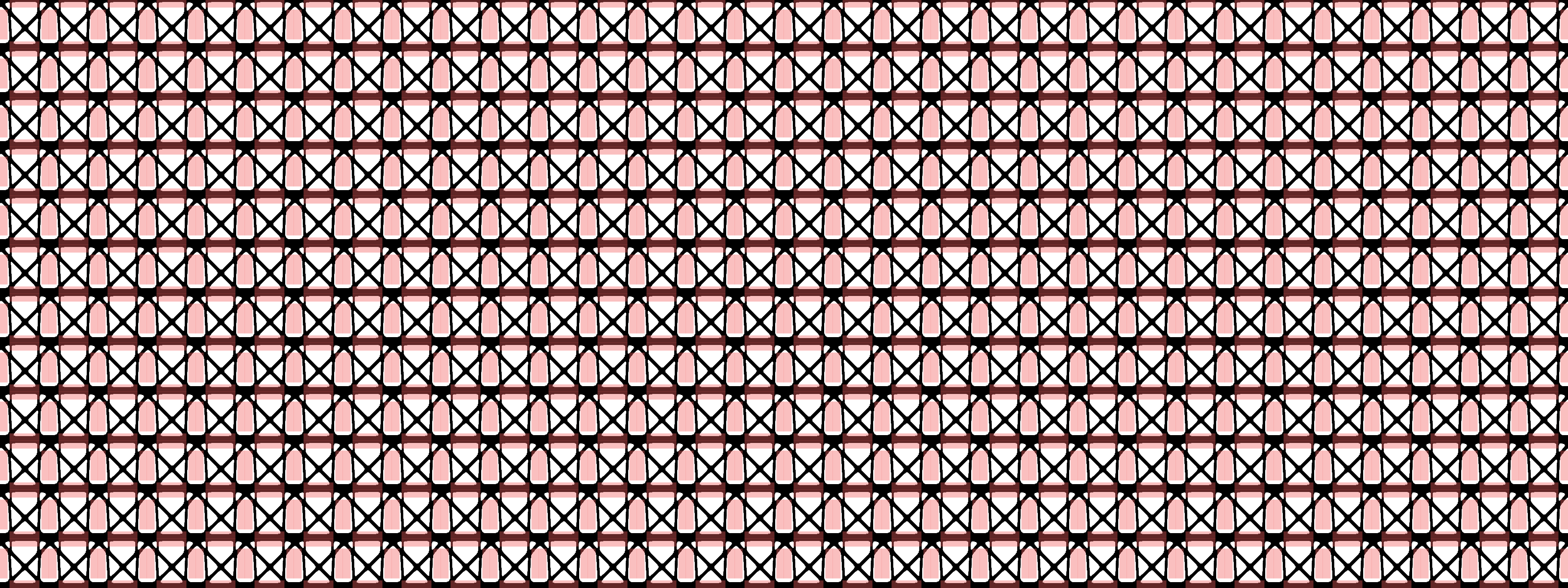}} & \imagetop{\includegraphics[width=0.18\linewidth]{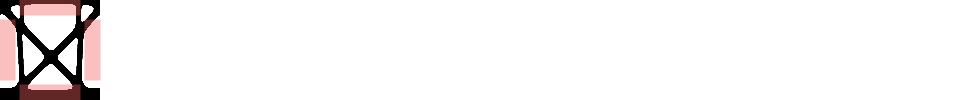}} \\
		\imagetop{\includegraphics[width=3.5cm]{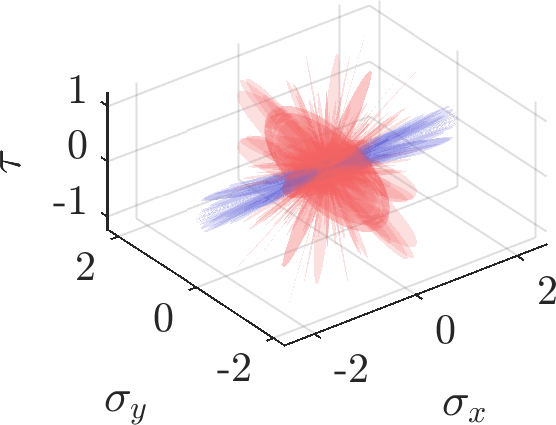}} & \imagetop{\includegraphics[width=3.5cm]{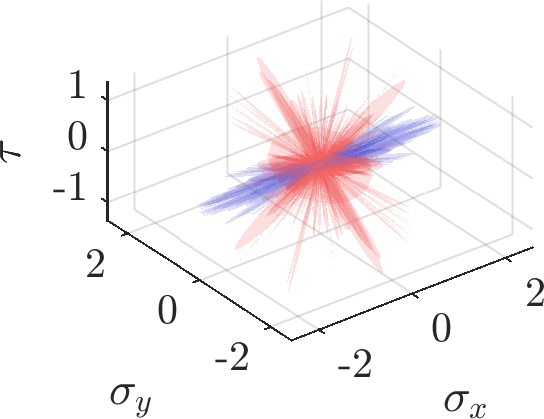}} & \imagetop{\includegraphics[height=0.135\linewidth]{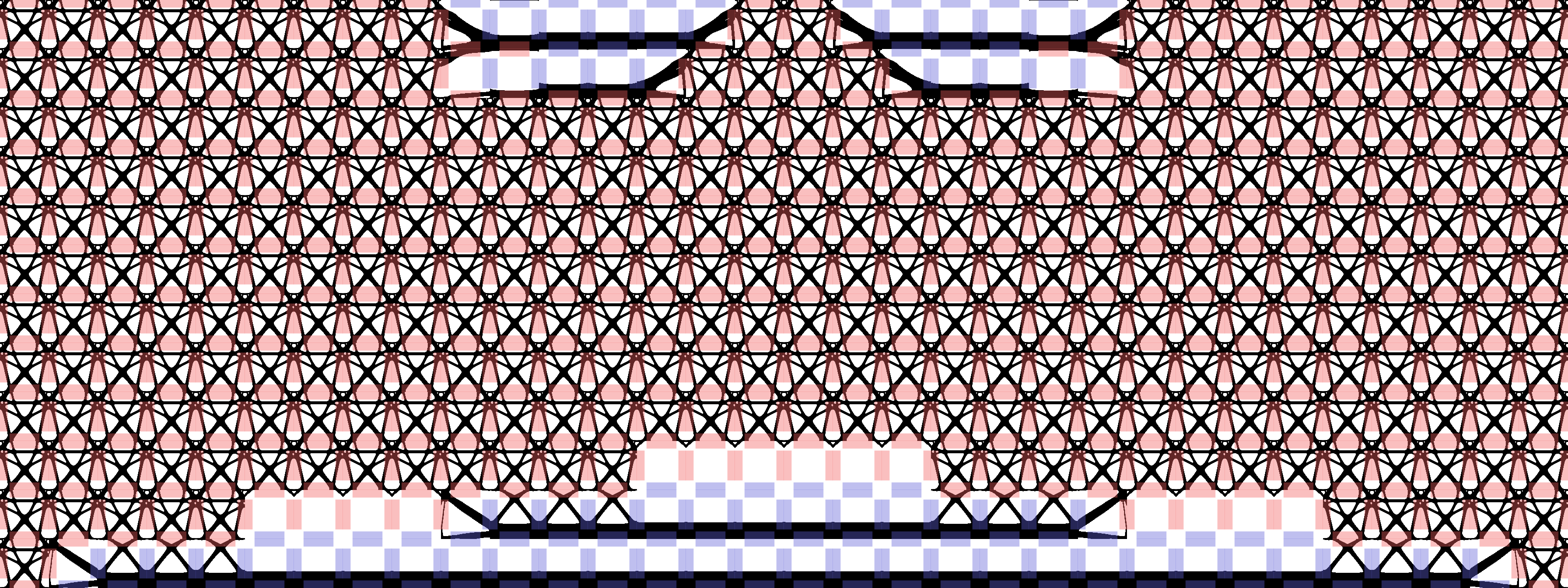}} & \imagetop{\includegraphics[width=0.18\linewidth]{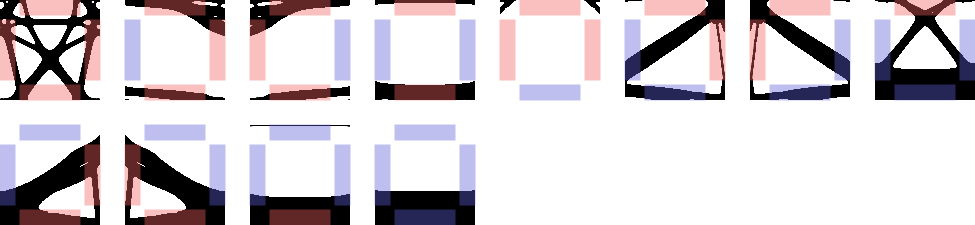}} \\
		\imagetop{\includegraphics[width=3.5cm]{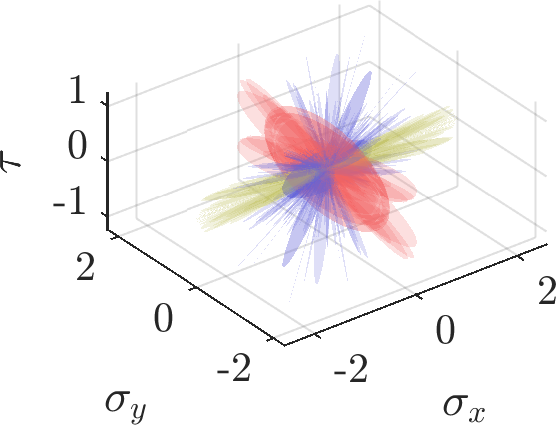}} & \imagetop{\includegraphics[width=3.5cm]{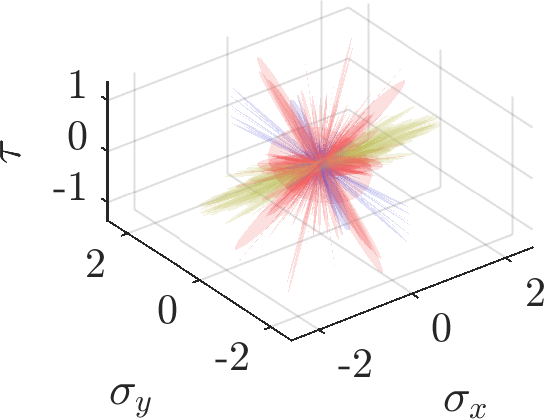}} & \imagetop{\includegraphics[height=0.135\linewidth]{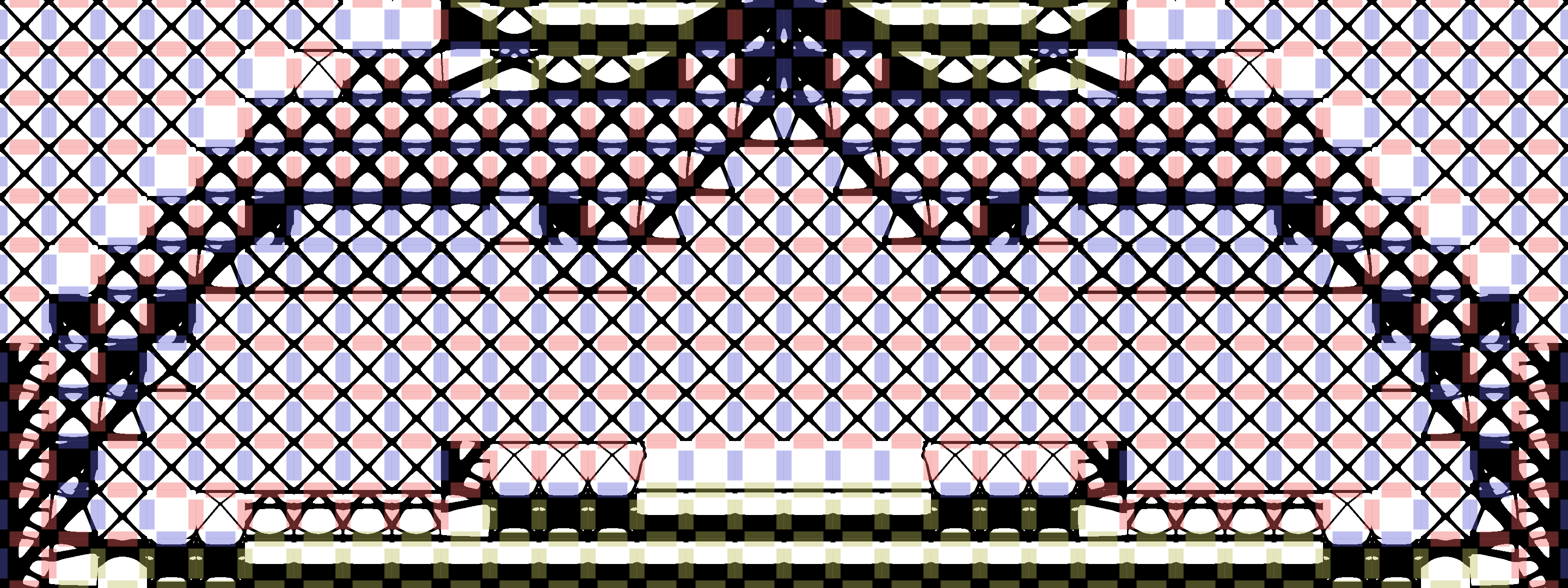}} & \imagetop{\includegraphics[width=0.18\linewidth]{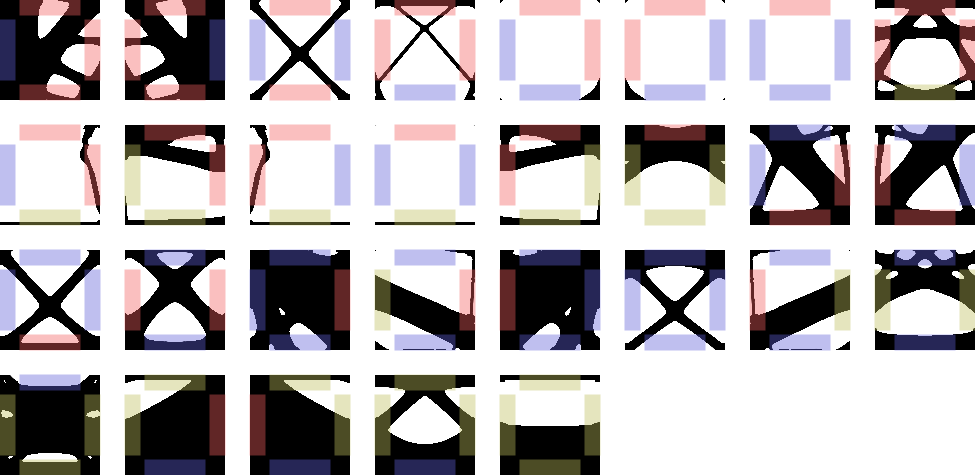}} \\
		\imagetop{\includegraphics[width=3.5cm]{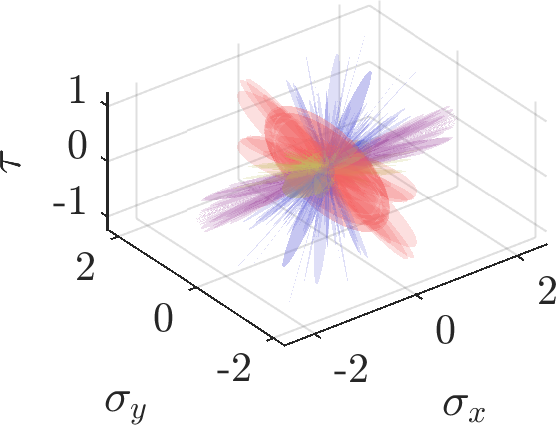}} & \imagetop{\includegraphics[width=3.5cm]{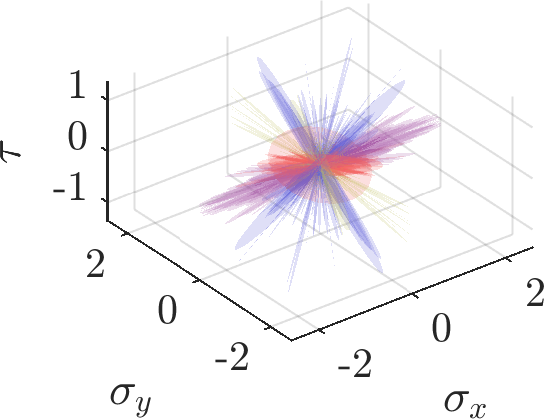}} & \imagetop{\includegraphics[height=0.135\linewidth]{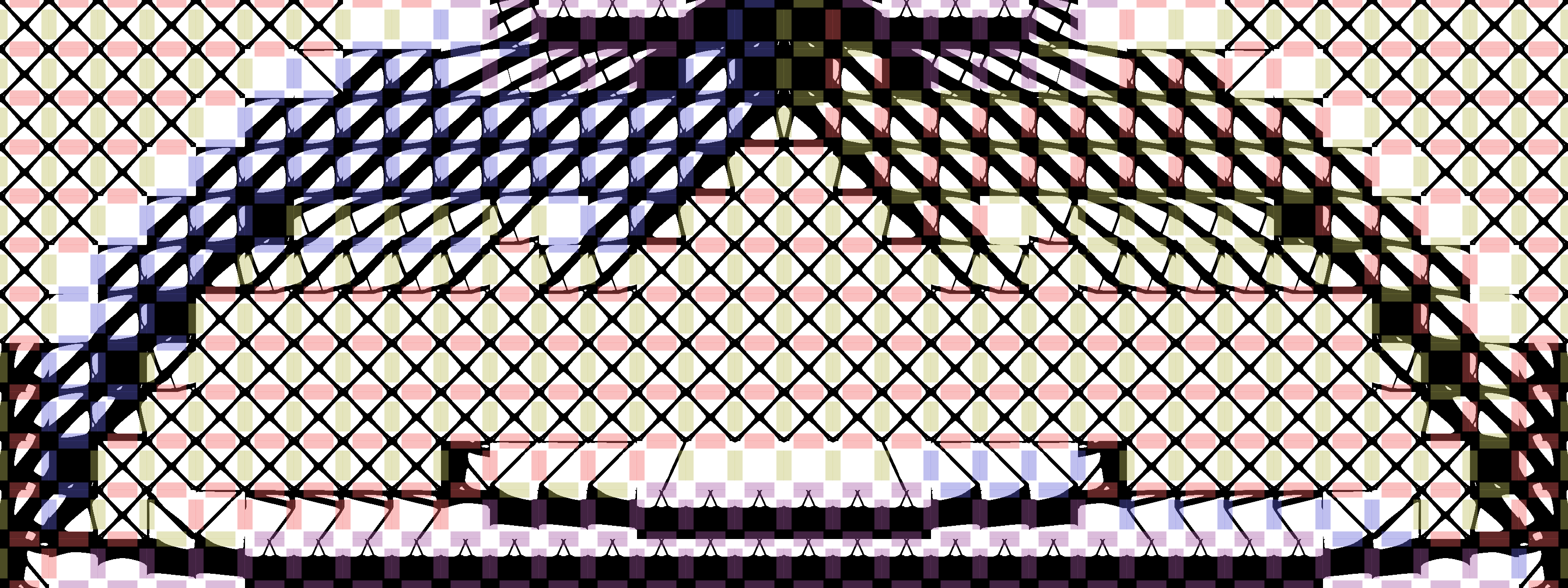}} & \imagetop{\includegraphics[width=0.18\linewidth]{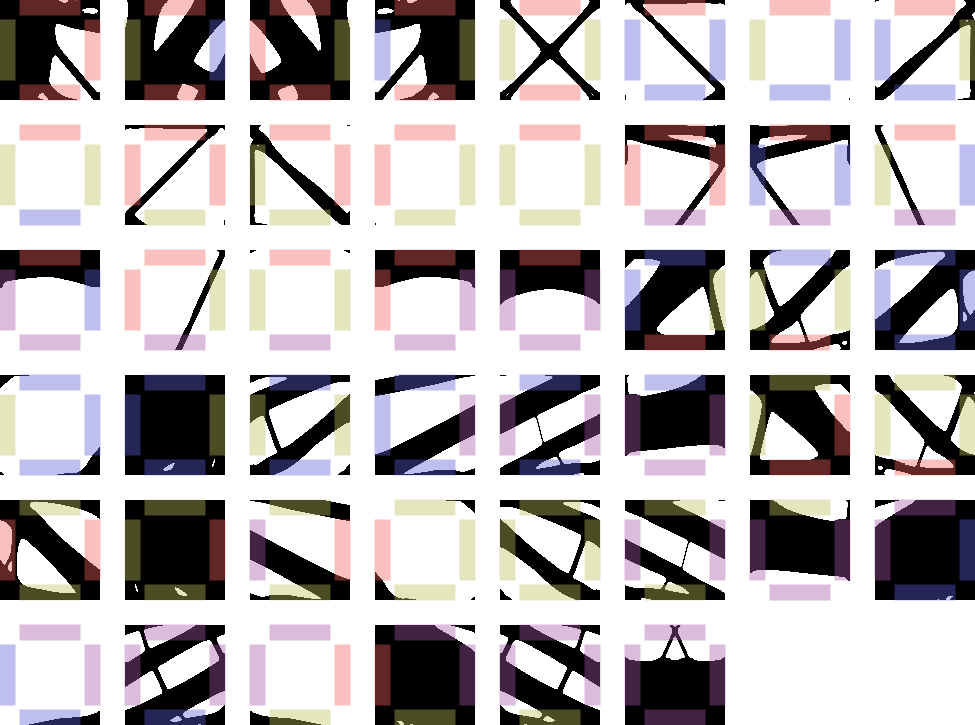}}
	\end{tabular}
	\caption{Modular designs of the MBB beam problem for the numbers of edge colors ranging from 1 to 4. The first two columns illustrate the partitioning produced by the hierarchical clustering algorithm, the third and fourth columns visualize optimized modular topologies and their assembly.}
	\label{fig:mbb}
\end{figure}

We start with discretizing the MBB beam with a rotated mesh, recall Section~\ref{sec:clustering}, and solve the FMO problem~\eqref{eq:fmo} that provides us with an optimal compliance of \replace{$s_\mathrm{FMO} = 5.60$. As is clearly visible in the plot of $\mathrm{Tr}\left(\semtrx{E}^{(\ell)}\right)$ in Fig.~\ref{fig:mbbnf}, the obtained design exhibits pronounced checkerboard pattern near the beam supports indicating that the optimization exploited the coarseness of the discretization and low order of the used elements. Indeed, when using the optimal elasticity matrices projected on a finer discretization containing $3,200 \times 1,200$ elements, i.e., a discretization that is approximately matching the one used subsequently in TO, the compliance significantly increases to $s_\mathrm{FMO}^{\mathrm{refined}}= 189.16$. When a simple post-processing step described in Section~\ref{sec:checkerboard} is performed, see Fig.~\ref{fig:mbbf}, the objective function increases to $s_\mathrm{FMO,f} = 6.92$ on a coarse mesh, but the refined-mesh compliance amounts to $s_\mathrm{FMO,f}^{\mathrm{refined}}= 16.15$ only. We believe that the latter increase is justifiable as the bilinear quadrilateral finite elements are prone to the shear locking phenomenon, so a very fine mesh is usually required in the case of structures under bending.}{$s_\mathrm{FMO}=6.11$, see Fig.~\ref{fig:mbbf}.} \replace{With}{Using} the \replace{post-processed optimized}{optimal} material stiffnesses, we proceed \replace{by removing the modules whose maximum eigenvalue of all edge stiffnesses is below $1\%$ of $\overline{\epsilon}$, and follow}{with} the clustering procedure explained in Section~\ref{sec:clustering} for a hierarchical merging of the varying material stiffness \replace{matrices}{tensors} into a predefined number of groups, one to four in our case. This clustering procedure is applied twice: once for horizontal and once for vertical edges. In the two-dimensional setting, this process can easily be visualized as follows: the stiffness matrices $\semtrx{E}^{(\ell)}$ are a linear mapping of the engineering strain vector $\sevek{\varepsilon}^{(\ell)}$ onto the stress vector $\sevek{\sigma}^{(\ell)}$, Eq. \eqref{eq:Hooke_tens}. Hence, a sphere of all unit engineering strains is mapped by the elasticity matrix onto a $3$-dimensional ellipsoid whose surface corresponds to all admissible stresses induced in the material by arbitrary unit-sized strains, determined using the $2$-norm. The clustering process partitions the ellipsoid of all elements in FMO into a given number of bins, trying to replace similar ellipsoids with an approximate one corresponding to the cluster centroid. Consequently, the lengths and orientations of the ellipsoid's axes characterize the material stiffness. Partitioning induced by the clustering is denoted by coloring of these ellipsoids, see first two columns in Fig.~\ref{fig:mbb}.

After clustering, each module position encompasses a quadruple of edge labels, defining therefore a Wang tile, and these tiles are placed in an assembly plan satisfying the Wang tiling formalism, recall Section~\ref{sec:wangtiles} and the developments in \citep{cohen2003}. For such modular assemblies, we solve a reduced-sized modular-FMO problem \eqref{eq:mfmo} to receive effective stiffnesses of individual modules, and use their traces to construct initial guesses for modular TO, recall Section~\ref{sec:modularityconstr}. 

Although the TO algorithm described in Section~\ref{sec:topopt} usually converges to a neighborhood of a locally-optimal solution, we believe that it is reasonable to assume that the performance of all modular designs shall be bounded by a PUC solution from above, and by a non-modular design from below. Please note, however, that despite the fact that we have not observed any violations of this assumption, it is not guaranteed in general due to the inherent non-convexity of the optimization problems. 
Using the parameters of TO listed at the beginning of the section, we discretize the interior of each module type using a regular $100\times100$ mesh, implying that the MBB beam contains $3.84$ millions bilinear quadrilateral finite elements. Using the filter radius of $3.5$ elements, standard TO converges to the design in Fig.~\ref{fig:mbb0} of compliance \replace{$s_{\mathrm{TO,\infty}}=29.08$}{$s_{\mathrm{TO},\infty} = 20.69$}. As expected, imposing modularity degrades structural performance: the worst-case single-module, i.e., PUC, design obtained for a single-color clustering possesses the compliance \replace{$s_\mathrm{TO,1} = 63.31$}{$s_\mathrm{TO,1} = 56.91$}, see the first row in Fig.~\ref{fig:mbb}. Using two colors, we obtain a design in the second row of Fig.~\ref{fig:mbb} composed of \replace{$14$}{$12$} modules, of an improved compliance of \replace{$s_\mathrm{TO,2} = 48.18$}{$s_\mathrm{TO,2} = 49.15$}. The three-color-based clustering produces a design with \replace{$25$}{$29$} modules, exhibiting the compliance \replace{$s_\mathrm{TO,3} = 38.26$}{$s_\mathrm{TO,3} = 31.24$}, see the third row in Fig.~\ref{fig:mbb}. Finally, the four-color clustering provides us with the design in the fourth row of Fig.~\ref{fig:mbb}, composed of $46$ modules and compliance \replace{$s_\mathrm{TO,4} = 34.30$}{$s_\mathrm{TO,4} = 29.09$}, which is \replace{only $18\%$}{$41\%$} more compliant than the non-modular design. We wish to emphasize that, because our hierarchical clustering algorithm respects the symmetric stiffness distribution, the resulting module sets contain multiple module pairs that have symmetric topologies.

\replace{}{In addition to structural performance, we have also investigated the influence of modularity on stress distribution. For the MBB beam problem, the maximum von Mises stress in optimized modular designs exceed the maximum stress value in the non-modular design, see Tab.~\ref{tab:mbb}, which is consistent with our findings in optimization of modular truss structures \cite{tyburec2020}.}

\begin{table}[!t]
	\centering
	\small
	\caption{\replace{}{Overview of the modular MBB beam performances, where $s_\mathrm{TO}$ denotes structural compliance and $\sigma_{\mathrm{v},\max}$ the maximum von Mises stress.}}
	\setlength\tabcolsep{0.5em}
	\begin{tabular}{lrr}
		Design & $s_\mathrm{TO}$ & $\sigma_{\mathrm{v},\max}$\\
		\hline
		$1$ module over $1$ color & $56.91$ & $303.001$ \\
		$12$ modules over $2$ colors & $49.15$ & $369.854$ \\
		$29$ modules over $3$ colors & $31.24$ & $285.621$ \\
		$46$ modules over $4$ colors & $29.09$ & $295.057$ \\
		Non-modular & $20.69$ & $243.935$
	\end{tabular}
	\label{tab:mbb}
\end{table}

\subsection{Force inverter}\label{sec:inverter}

\begin{figure}[!b]
	\centering
	\begin{subfigure}[t]{0.300\linewidth}
		\begin{tikzpicture}[scale=0.74]
		\scaling{0.37}
		\point{a}{0}{4};
		\point{b}{0}{0};
		\point{c}{8}{0};
		\point{d}{8}{4};
		\point{e}{8}{8};
		\point{f}{0}{8};
		\point{se}{8.35}{4};\point{sb}{9.5}{4};
		\draw[gray, dashdotted, xstep=0.4cm, ystep=0.4cm] (0,0) grid (4,4);
		\beam{2}{a}{b};\beam{2}{b}{c};\beam{2}{c}{d};\beam{2}{d}{e};\beam{2}{e}{f};\beam{2}{f}{a};
		\support{1}{b}[-90];\support{1}{f}[-90];
		\dimensioning{1}{b}{c}{-0.75}[$1$];
		\dimensioning{2}{b}{a}{-1.25}[$0.5$];
		\dimensioning{2}{a}{f}{-1.25}[$0.5$];
		\load{1}{a}[180][0.75][0];
		\draw[black,fill=black] (a) circle (1mm);
		\draw[black,fill=black] (d) circle (1mm);
		\draw[style={double, thick},-stealth] (se) -- (sb);
		\notation{1}{a}{$1$}[above left=0mm and 4mm];
		\notation{1}{a}{$k_\mathrm{in}$}[below left=0mm and -0.5mm];
		\notation{1}{d}{$k_\mathrm{out}$}[below right=0mm and -0.5mm];
		\notation{1}{d}{$u_\mathrm{out}$}[above right=0mm and 2mm];
		\end{tikzpicture}
		\caption{}
		\label{fig:inv_bc}
	\end{subfigure}%
	\hspace{1cm}\begin{subfigure}[t]{0.185\linewidth}
		\raisebox{6.5mm}{\includegraphics[width=\linewidth]{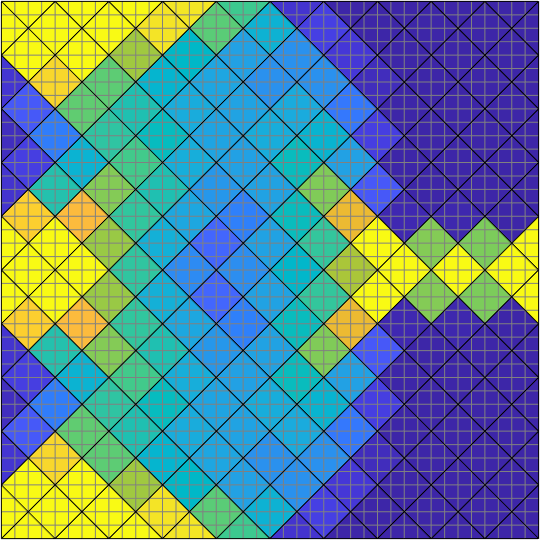}}
		\caption{}
		\label{fig:invf}
	\end{subfigure}%
	\hspace{1cm}\begin{subfigure}[t]{0.185\linewidth}
		\raisebox{6.5mm}{\includegraphics[width=\linewidth]{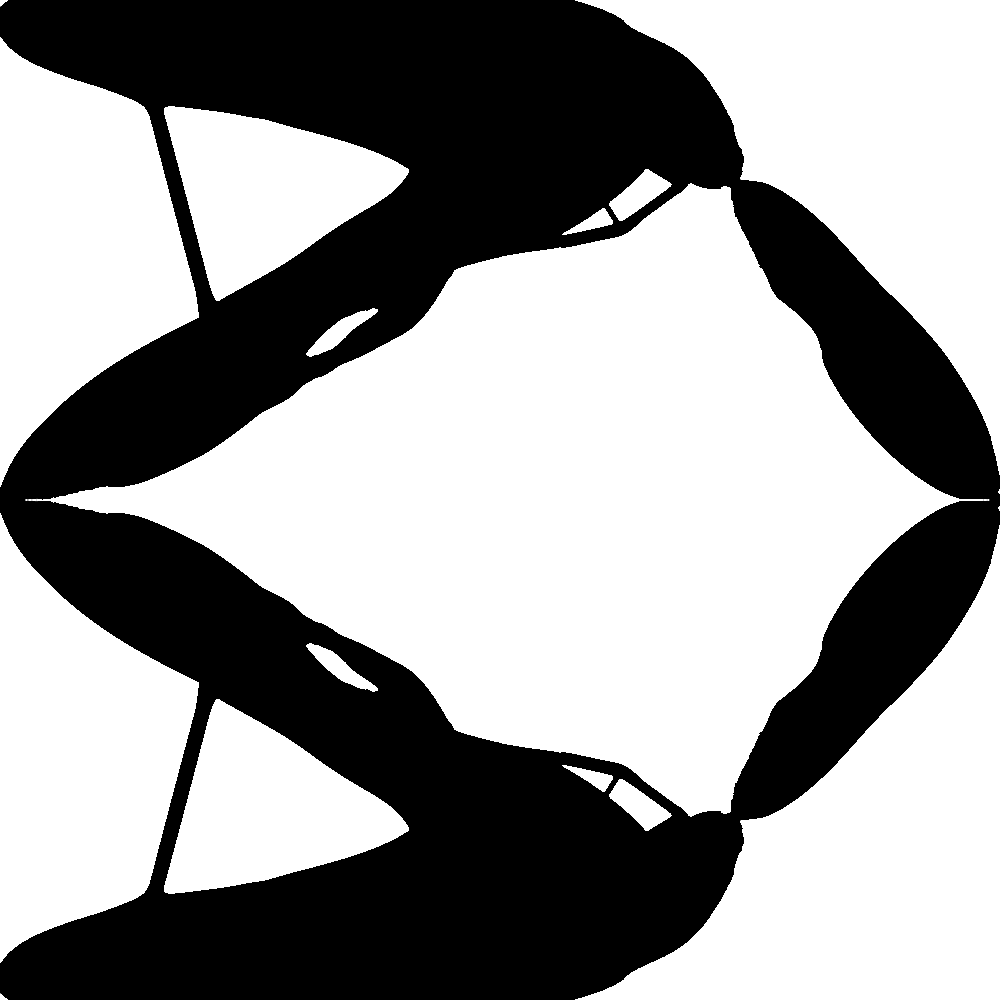}}
		\caption{}
		\label{fig:inv0}
	\end{subfigure}
	\caption{(a) Boundary conditions of the inverter problem and assumed modular discretization, and lower-bound designs: (b) optimal traces of material stiffness matrices \replace{($s_\mathrm{FMO}=-1.000$), (c) post-processed material stiffness distribution ($s_\mathrm{FMO,f}=-0.282$), and (d)}{($s_{\mathrm{FMO}}=-0.892$), and (c)} non-modular design obtained by TO \replace{($s_{\mathrm{TO},\infty} =-0.354$)}{($s_{\mathrm{TO},\infty} = -0.398$)}.}
\end{figure}

Our second illustrative problem involves designing a modular force inverter mechanism. We consider the design domain to be $1.0$ both in width and height and to consist of $10 \times 10$ square-sized modules, as shown in Fig.~\ref{fig:inv_bc}. In this domain, \replace{$0.01$}{$0.025$}-long segments at the top and bottom of the left edge are fixed. \replace{For a coarse, module-based mesh adopted in the FMO problems \eqref{eq:fmo} and \eqref{eq:mfmo}, these boundary conditions can not be reproduced exactly, so we collapse these segments into single corner points.}{}The loading is posed by a horizontal unit force acting in the middle of the left edge. Similarly to~\cite{sigmund1997}, we add a horizontal spring of the stiffness $k_\mathrm{in} = 1$ to this point of loading and another, output horizontal spring of the stiffness $k_\mathrm{out} = 0.05$ is placed at the mid-height of the right edge of the domain.

For these boundary conditions, we search \replace{}{for} a material distribution that makes the unit input force minimize the motion of the output point to the right, denoted by $u_\mathrm{out}$ in Fig.~\ref{fig:inv_bc}. When solving the FMO formulation \eqref{eq:fmo} on a rotated mesh\replace{and without post-processing,}{} we receive \replace{$s_\mathrm{FMO} = -1.000$ on a coarse mesh, Fig.~\ref{fig:invnf}, and $s_\mathrm{FMO}^{\mathrm{refined}} = -0.072$ on a $100$-times refined mesh with accurate boundary conditions}{$s_{\mathrm{FMO}}=-0.892$, see Fig.~\ref{fig:invf}}. \replace{After application of the post-processing step, we obtain $s_\mathrm{FMO,f} = -0.282$ on the coarse mesh, Fig.~\ref{fig:invf}, and $s_\mathrm{FMO,f}^{\mathrm{refined}} = -0.128$ on the refined one. These values indicate that even though any visual checkerboard-like patterns is not present in the traces of the stiffness matrices in Fig.~\ref{fig:invnf}, the design performance is not physical due to the enormous increase of the output displacement with discretization refinement of optimal design. With the post-processing step, the output displacement inevitably increases, but the mesh shows a much more stable response with refinement.}{}

\begin{figure}[!t]
	\centering
	\setlength{\tabcolsep}{2pt}
	\begin{tabular}{cccc}
		Horizontal clustering & Vertical clustering & Optimized structure & Wang module set\\
		\imagetop{\includegraphics[width=3.5cm]{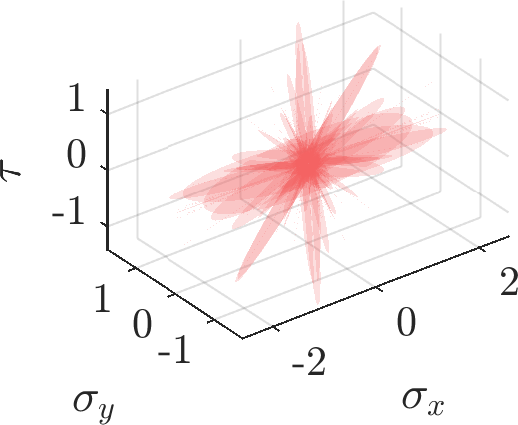}} & \imagetop{\includegraphics[width=3.5cm]{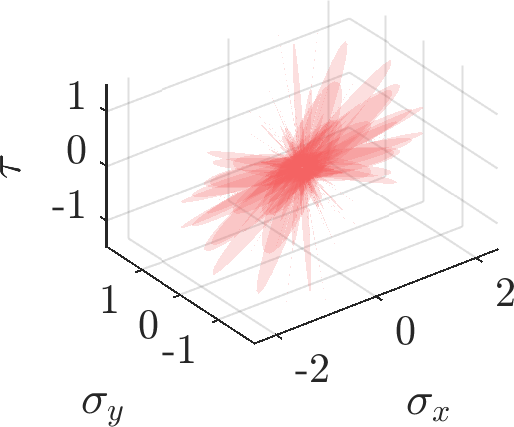}} & \imagetop{\includegraphics[height=0.16\linewidth]{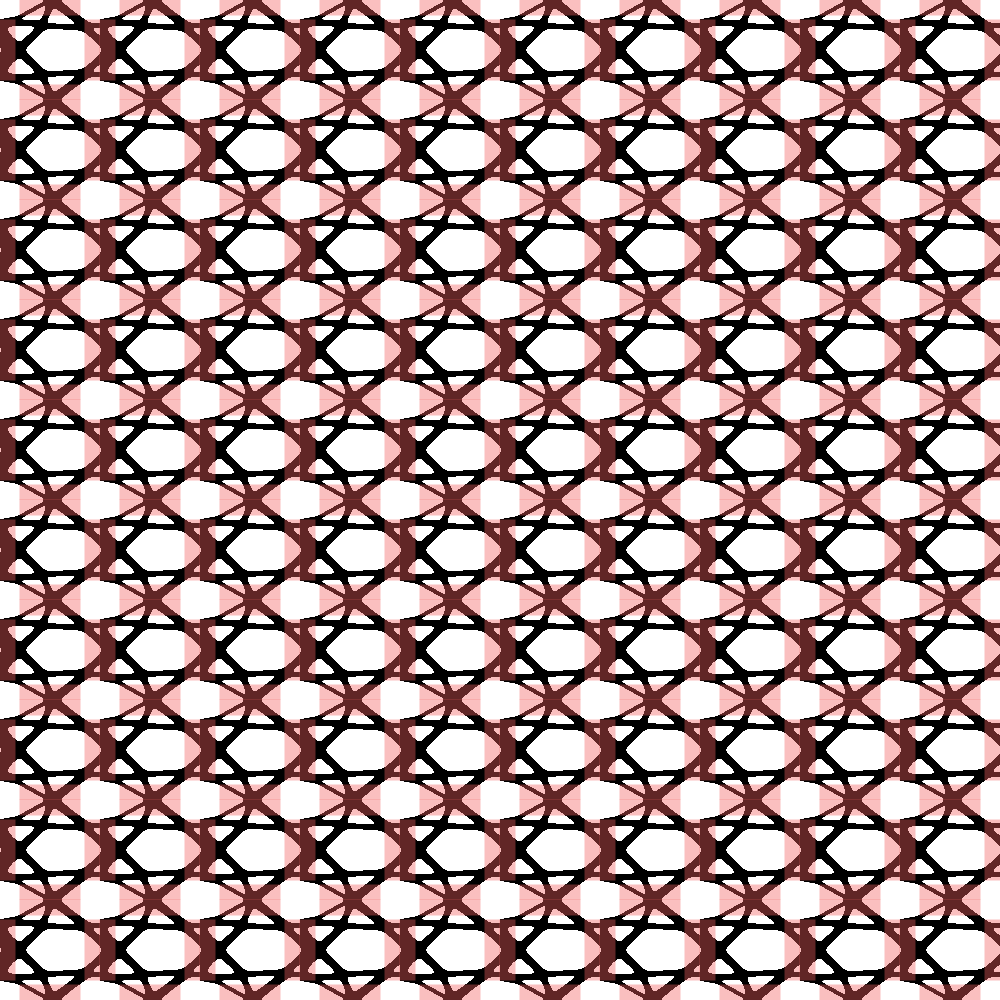}} & \imagetop{\includegraphics[width=0.215\linewidth]{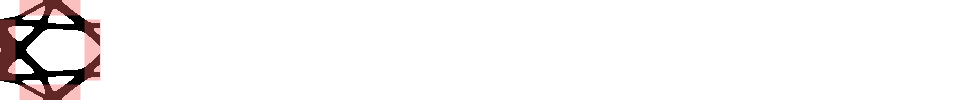}} \\
		\imagetop{\includegraphics[width=3.5cm]{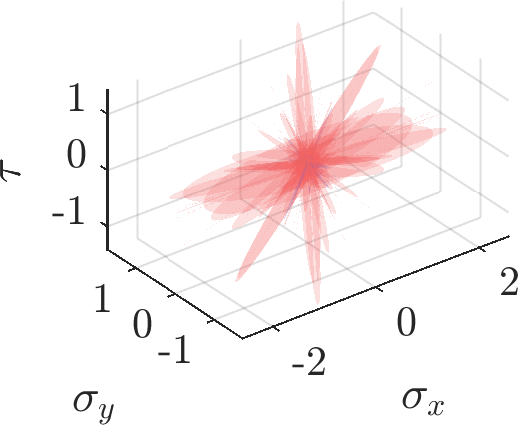}} & \imagetop{\includegraphics[width=3.5cm]{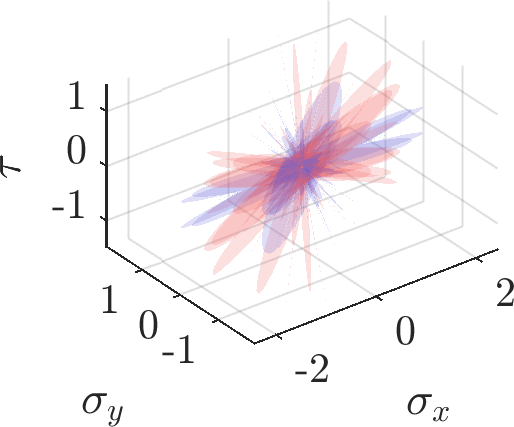}} & \imagetop{\includegraphics[height=0.16\linewidth]{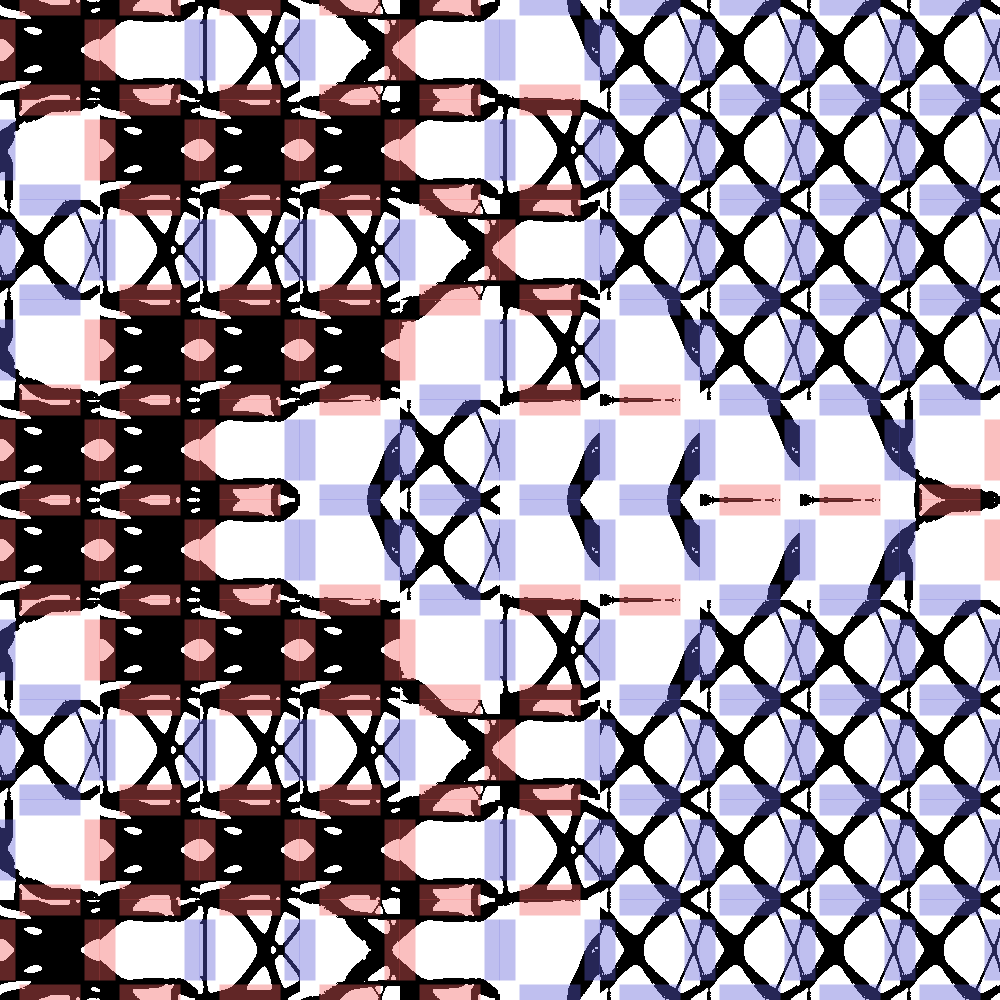}} & \imagetop{\includegraphics[width=0.215\linewidth]{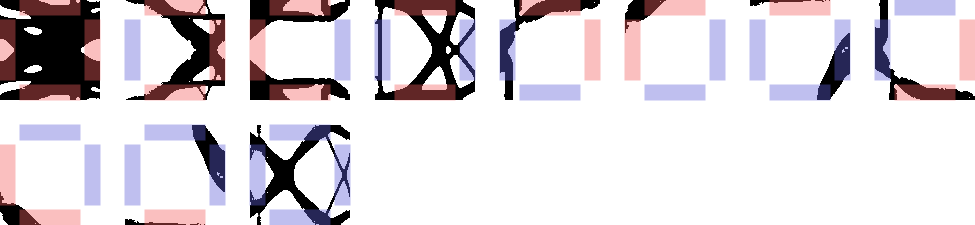}} \\
		\imagetop{\includegraphics[width=3.5cm]{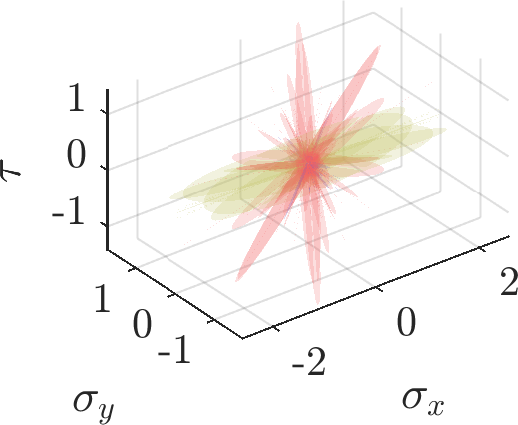}} & \imagetop{\includegraphics[width=3.5cm]{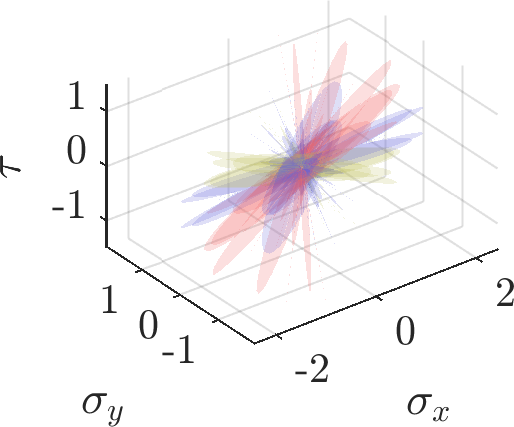}} & \imagetop{\includegraphics[height=0.16\linewidth]{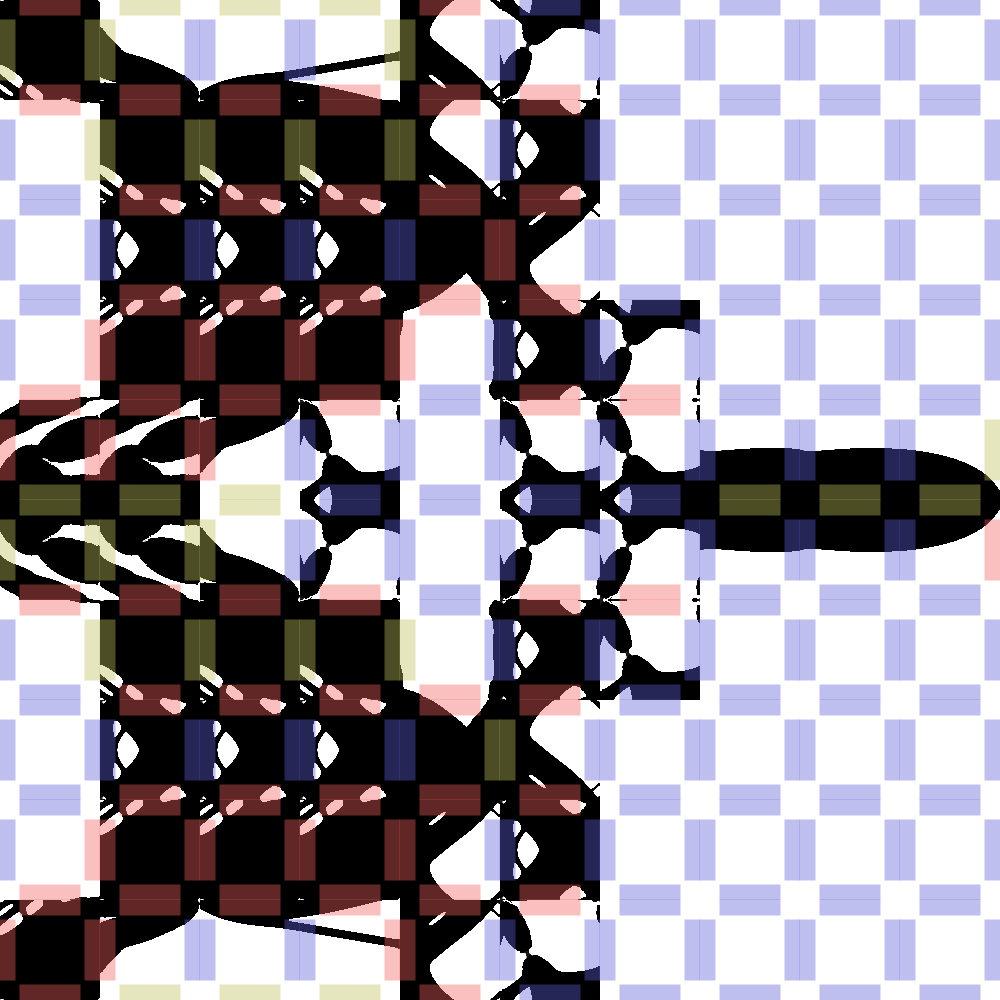}} & \imagetop{\includegraphics[width=0.215\linewidth]{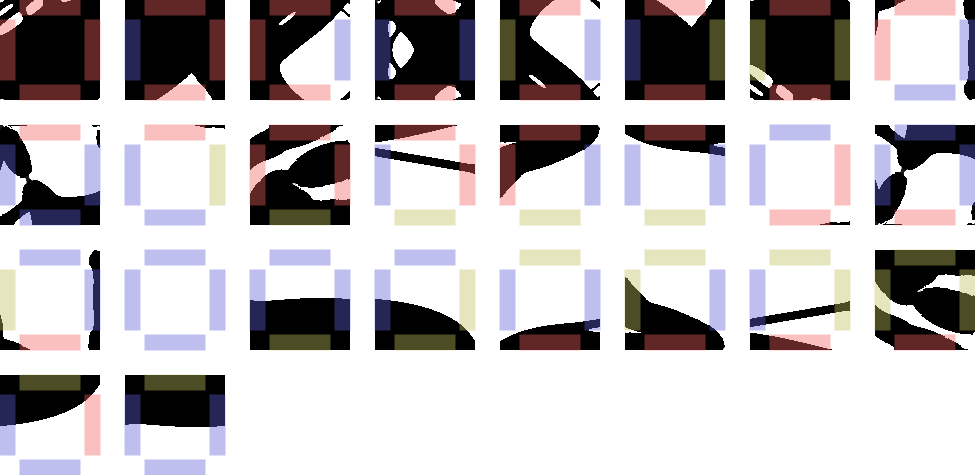}} \\
		\imagetop{\includegraphics[width=3.5cm]{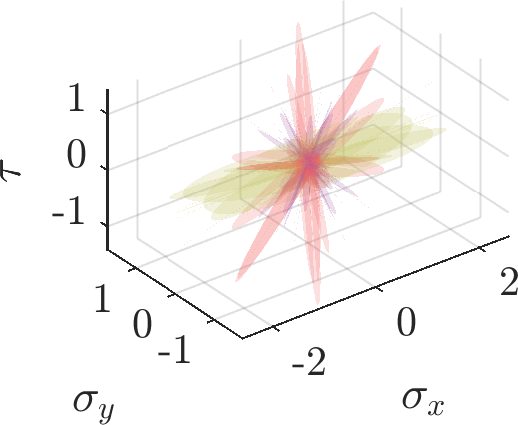}} & \imagetop{\includegraphics[width=3.5cm]{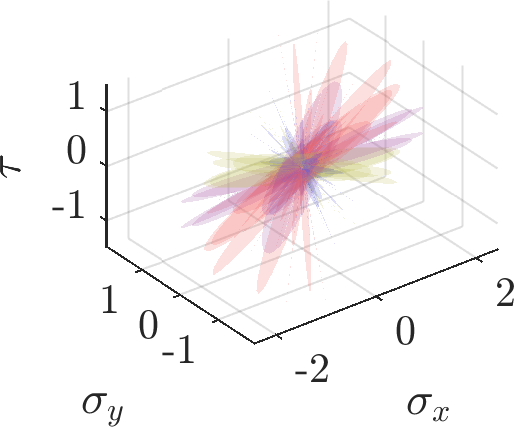}} & \imagetop{\includegraphics[height=0.16\linewidth]{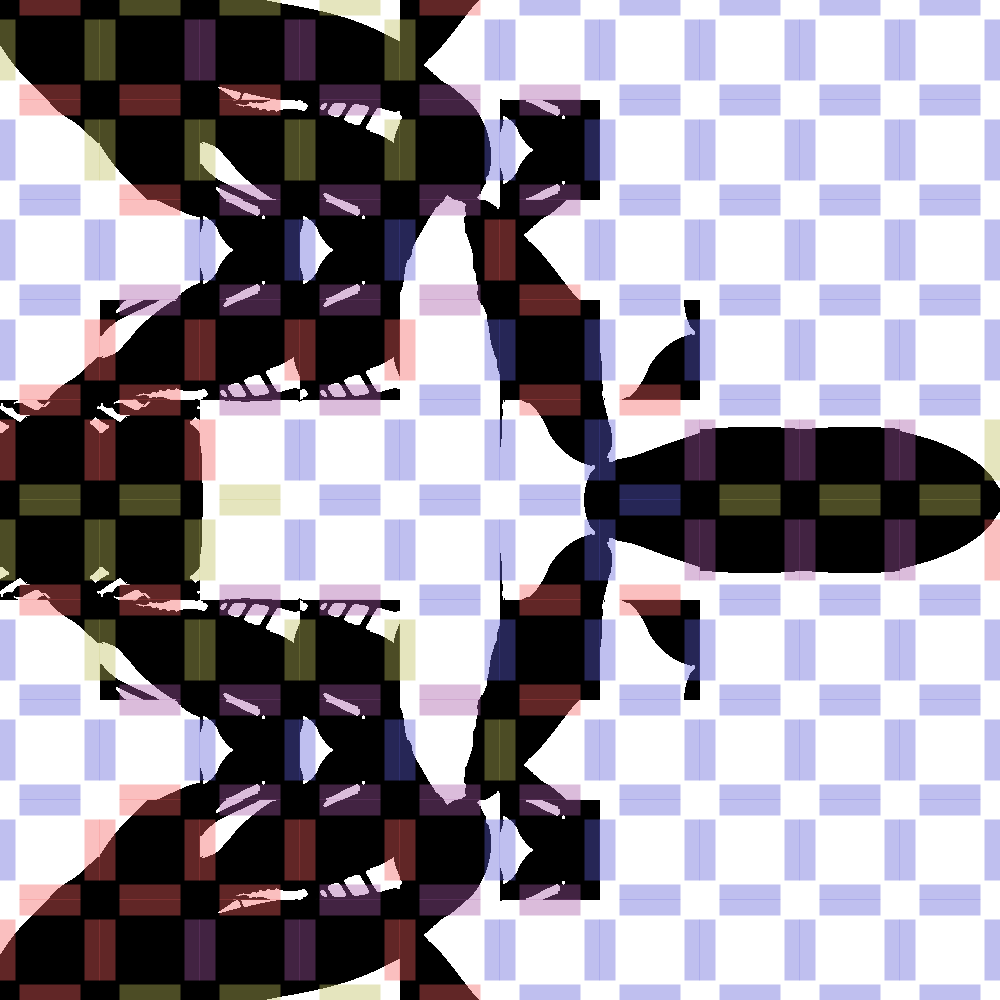}} & \imagetop{\includegraphics[width=0.215\linewidth]{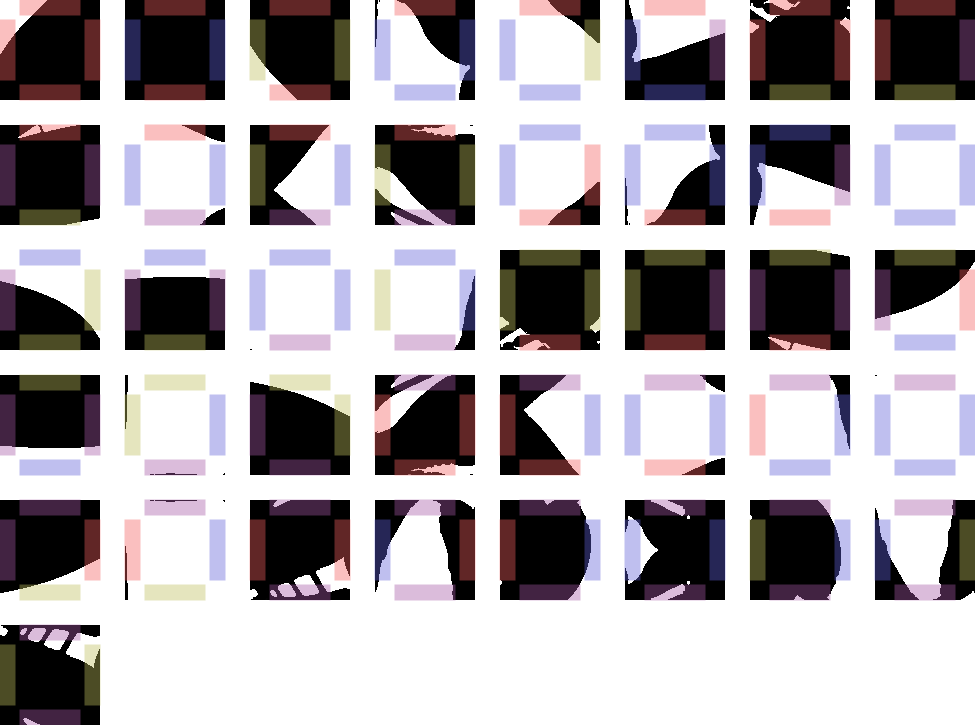}}
	\end{tabular}
	\caption{Modular designs of the force inverter problem for the numbers of edge colors ranging from 1 to 4. The first two columns illustrate the partitioning produced by the hierarchical clustering algorithm, the third and fourth columns visualize optimized modular topologies and their assembly.}
	\label{fig:inv}
\end{figure}

The clustering procedure, illustrated within the first two columns in Fig.~\ref{fig:inv}, produced assembly plans shown in the third column of the same figure. For these assembly plans, we solved the modular-FMO problem \eqref{eq:mfmo} to receive the initial guesses for TO. Similarly to the MBB beam design, we may provide performance bounds of topologically-optimized designs established by a non-modular design, \replace{$s_{\mathrm{TO},\infty} =-0.354$}{$s_{\mathrm{TO},\infty} =-0.398$} in Fig.~\ref{fig:inv0}, and a single color clustering PUC, $s_{\mathrm{TO},1} = 0.000$, see the top row of Fig.~\ref{fig:inv} showing that the PUC design does not connect the unitary input force to the output properly. However, based on our tests, we believe that such a PUC inverter  design does not exist for the selected combination of materials used in TO and for the given modular discretization. Please note that although a PUC force inverter was presented in \citep{wu2021b} for a symmetric part of the design domain, its extension to the full domain would require a second module type in our formalism.

As shown in Fig.~\ref{fig:inv}, the two-color-based clustering provides a design composed of \replace{$8$}{$11$} modules with the performance \replace{$s_{\mathrm{TO},2} = -0.211$}{$s_{\mathrm{TO},2} = -0.125$}. Three color-based design possesses \replace{$s_{\mathrm{TO},3}=-0.288$}{$s_{\mathrm{TO},3} = -0.293$}, and requires \replace{$23$}{$26$} different module types. Finally, \replace{$25$}{$41$} module types in the four-color based clustering leads to the output displacement \replace{$s_{\mathrm{TO},4}=-0.290$}{$s_{\mathrm{TO},4} = -0.353$}. From the performance point of view, \replace{$3$- and}{the} $4$-color design\replace{s}{} result\replace{}{s} in \replace{}{an} objective\replace{s}{} that \replace{are}{is} only by \replace{$18\%$}{$11\%$} worse than that of the non-modular design. \replace{}{Finally, contrary to the MBB beam problem, modular inverters tend to exhibit a lower maximum von Mises stress, see Tab.~\ref{tab:inv}.}

\begin{table}[!t]
	\centering
	\small
	\caption{\replace{}{Overview of the modular inverter performances, where $u_\mathrm{in}$ denotes the horizontal displacement at the force input, $s_\mathrm{TO}$ the mechanism performance, and $\sigma_{\mathrm{v},\max}$ the maximum von Mises stress. Values marked by $^\dagger$ express a non-functional mechanism. Further, GA abbreviates geometrical advantage, evaluated as $s_\mathrm{TO}/u_\mathrm{in}$.}}
	\setlength\tabcolsep{0.5em}
	\begin{tabular}{lrrrr}
		Design & $u_\mathrm{in}$ & $s_\mathrm{TO}$ & GA & $\sigma_{\mathrm{v},\max}$\\
		\hline
		$1$ module over $1$ color & $0.000$ & $^\dagger0.000$ & $0.000$ & - \\
		$11$ modules over $2$ colors & $0.982$ & $-0.125$ & $-0.127$ & $7.627$ \\
		$26$ modules over $3$ colors & $0.970$ & $-0.293$ & $-0.302$ & $12.822$ \\
		$41$ modules over $4$ colors & $0.963$ & $-0.353$ & $-0.366$ & $15.981$ \\
		Non-modular & $0.968$ & $-0.398$ & $-0.411$ & $13.842$
	\end{tabular}
	\label{tab:inv}
\end{table}

\subsection{Gripper}\label{sec:gripper}

\begin{figure}[!b]
	\centering
	\begin{subfigure}[t]{0.300\linewidth}
		\begin{tikzpicture}[scale=0.74]
		\scaling{0.37}
		\point{a}{0}{4};
		\point{b}{0}{0};
		\point{c}{8}{0};
		\point{d}{8}{2.4};
		\point{e}{6.4}{2.4};
		\point{f}{6.4}{5.6};
		\point{g}{8}{5.6};
		\point{h}{8}{8};
		\point{i}{0}{8};
		\point{dh}{6.4}{0};
		\point{se1}{8}{2.6};\point{sb1}{8}{3.6};
		\point{se2}{8}{5.4};\point{sb2}{8}{4.4};
		\draw[gray, dashdotted, xstep=0.4cm, ystep=0.4cm] (b) grid (d);
		\draw[gray, dashdotted, xstep=0.4cm, ystep=0.4cm] (0,1.2) grid (f);
		\draw[gray, dashdotted, xstep=0.4cm, ystep=0.4cm] (0,2.8) grid (4,4);
		\fill [black] (e) rectangle (4,1.3);
		\fill [black] (f) rectangle (4,2.7);
		\beam{2}{a}{b};\beam{2}{b}{c};\beam{2}{c}{d};\beam{2}{d}{e};\beam{2}{e}{f};\beam{2}{f}{g};\beam{2}{g}{h};\beam{2}{h}{i};\beam{2}{i}{a};
		\support{1}{b}[-90];\support{1}{i}[-90];
		\dimensioning{1}{b}{dh}{-0.75}[$0.8$];
		\dimensioning{1}{dh}{c}{-0.75}[$0.2$];
		\dimensioning{2}{b}{a}{-1.25}[$0.5$];
		\dimensioning{2}{a}{i}{-1.25}[$0.5$];
		\dimensioning{2}{d}{g}{4.75}[$0.4$];
		\load{1}{a}[180][0.75][0];
		\draw[black,fill=black] (a) circle (1mm);
		\draw[black,fill=black] (d) circle (1mm);
		\draw[black,fill=black] (g) circle (1mm);
		\draw[style={double, thick},-stealth] (sb1) -- (se1);
		\draw[style={double, thick},-stealth] (sb2) -- (se2);
		\notation{1}{a}{$1$}[above left=0mm and 4mm];
		\notation{1}{a}{$k_\mathrm{in}$}[below left=0mm and -0.5mm];
		\notation{1}{d}{$k_\mathrm{out}$}[below right=0mm and -0.5mm];
		\notation{1}{d}{$u_{\mathrm{o}1}$}[above left=0mm and -0.5mm];
		\notation{1}{g}{$k_\mathrm{out}$}[above right=0mm and -0.5mm];
		\notation{1}{g}{$u_{\mathrm{o}2}$}[below left=0mm and -0.5mm];
		\end{tikzpicture}
		\caption{}
		\label{fig:grip_bc}
	\end{subfigure}%
	\hspace{1cm}\begin{subfigure}[t]{0.185\linewidth}
		\raisebox{6.5mm}{\includegraphics[width=\linewidth]{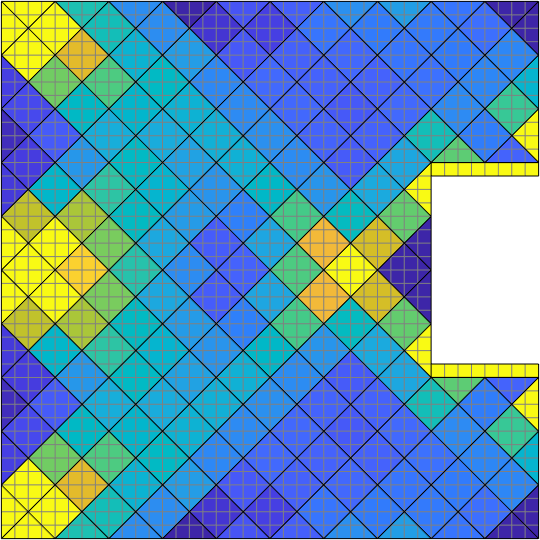}}
		\caption{}
		\label{fig:gripf}
	\end{subfigure}%
	\hspace{1cm}\begin{subfigure}[t]{0.185\linewidth}
		\raisebox{6.5mm}{\includegraphics[width=\linewidth]{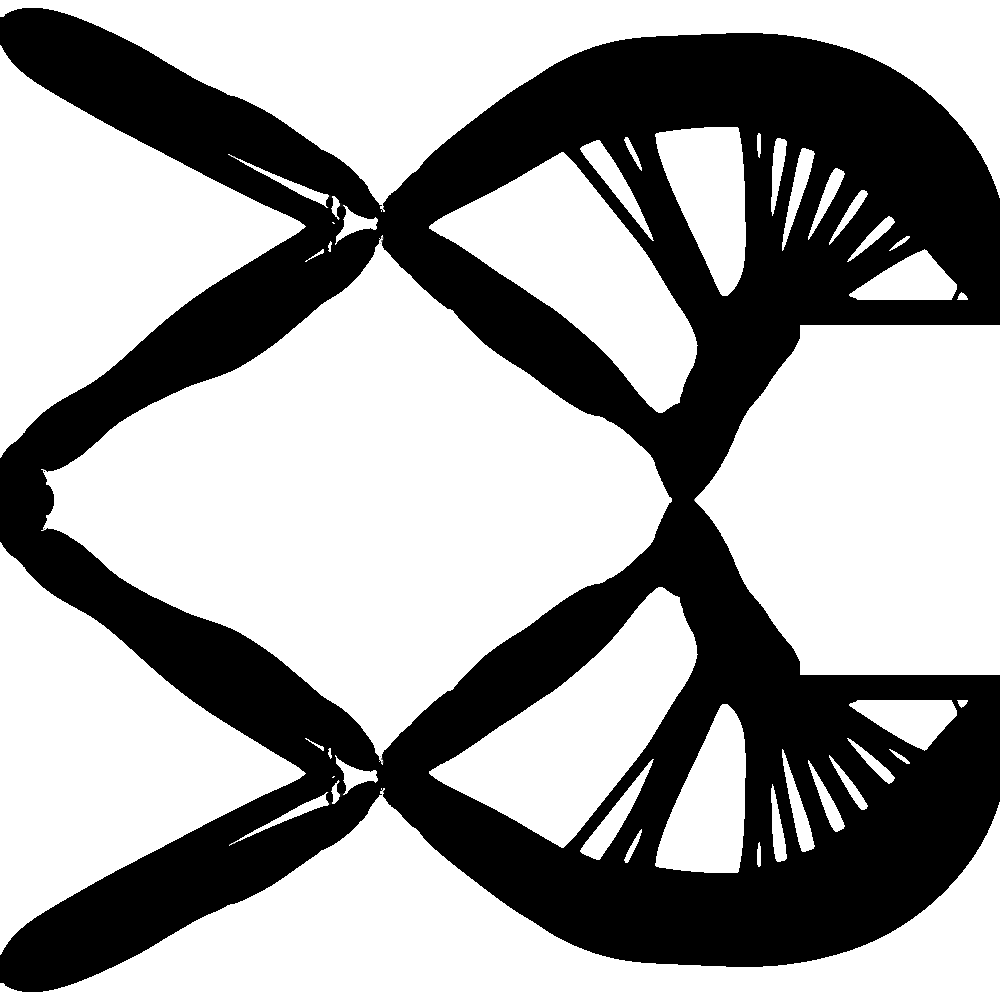}}
		\caption{}
		\label{fig:grip0}
	\end{subfigure}
	\caption{Boundary conditions of the gripper problem and assumed modular discretization, and lower-bound designs: (b) optimal traces of material stiffness matrices \replace{($s_\mathrm{FMO}=-2.096$), (c) post-processed material stiffness distribution ($s_\mathrm{FMO,f}=-0.665$), and (d)}{($s_\mathrm{FMO} = -1.838$), and (c)} non-modular design obtained by TO \replace{($s_{\mathrm{TO},\infty} =-0.616$)}{($s_{\mathrm{FMO},\infty} = -0.636$)}.}
\end{figure}

As the third problem, we consider the traditional gripper mechanism design \citep{sigmund1997}. Similarly to the force inverter, we adopt a square design domain of the size $1.0 \times 1.0$ and split it \replace{using}{into} $10 \times 10$ equisized square modules, see Fig.~\ref{fig:grip_bc}. \replace{Following}{Similarly to the force inverter and}~\citep{sigmund1997}, we \replace{also}{}fix the top and bottom \replace{$0.01$}{$0.025$}-long segments of the left edge of the domain\replace{, and collapse these supports into single points for coarse discretizations adopted in the FMO problems~\eqref{eq:fmo} and \eqref{eq:mfmo}}{}.

At the center of the left edge, we place a spring with the stiffness $k_{\mathrm{in}}=1$ and an input force of a magnitude $1$, both aligned with the horizontal direction. In the central part of the opposite, right edge, we prescribe an empty space of $2\times4$ modules to facilitate a placement of an item that shall be grasped by the mechanism. The grasping occurs through fixed jaws placed at the top and bottom of the empty space. These jaws are $0.025$ in height and $0.2$ in width and made of a solid isotropic material with $E=1$ and $\nu=0.3$, matching so the material adopted in TO and the maximum trace constraint in FMO \eqref{eq:fmo_eltrace}. \replace{We note here that to use the edge/module-based discretization and these differently-sized fixed elements in FMO problems, we interpolate their stiffness into the neighboring nodes.}{}

For the mechanism functionality, the input force must trigger closing of the gripper's jaws, which we quantify by minimizing the difference of the displacements $u_\mathrm{o2}-u_\mathrm{u1}$ of the rightmost points of the jaws, each equipped with a stiffness $k_\mathrm{out}= 0.025$ in the vertical direction.

\begin{figure}[!t]
	\centering
	\setlength{\tabcolsep}{3.5pt}
	\begin{tabular}{cccc}
		Horizontal clustering & Vertical clustering & Optimized structure & Wang module set\\
		\imagetop{\includegraphics[width=3.5cm]{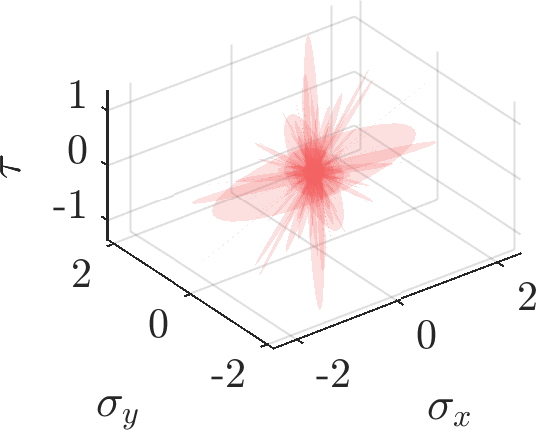}} & \imagetop{\includegraphics[width=3.5cm]{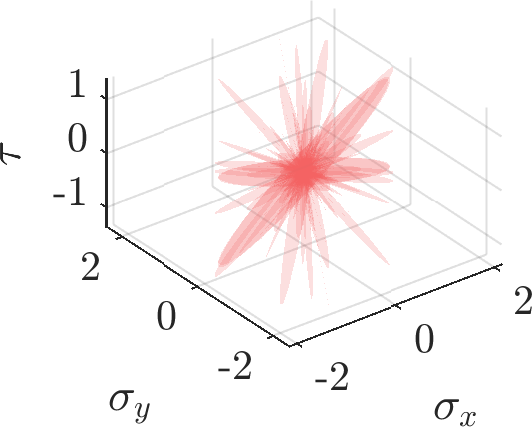}} & \imagetop{\includegraphics[height=0.16\linewidth]{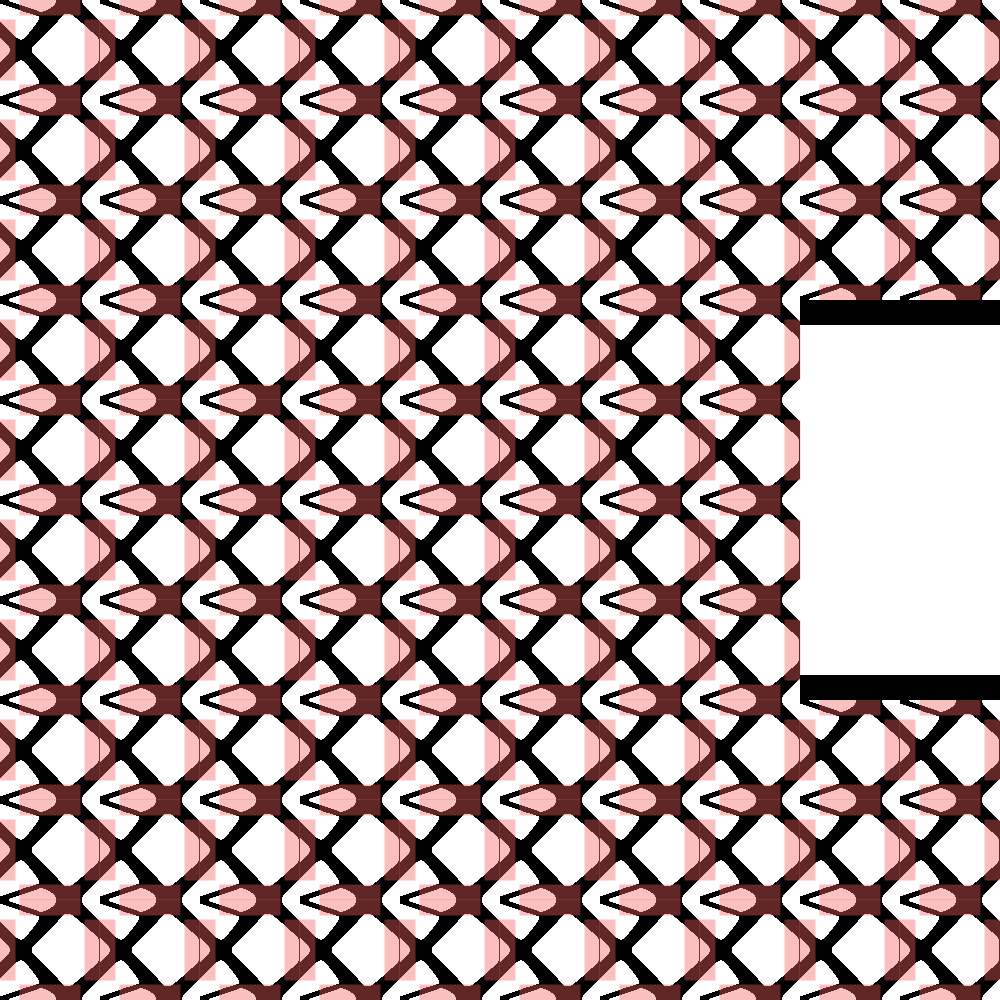}} & \imagetop{\includegraphics[width=0.215\linewidth]{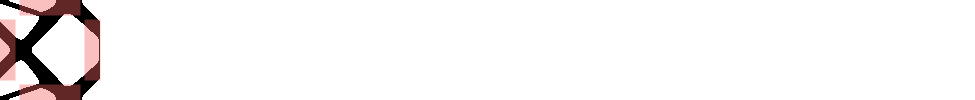}} \\
		\imagetop{\includegraphics[width=3.5cm]{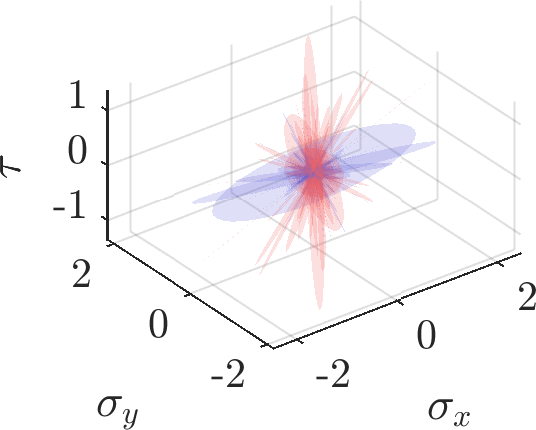}} & \imagetop{\includegraphics[width=3.5cm]{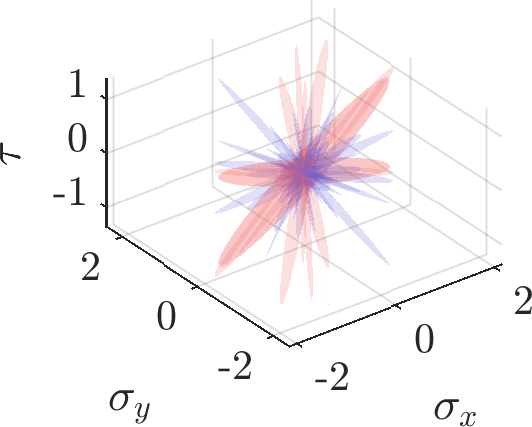}} & \imagetop{\includegraphics[height=0.16\linewidth]{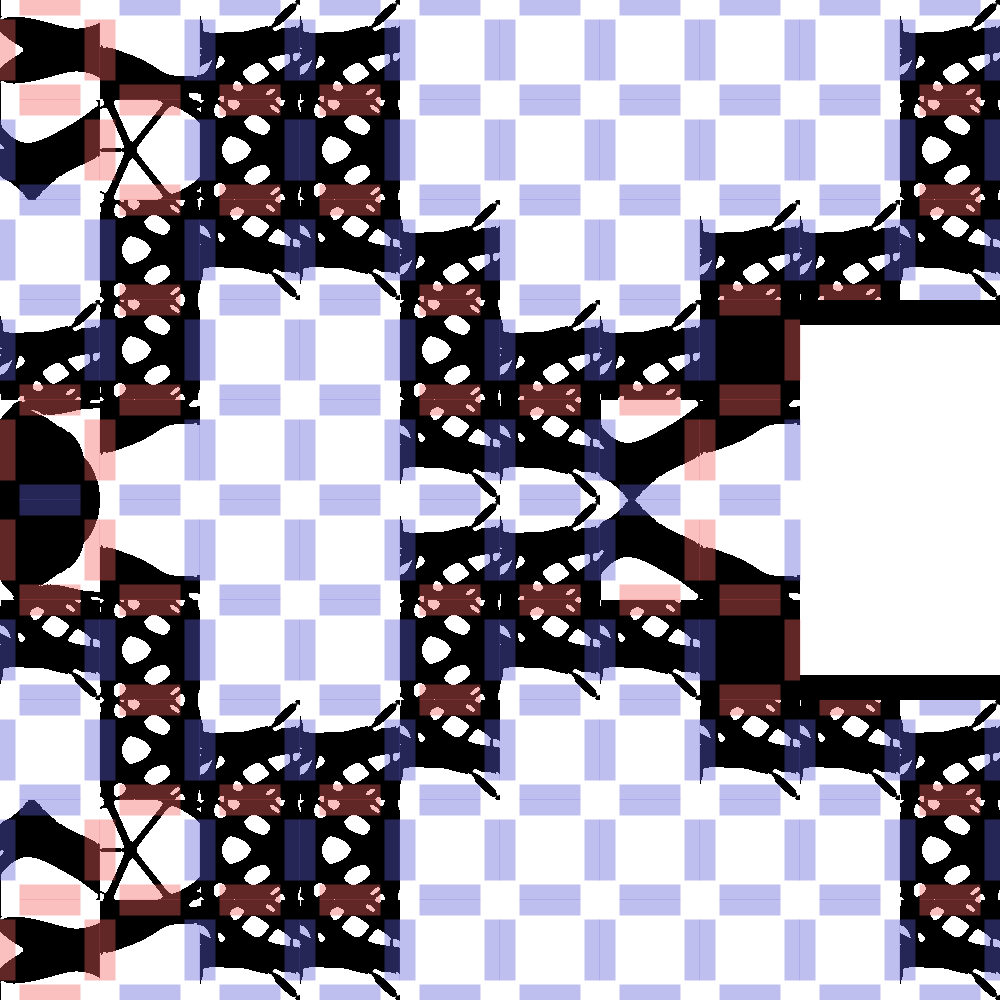}} & \imagetop{\includegraphics[width=0.215\linewidth]{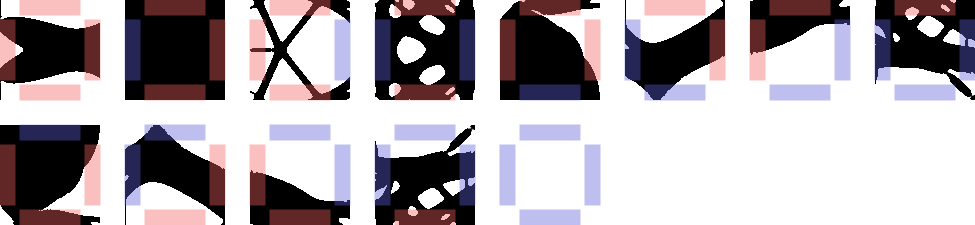}} \\
		\imagetop{\includegraphics[width=3.5cm]{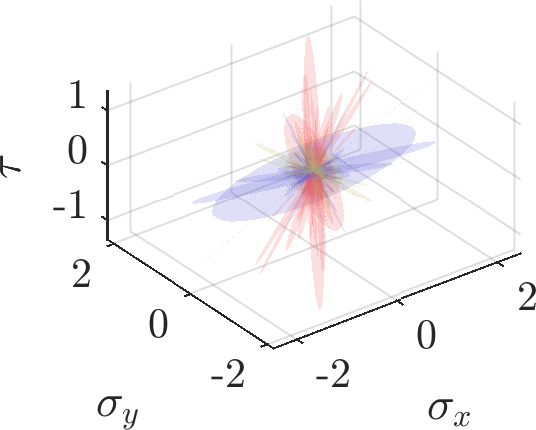}} & \imagetop{\includegraphics[width=3.5cm]{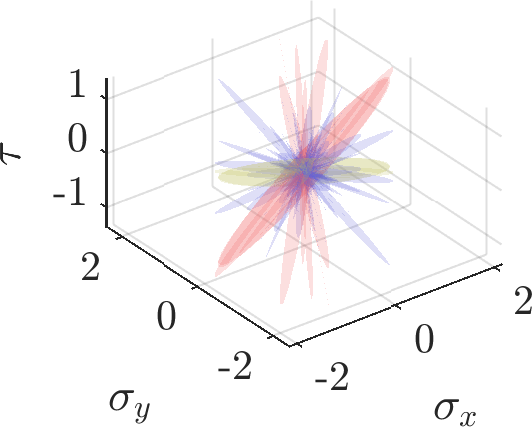}} & \imagetop{\includegraphics[height=0.16\linewidth]{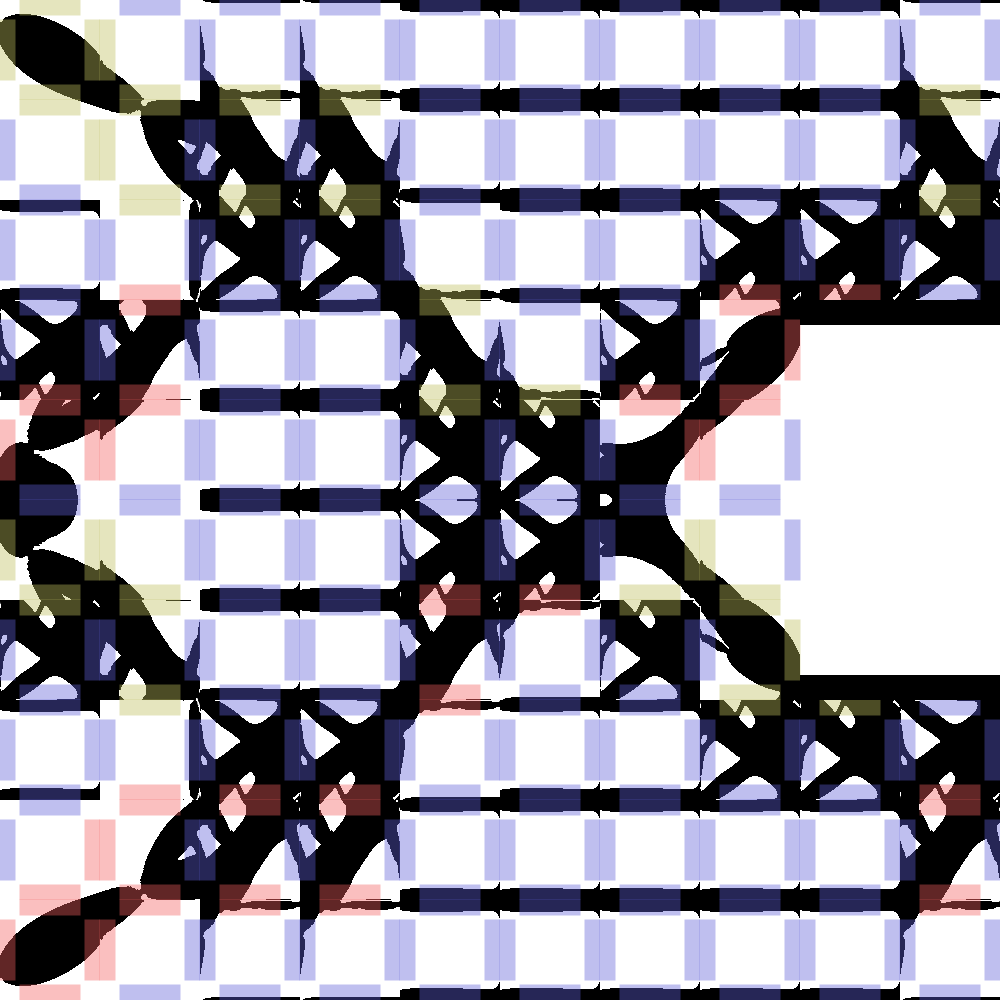}} & \imagetop{\includegraphics[width=0.215\linewidth]{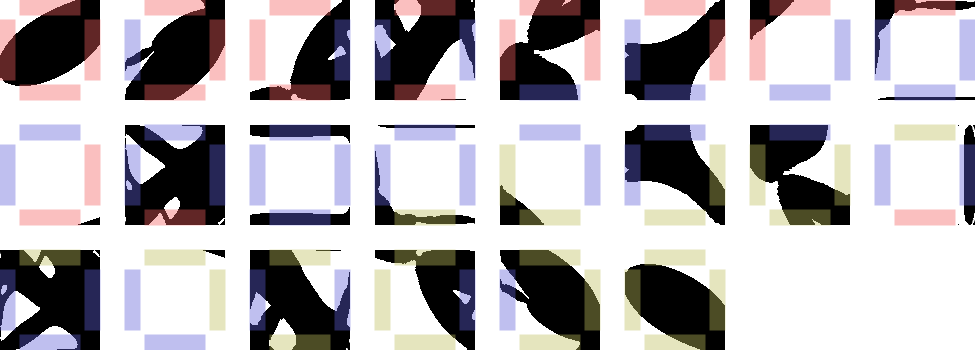}} \\
		\imagetop{\includegraphics[width=3.5cm]{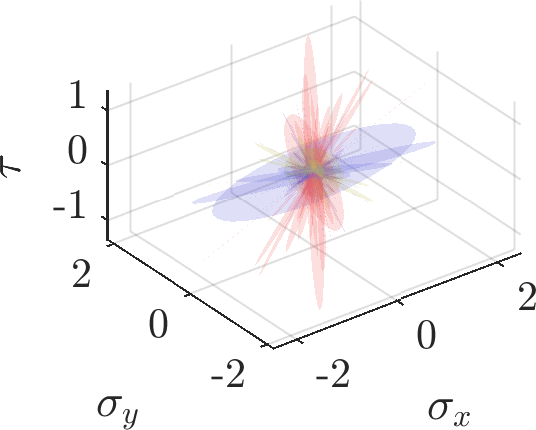}} & \imagetop{\includegraphics[width=3.5cm]{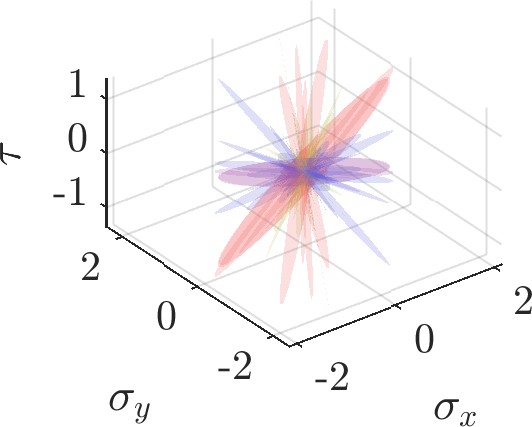}} & \imagetop{\includegraphics[height=0.16\linewidth]{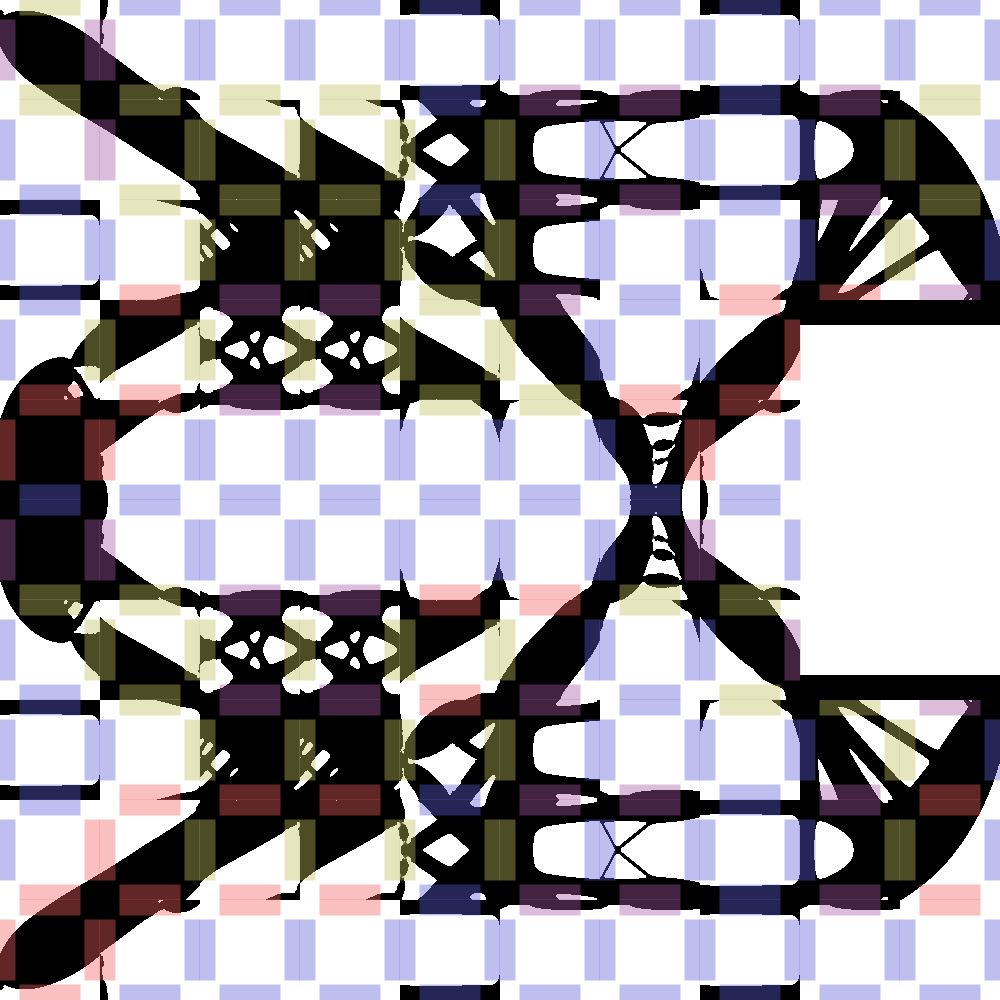}} & \imagetop{\includegraphics[width=0.215\linewidth]{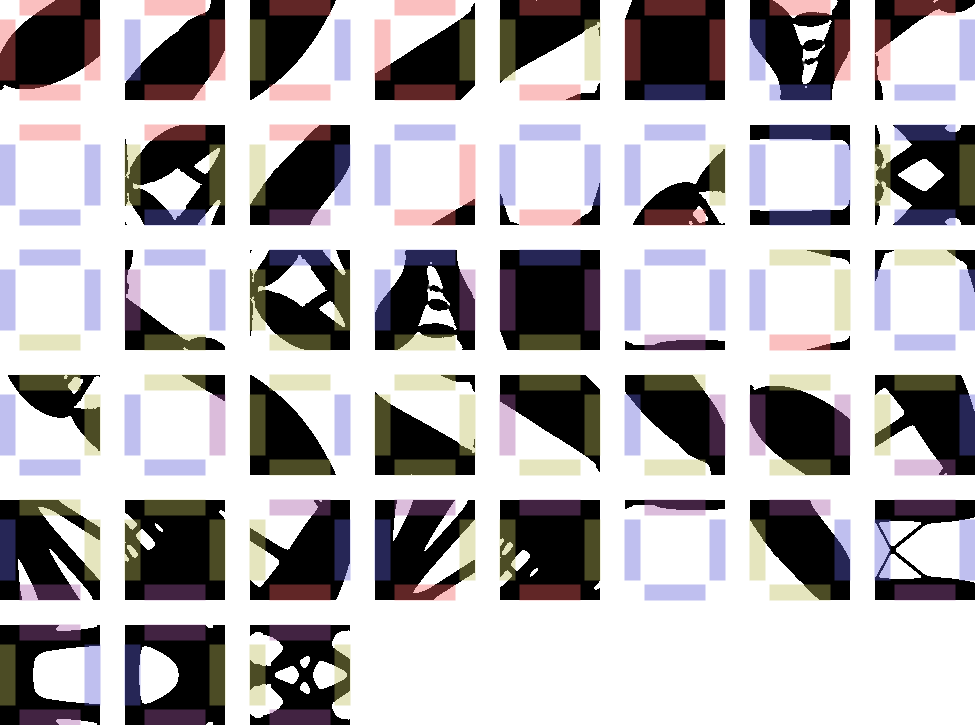}}
	\end{tabular}
	\caption{Modular designs of the gripper problem for the numbers of edge colors ranging from 1 to 4. The first two columns illustrate the partitioning produced by the hierarchical clustering algorithm, the third and fourth columns visualize optimized modular topologies and their assembly.}
	\label{fig:grip}
\end{figure}

Again, we first solve the FMO optimization problem \eqref{eq:fmo} on the rotated mesh, providing us with \replace{$s_\mathrm{FMO} = -2.096$ on the coarse mesh and $s_\mathrm{FMO}^{\mathrm{refined}} = -0.098$ after $100$ times refinement. After the post-processing step, we obtain $s_\mathrm{FMO,f} = -0.665$ on the coarse mesh and $s_\mathrm{FMO,f}^{\mathrm{refined}} = -0.411$ on the refined mesh; compare Figs. \ref{fig:gripnf} and \ref{fig:gripf} for a visualization of the post-processing step. Similarly to the first two examples, also here the refined mesh certifies the importance of filtering or post-processing phase to avoid nonphysical designs}{$s_\mathrm{FMO} = -1.838$, see Fig.~\ref{fig:gripf}}.
\replace{For the case of the gripper mechanism}{Exploring the modularity bounds in TO}, the non-modular design produces \replace{$s_{\mathrm{TO},\infty}=-0.616$}{$s_{\mathrm{TO},\infty}=-0.636$} and \replace{the design}{topology} shown in Fig.~\ref{fig:grip0}. The other extreme case, PUC single-color-based clustering at the top row of Fig.~\ref{fig:grip}, possesses \replace{$s_{\mathrm{TO},1}=-0.080$}{$s_{\mathrm{T0},1} = -0.070$}. While a two-color-based design contains \replace{$7$}{$13$} module types and an objective function \replace{$s_{\mathrm{TO},2}=-0.151$}{$s_{\mathrm{TO},2}=-0.225$}, a three-color design encompasses \replace{$13$}{$22$} modules that induce \replace{$s_{\mathrm{TO},3}=-0.196$}{$s_{\mathrm{TO},3}=-0.405$}. Finally, the four-color clustering possesses \replace{$s_{\mathrm{TO},4}=-0.399$}{$s_{\mathrm{TO},4}=-0.606$} and consists of \replace{$23$}{$43$} distinct modules; see Fig.~\ref{fig:grip} for a visualization of all modular designs. Notice that for this problem, four-color clustering led to a \replace{$35\%$}{less than $2\%$} loss in the performance compared to the non-modular design. \replace{}{Finally, consistently with the findings for the inverter mechanism, modularity in this case also decreases the maximum von Mises stress in the structure, see Tab.~\ref{tab:grip}.}

\begin{table}[!b]
	\centering
	\small
	\caption{\replace{}{Overview of the modular gripper performances, where $u_\mathrm{in}$ denotes the horizontal displacement at the force input, $s_\mathrm{grip}$ the mechanism performance, and $\sigma_{\mathrm{v},\max}$ the maximum von Mises stress. Further, GA abbreviates geometrical advantage, evaluated as $s_\mathrm{TO}/u_\mathrm{in}$.}}
	\setlength\tabcolsep{0.5em}
	\begin{tabular}{lrrrr}
		Design & $u_\mathrm{in}$ & $s_\mathrm{TO}$ & GA & $\sigma_{\mathrm{v},\max}$\\
		\hline
		$1$ module over $1$ color & $0.985$ & $-0.070$ & $-0.071$ & $6.452$ \\
		$13$ modules over $2$ colors & $0.987$ & $-0.225$ & $-0.228$ & $5.547$ \\
		$22$ modules over $3$ colors & $0.983$ & $-0.405$ & $-0.412$ & $7.226$ \\
		$43$ modules over $4$ colors & $0.977$ & $-0.606$ & $-0.621$ & $10.024$ \\
		Non-modular & $0.984$ & $-0.636$ & $-0.646$ & $10.420$
	\end{tabular}
	\label{tab:grip}
\end{table}

\subsection{Module reusability in gripper and inverter}\label{sec:reusable}

\begin{figure}[!t]
	\centering
	\includegraphics[height=0.16\linewidth]{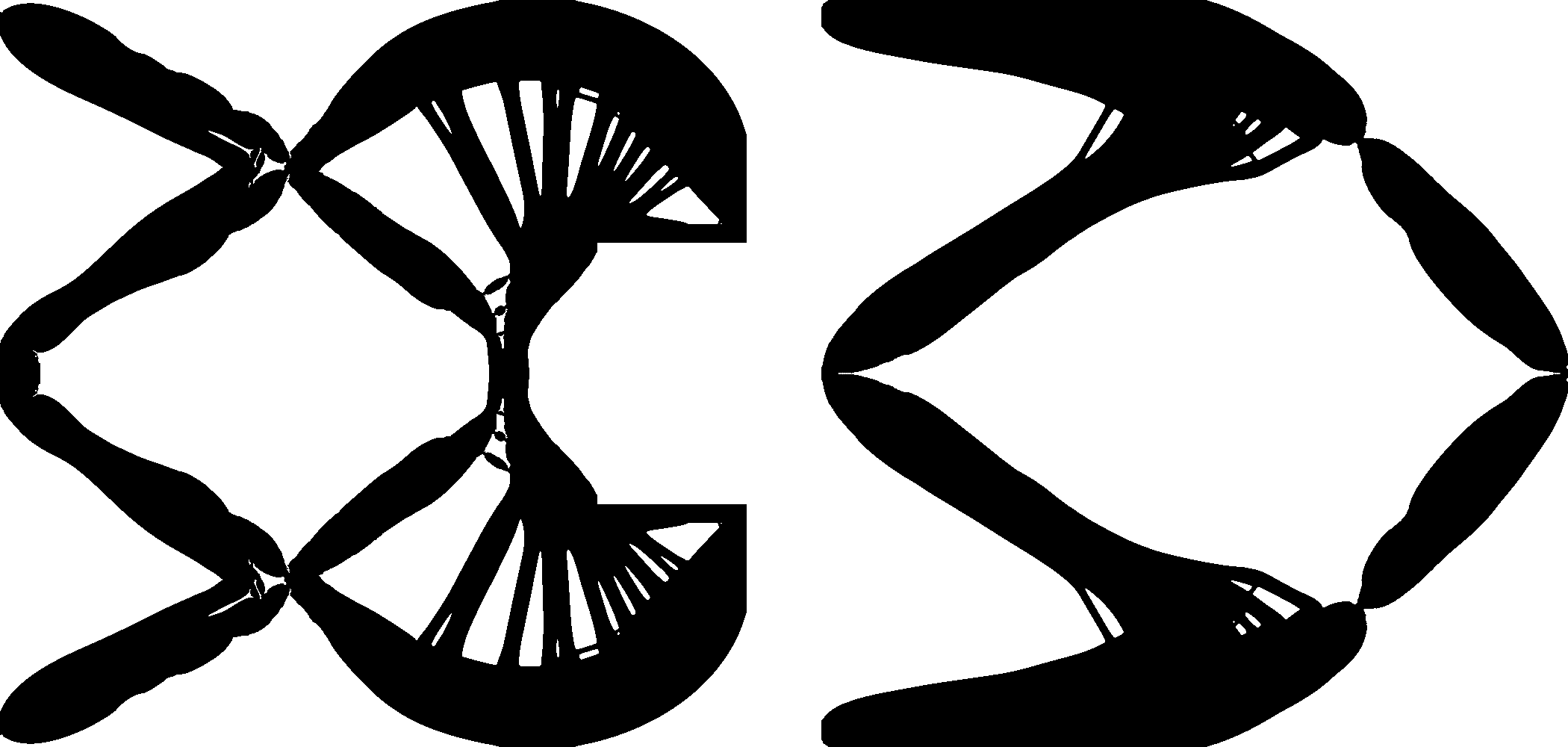}
	\caption{Non-modular design of coupled inverter-gripper mechanisms obtained by TO.}
	\label{fig:invgrip0}
\end{figure}

To demonstrate \replace{}{the} merits of modularity, we address a reusable simultaneous design of the previous\replace{}{ly} described inverter and gripper mechanisms as our last example. Based on the multi-functionality of the resulting module designs, it becomes apparent that modularity paves a way from structure-based designs to meta-material based ones.

\begin{figure}[!b]
	\centering
	\setlength{\tabcolsep}{2pt}
	\begin{tabular}{cccc}
		Horizontal clustering & Vertical clustering & Optimized structure & Wang module set\\
		\imagetop{\includegraphics[width=3.45cm]{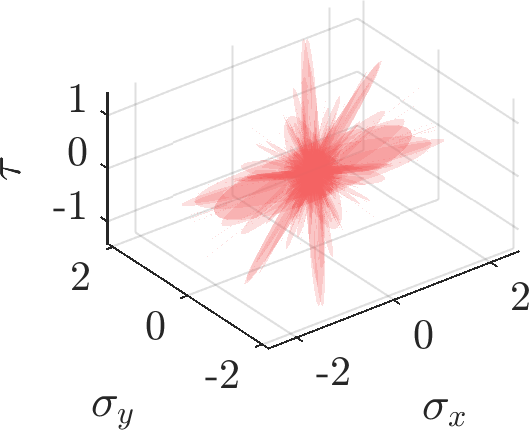}} & \imagetop{\includegraphics[width=3.45cm]{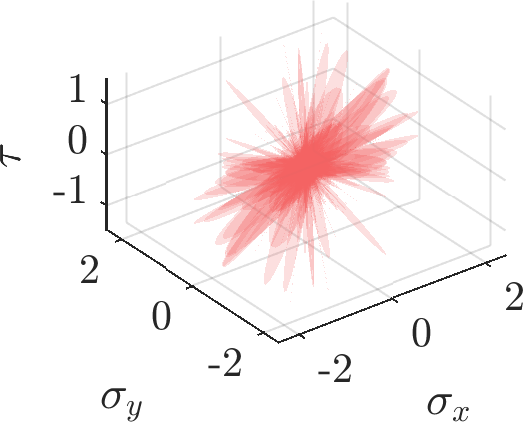}} & \imagetop{\includegraphics[height=0.16\linewidth]{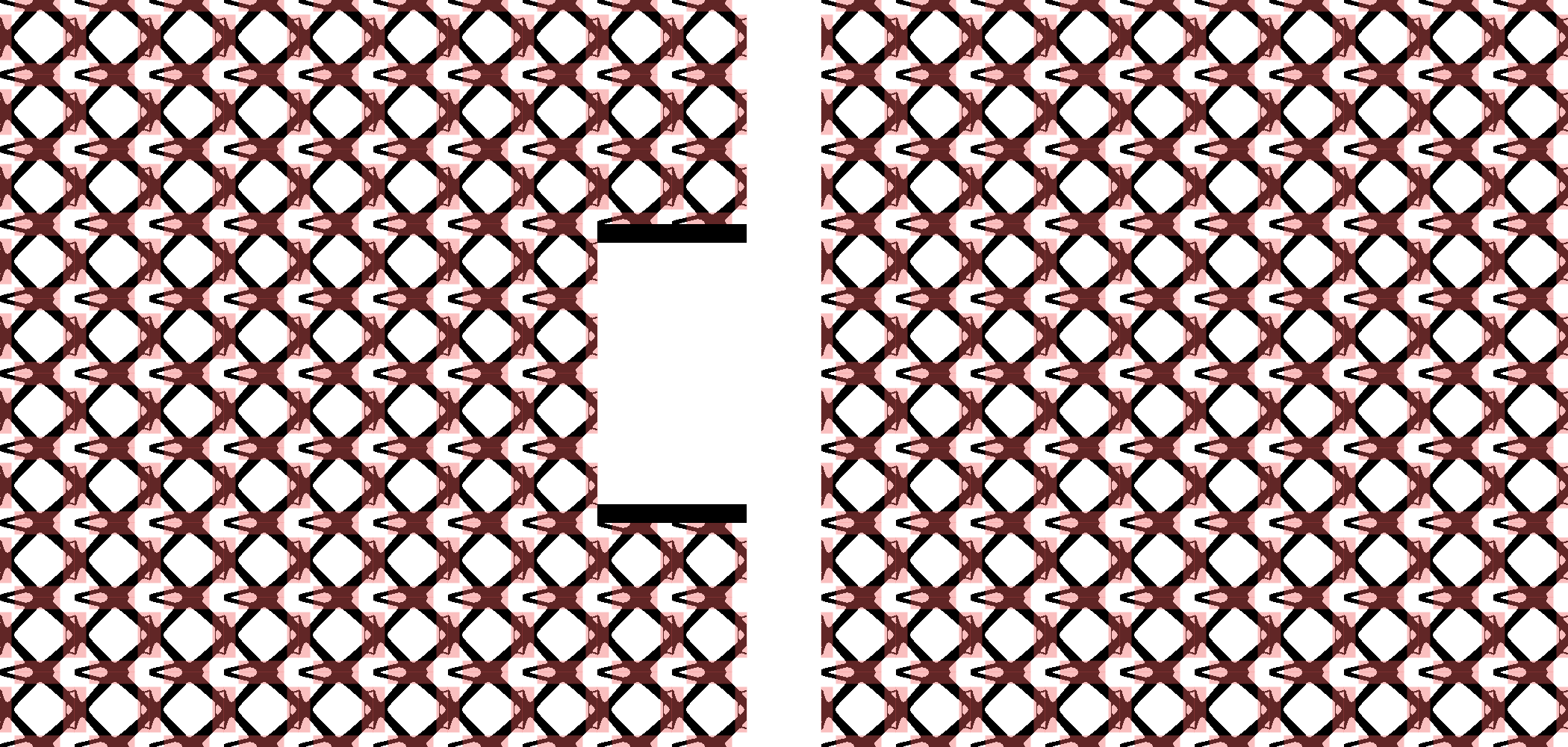}} & \imagetop{\includegraphics[width=0.215\linewidth]{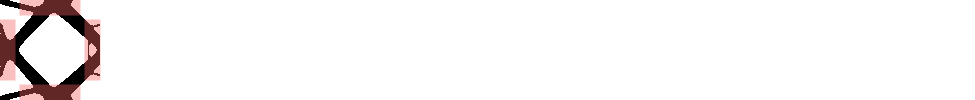}} \\
		\imagetop{\includegraphics[width=3.45cm]{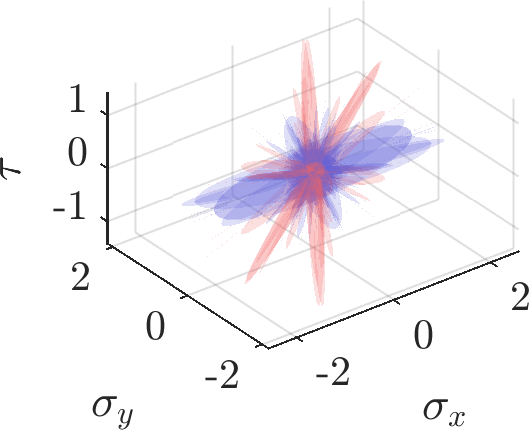}} & \imagetop{\includegraphics[width=3.45cm]{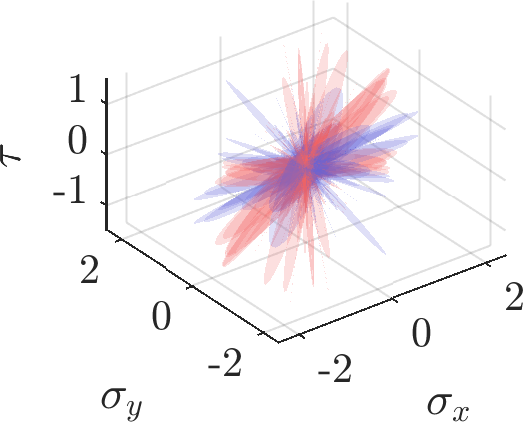}} & \imagetop{\includegraphics[height=0.16\linewidth]{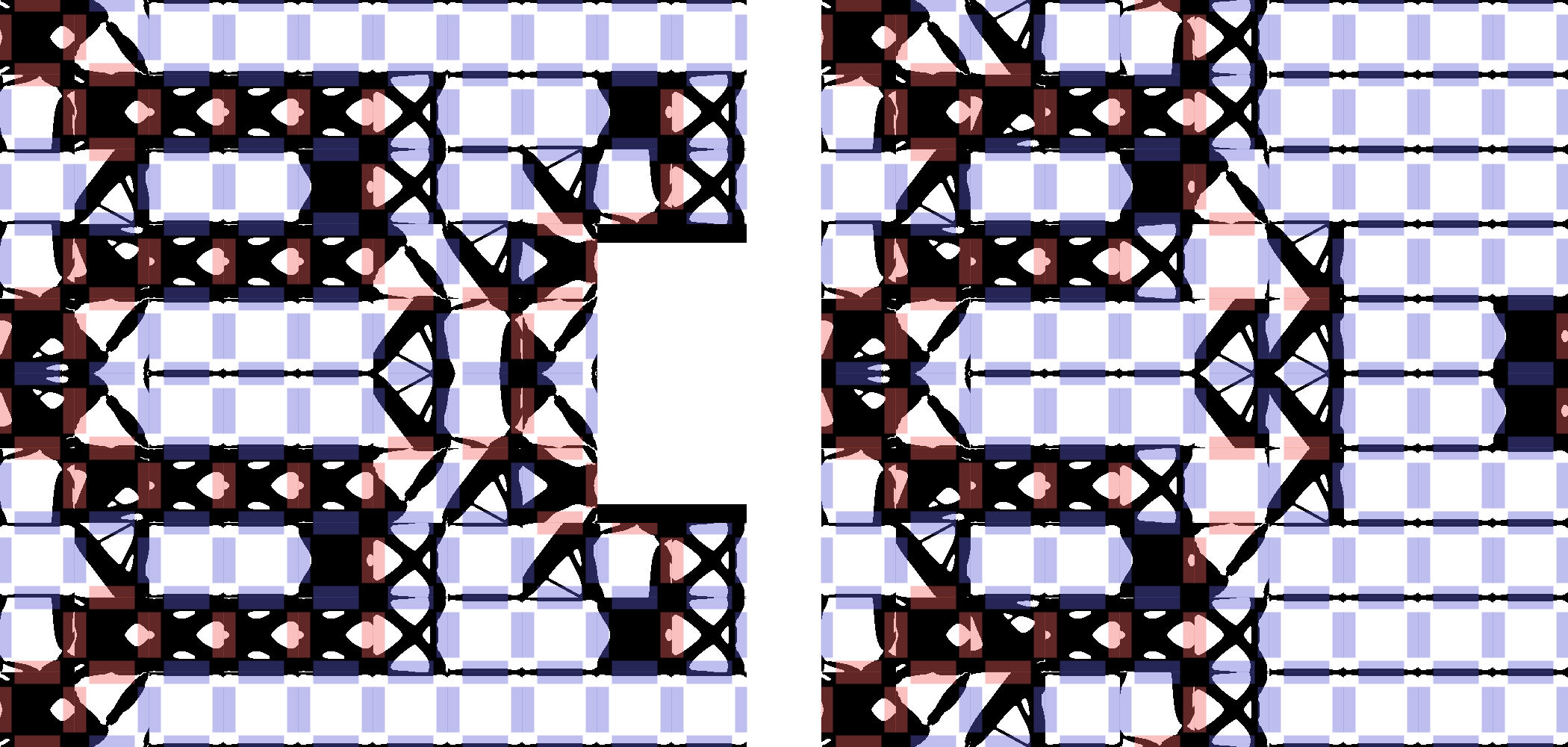}} & \imagetop{\includegraphics[width=0.215\linewidth]{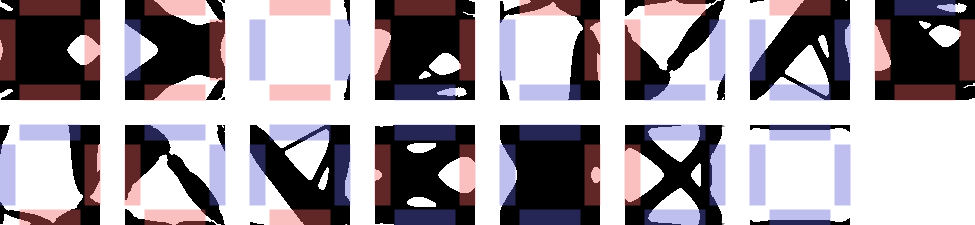}} \\
		\imagetop{\includegraphics[width=3.45cm]{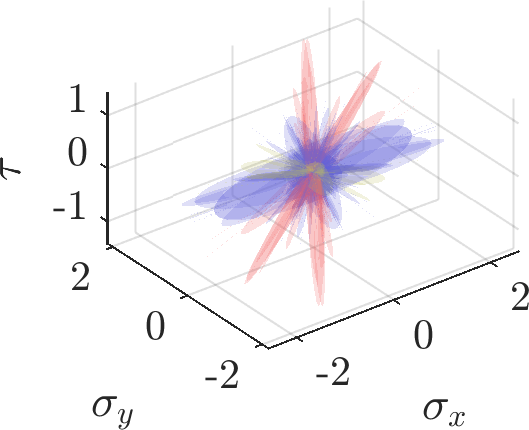}} & \imagetop{\includegraphics[width=3.45cm]{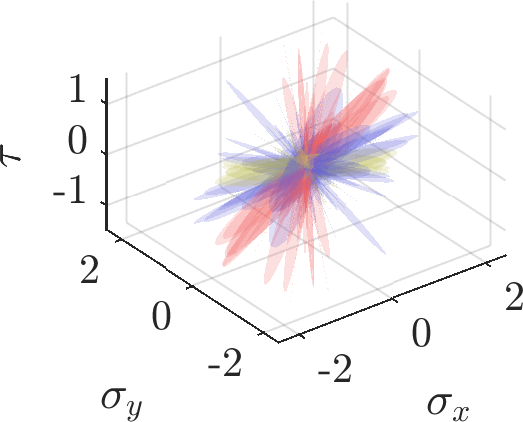}} & \imagetop{\includegraphics[height=0.16\linewidth]{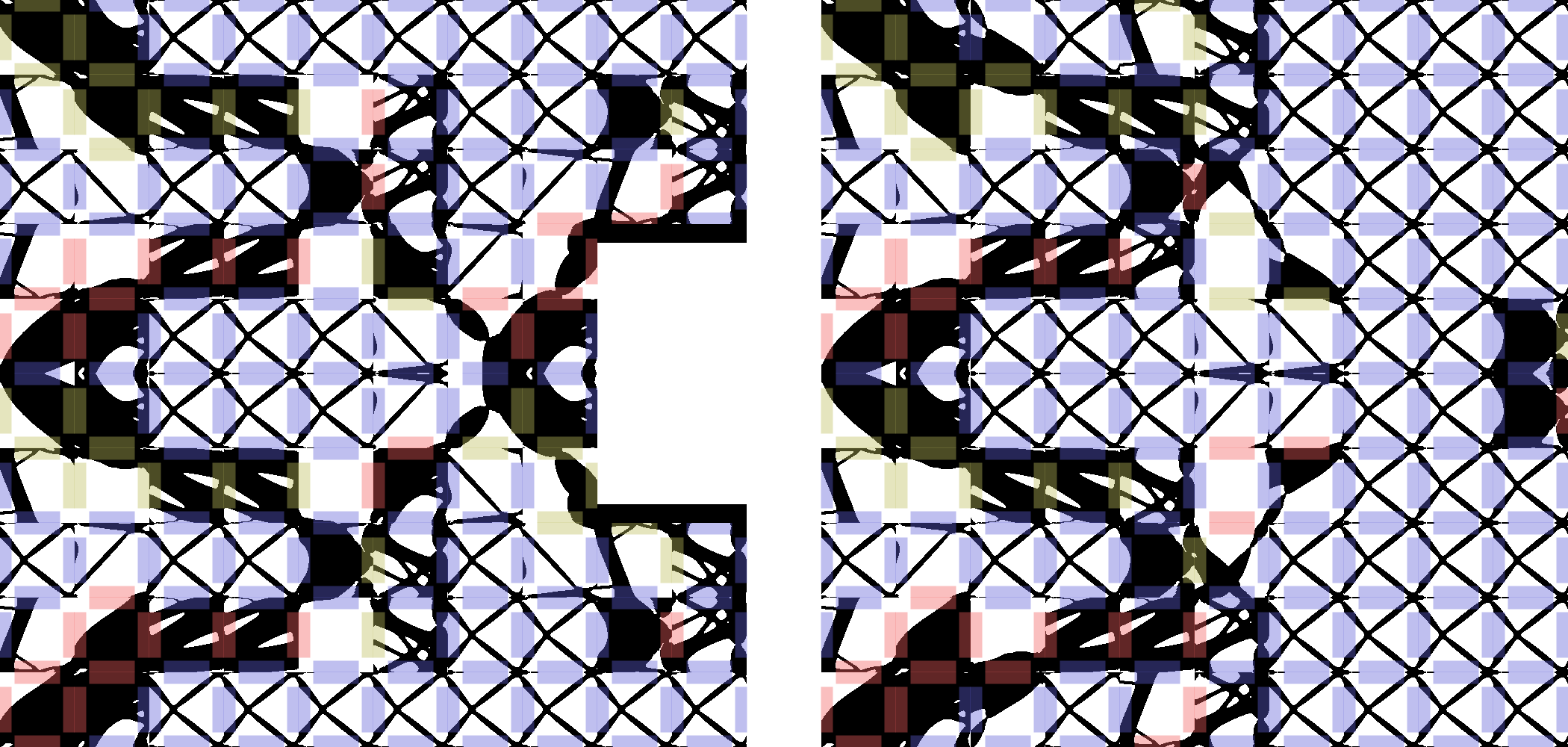}} & \imagetop{\includegraphics[width=0.215\linewidth]{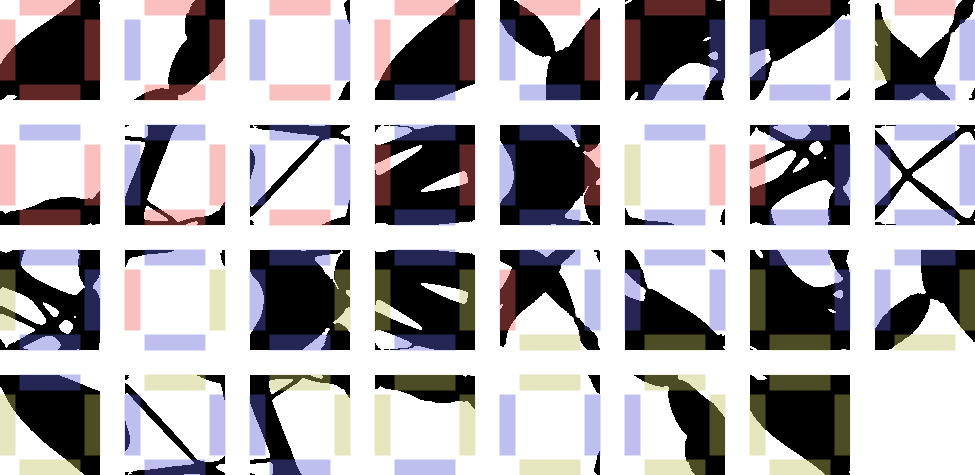}} \\
		\imagetop{\includegraphics[width=3.45cm]{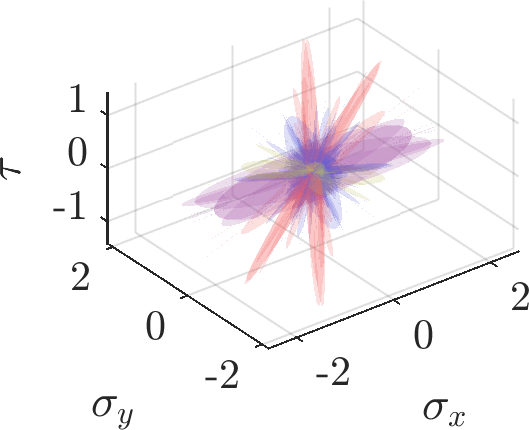}} & \imagetop{\includegraphics[width=3.45cm]{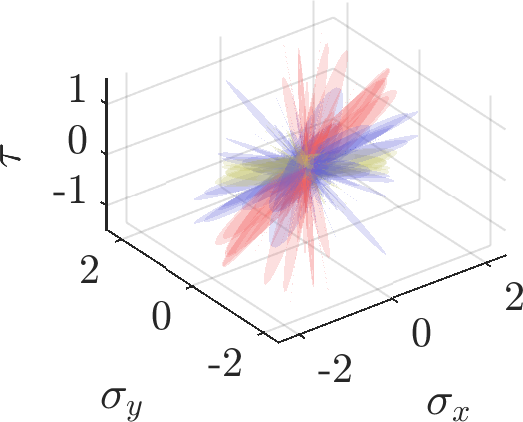}} & \imagetop{\includegraphics[height=0.16\linewidth]{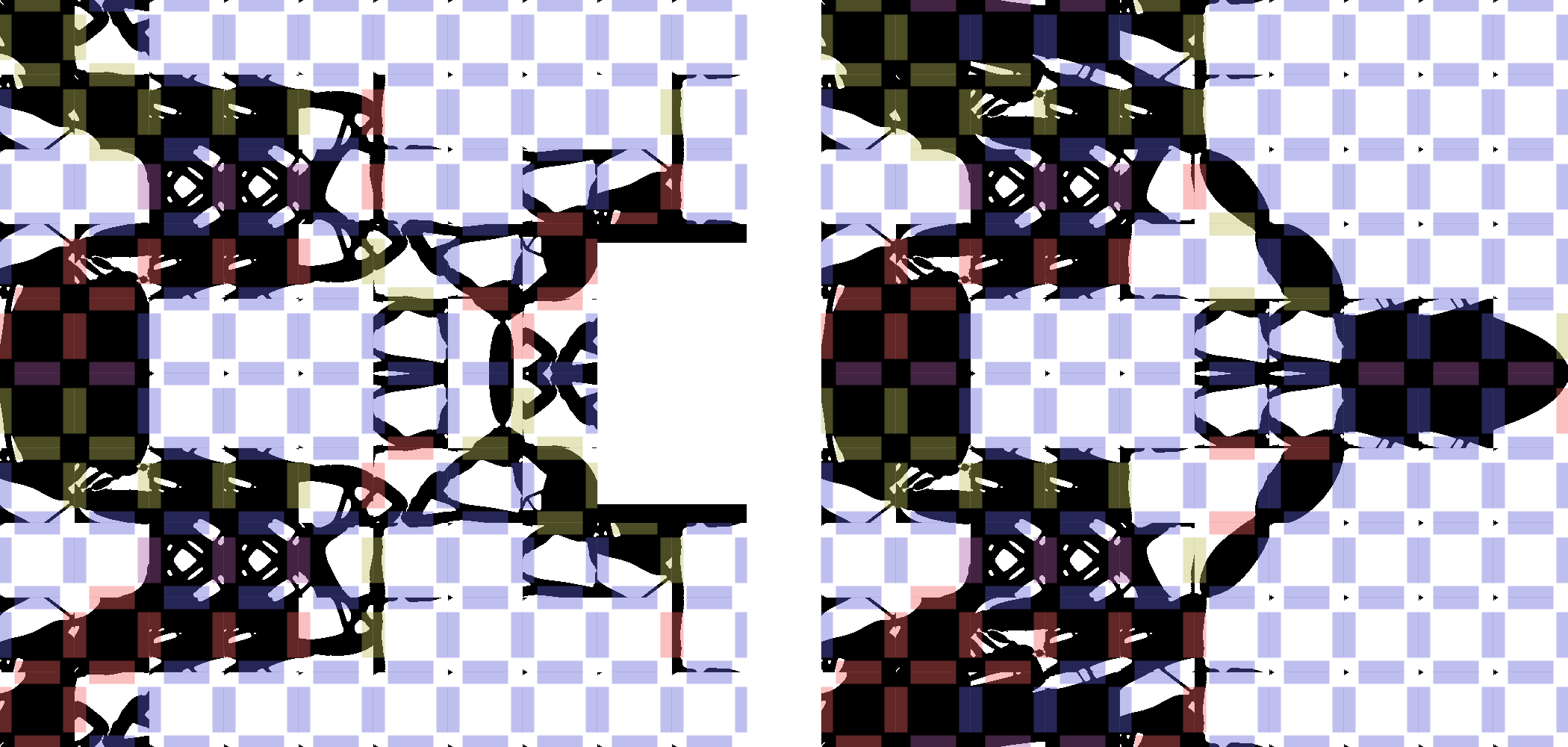}} & \imagetop{\includegraphics[width=0.215\linewidth]{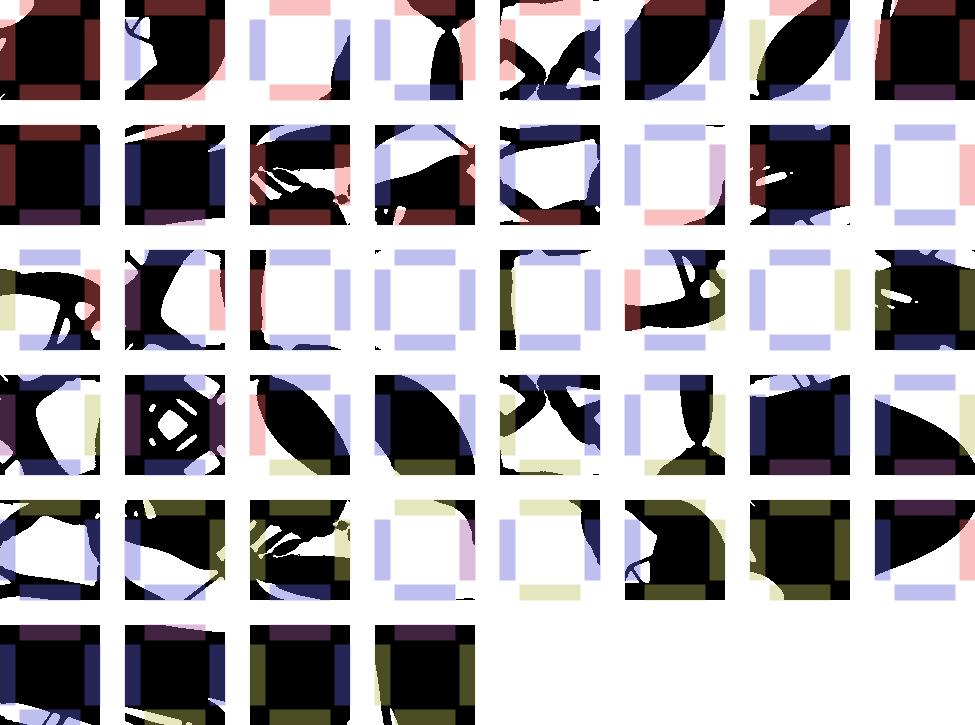}}
	\end{tabular}
	\caption{Modular designs of reusable inverter-gripper problem for the numbers of edge colors ranging from 1 to 4. While the first two columns illustrate the partitioning produced by the hierarchical clustering algorithm, the third and fourth columns visualize optimized modular topologies and their assembly.}
	\label{fig:invgrip}
\end{figure}

For the reusability design problem, we adopt \replace{}{the} discretizations and boundary conditions described in Sections \ref{sec:inverter} and \ref{sec:gripper}. However, the reusable designs add another level of complexity as they allow for prioritizing of individual subproblems, i.e., they generate a multi-objective optimization problem. In what follows, we reduce this setting to a single-objective one by introducing weights $w_\mathrm{inv}$ and $w_\mathrm{grip}$, and write the resulting objective as
\begin{equation}
s = w_\mathrm{inv} s_\mathrm{inv} + w_\mathrm{grip} s_\mathrm{grip},
\end{equation}
with $s_\mathrm{inv}$ and $s_\mathrm{grip}$ denoting the objective values of the inverter and gripper problems, respectively. Based on a comparison of the objectives of modular gripper and inverter designs and \replace{extensive}{numerical} tests (not included), we believe that the choice of $w_\mathrm{inv}=1.0$ and $w_\mathrm{grip} = 0.60$ is reasonable, because it balances the terms $w_\mathrm{inv} s_\mathrm{inv}$ and $w_\mathrm{grip} s_\mathrm{grip}$ for the non-modular designs, recall Sections \ref{sec:inverter} and \ref{sec:gripper}.

The first step in all previous examples was a solution to the \replace{FEM}{FMO} problem on the rotated FE mesh; however, we have already solved these \replace{initial steps for both the}{}subproblems in Sections~\ref{sec:inverter} and \ref{sec:gripper} and, hence, we can recycle them here. Thus, instead of another solution, we cluster the former material distributions concurrently, Figs.~\ref{fig:invf} and \ref{fig:gripf}. In the clustering procedure, recall Section \ref{sec:clustering}, we weight the stiffness matrices of the gripper mechanism by $w_\mathrm{grip}$ and of inverter by $w_\mathrm{inv}$. Similarly to TO, these weights maintain more balanced performance. Finally, we reach the assembly plans and module sets shown in Fig.~\ref{fig:invgrip}.

For TO, we aggregate the independent volume constraints and adopt $\overline{V}=0.4$ for the union of the domains. Then, the optimized non-modular designs in Fig.~\ref{fig:invgrip0} possess \replace{$s_{\mathrm{TO},\infty} = -0.731$}{$s_{\mathrm{TO},\infty} = -0.772$} (with \replace{$s_{\mathrm{TO},\infty,\mathrm{inv}}=-0.320$}{$s_{\mathrm{TO},\infty,\mathrm{inv}}=-0.370$} and \replace{$s_{\mathrm{TO},\infty,\mathrm{grip}}=0.687$}{$s_{\mathrm{TO},\infty,\mathrm{grip}}=-0.669$}), and the single-color-based PUC design in the first row of Fig.~\ref{fig:invgrip} provides \replace{$s_{\mathrm{TO}, 1} =-0.043$}{$s_{\mathrm{TO}, 1} =-0.040$} (with $s_{\mathrm{TO},1,\mathrm{inv}}=0$ and \replace{$s_{\mathrm{TO},1,\mathrm{grip}}=-0.072$}{$s_{\mathrm{TO},1,\mathrm{grip}}=-0.067$}). In both these \replace{settings}{extreme cases}, the performances of reusable designs \replace{naturally approach}{are very similar to} the independent ones\replace{}{, recall Sections \ref{sec:inverter} and \ref{sec:gripper}, certifying an appropriate choice of the weights $w_\mathrm{grip}$ and $w_\mathrm{inv}$}.

\begin{figure}[!b]
	\centering
	\begin{subfigure}{0.4\linewidth}
		\includegraphics[width=\linewidth]{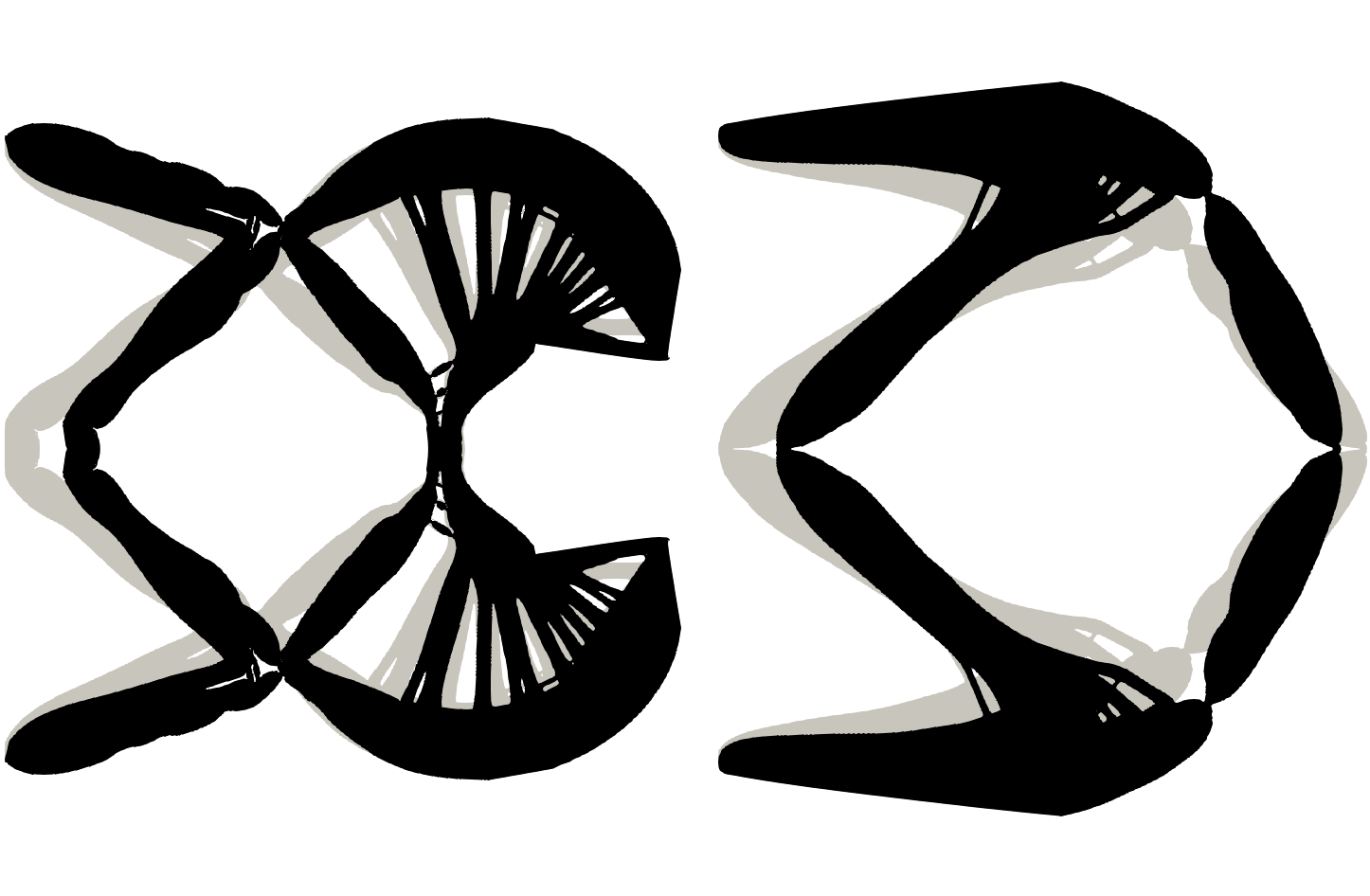}
		\caption{}
		\label{fig:invgrip_deform_nonmodular}
	\end{subfigure}%
	\hspace{1cm}\begin{subfigure}{0.4\linewidth}
		\includegraphics[width=\linewidth]{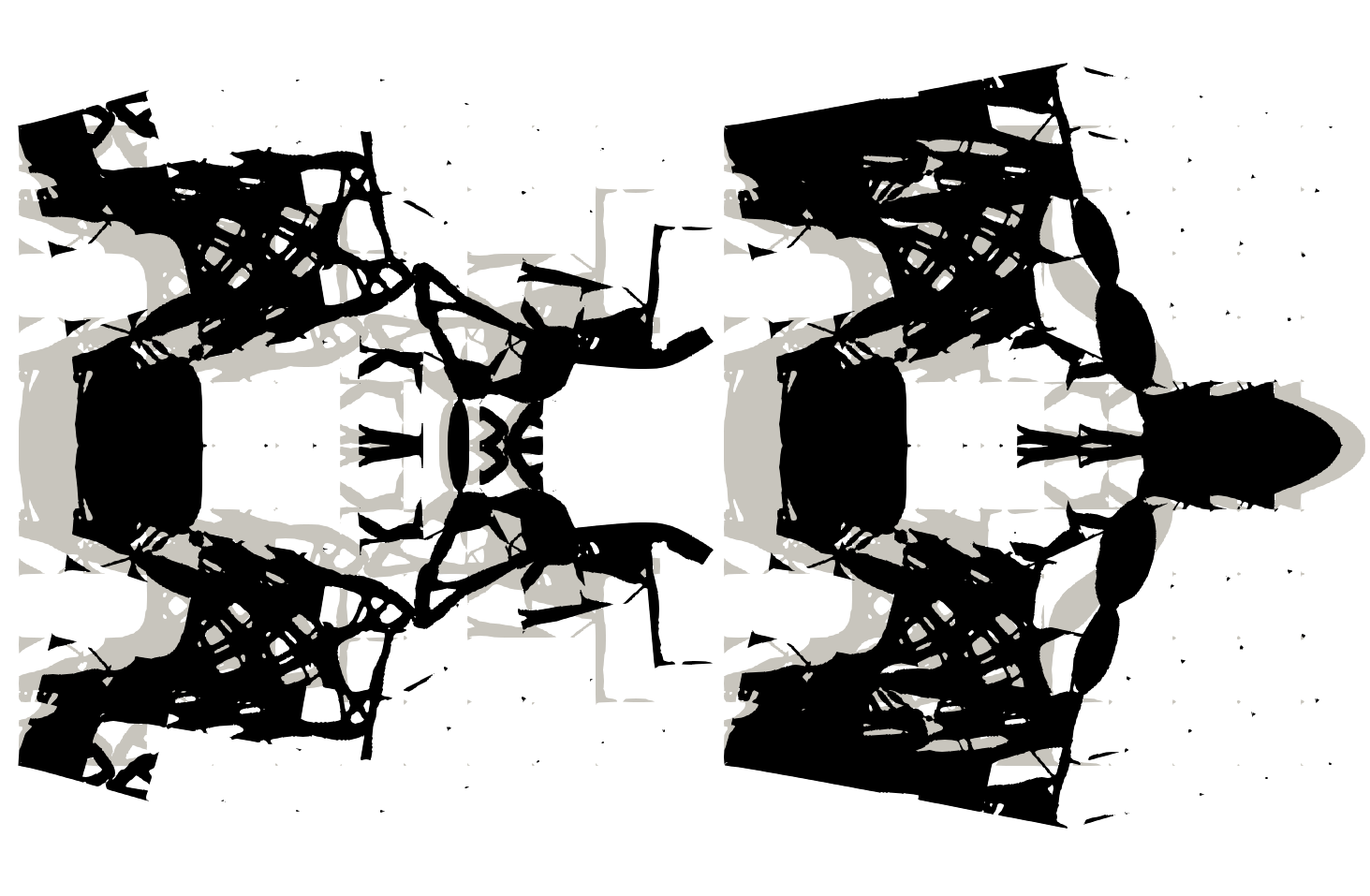} 
		\caption{}
		\label{fig:invgrip_deform_reusable}
	\end{subfigure}
	\caption{\replace{}{Deformed shape of (a) the non-modular design of a gripper and an inverter, and (b) the corresponding reusable modular design with 44 modules over 4 colors. Displacements are scaled down by the factor 0.1. Undeformed shapes are plotted in light gray color.}}
	\label{fig:invgrip_deform}
\end{figure}

\begin{figure}[!t]
	\centering
	\begin{tabular}{ccc}
		\includegraphics[height=3.3cm]{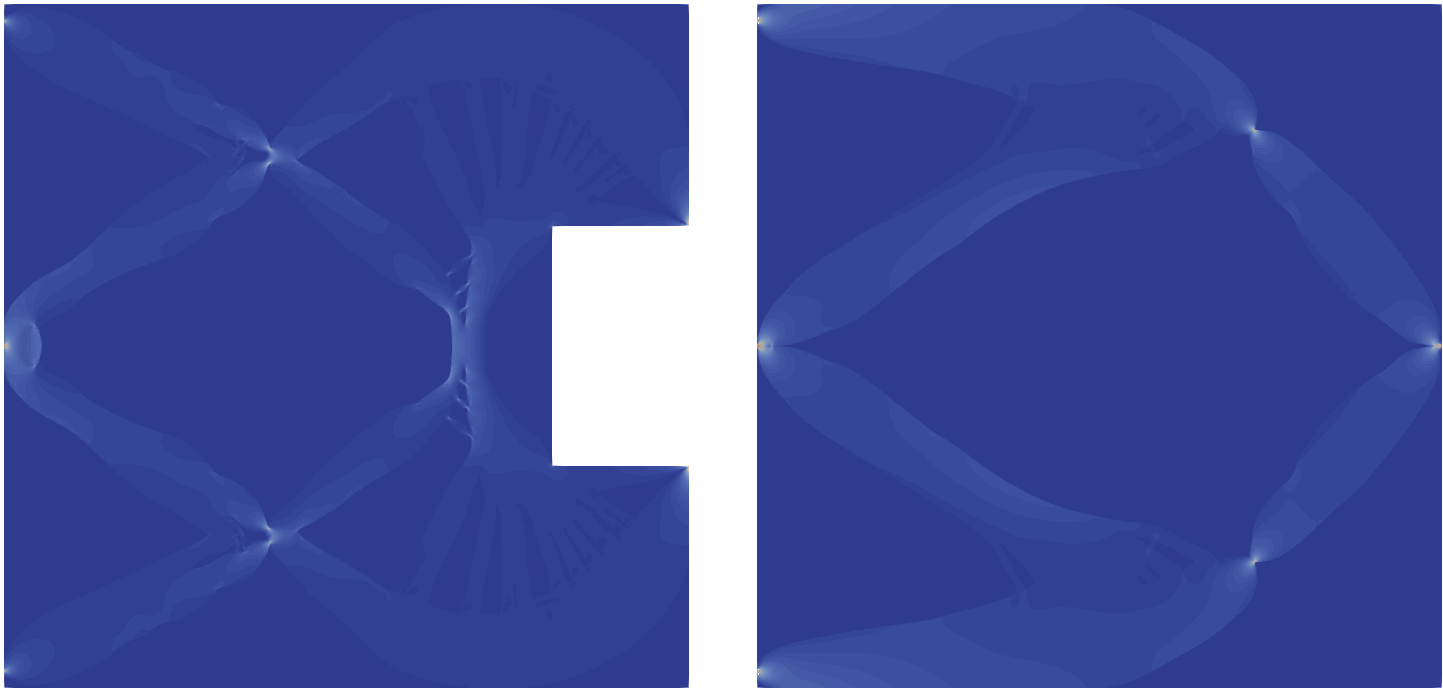} &
		\includegraphics[height=3.3cm]{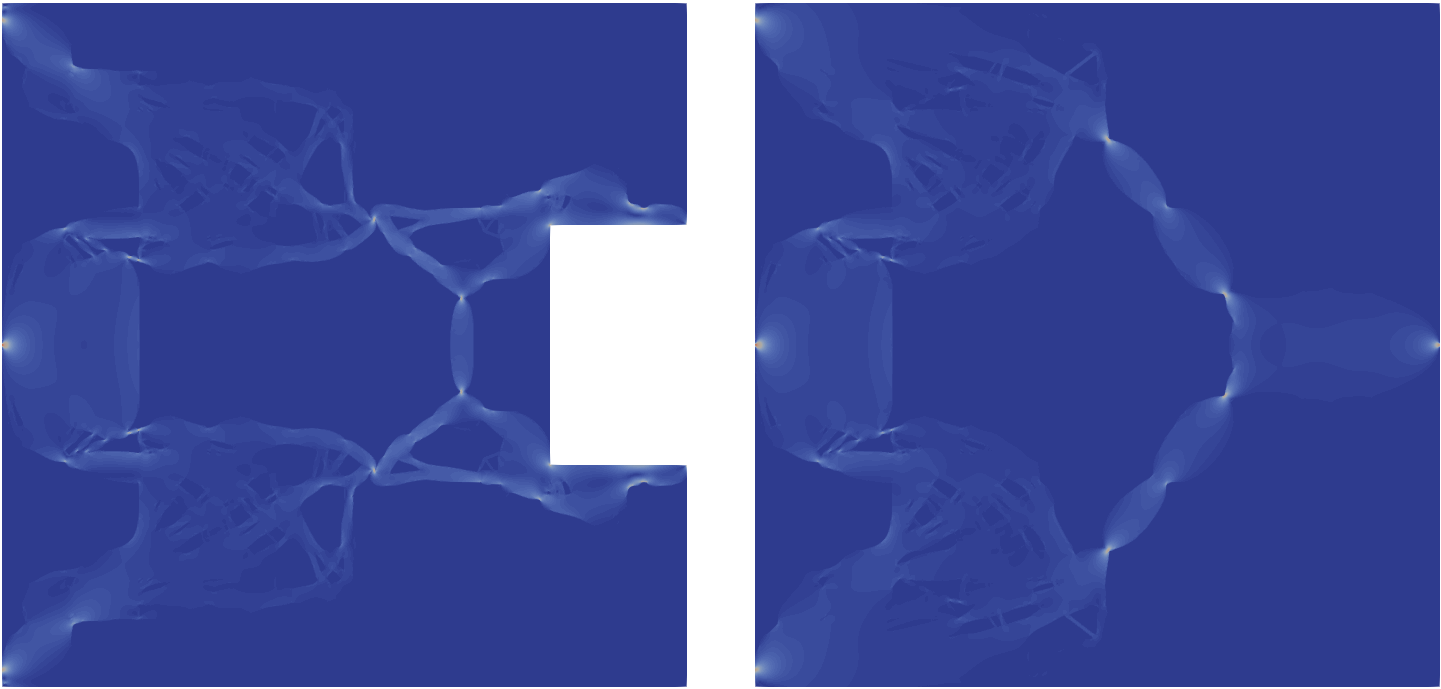} &
		\scalebox{0.35}{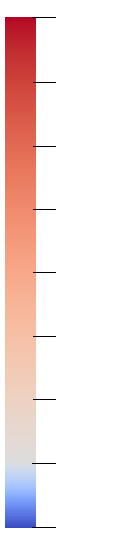} \\
		(a) & (b) &
	\end{tabular}
	\caption{\replace{}{Comparison of von Mises stress field $\sigma_\mathrm{v}$ in (a) the non-modular design of a gripper and an inverter, and (b) the corresponding reusable modular design with 44 modules over 4 colors.}}
	\label{fig:invgrip_vm}
\end{figure}

The two-color-based design in the second row of Fig.~\ref{fig:invgrip} contains \replace{$12$}{$15$} modules and exhibits \replace{$s_{\mathrm{TO}, 2} =-0.225$}{$s_{\mathrm{TO}, 2} =-0.451$} (with \replace{$s_{\mathrm{TO},2,\mathrm{inv}}=-0.189$}{$s_{\mathrm{TO},2,\mathrm{inv}}=-0.203$} and \replace{$s_{\mathrm{TO},2,\mathrm{grip}}=-0.060$}{$s_{\mathrm{TO},2,\mathrm{grip}}=-0.414$}). \replace{Here, the resulting assembly seem to be slightly more effective for the inverter mechanism.}{}\replace{Final}{Another} improvement is obtained for $3$-color clustering, which allows for a design in the third row of Fig.~\ref{fig:invgrip} composed of \replace{$21$}{$31$} modules and \replace{$s_{\mathrm{TO}, 3} =-0.275$}{$s_{\mathrm{TO}, 3} =-0.481$} (with \replace{$s_{\mathrm{TO},3,\mathrm{inv}}=-0.205$}{$s_{\mathrm{TO},3,\mathrm{inv}}=-0.225$} and \replace{$s_{\mathrm{TO},3,\mathrm{grip}}=-0.118$}{$s_{\mathrm{TO},3,\mathrm{grip}}=-0.427$}). Interestingly, \replace{almost the same}{the $4$-color-based $44$ module design exhibits a slightly worse overall} performance \replace{$s_{\mathrm{TO},4} =-0.279$}{$s_{\mathrm{TO}, 4} =-0.479$} (with \replace{$s_{\mathrm{TO},4,\mathrm{inv}}=-0.208$}{$s_{\mathrm{TO},4,\mathrm{inv}}=-0.345$} and \replace{$s_{\mathrm{TO},4,\mathrm{grip}}=-0.118$}{$s_{\mathrm{TO},4,\mathrm{grip}}=-0.224$})\replace{ results from $4$-color clustering, because the $3$- and $4$-color module sets result in the same assembly plan}{, while performing acceptably well when compared to the non-modular designs, see Fig.~\ref{fig:invgrip_deform}, reaching a similar maximum von Mises stress value, see Tab.~\ref{tab:reusable} and Fig.~\ref{fig:invgrip_vm}}.

\begin{table}[!t]
	\centering
	\small
	\caption{\replace{}{Overview of the modular reusable mechanism performances, where $u_\mathrm{in, inv}$ and $u_\mathrm{grip}$ denote the horizontal displacement at the force input of the inverter and gripper mechanisms, respectively, and $s_\mathrm{TO,inv}$ with $s_\mathrm{TO,grip}$ the corresponding objective function of individual mechanisms, $s_\mathrm{TO}$ the combined one, and $\sigma_{\mathrm{v},\max}$ the maximum von Mises stress. Values preceded by $^\dagger$ symbol denote a non-functional mechanism. Further, GA abbreviates geometrical advantage, evaluated as the sum of the mechanism performances $s_\mathrm{TO}$ to the sum of mechanisms inputs.}}
	\setlength\tabcolsep{0.5em}
	\begin{tabular}{lrrrrrrrrr}
		Design & $u_\mathrm{in,inv}$ & $u_\mathrm{in,grip}$& $s_\mathrm{TO,inv}$ & $s_\mathrm{TO,grip}$ & GA$_\mathrm{inv}$ & GA$_\mathrm{grip}$ & $s_\mathrm{TO}$ & GA & $\sigma_{\mathrm{v},\max}$\\
		\hline
		$1$ module over $1$ color & $0.993$ & $0.986$ & $^\dagger0.000$ & $-0.067$ & $0.000$ & $-0.068$ & $-0.040$ & $-0.334$ & $8.630$ \\
		$15$ modules over $2$ colors & $0.992$ & $0.975$ & $-0.203$ & $-0.414$ & $-0.204$ & $-0.425$ & $-0.451$ & $-0.314$ & $10.737$ \\
		$31$ modules over $3$ colors & $0.989$ & $0.975$ & $-0.225$ & $-0.427$ & $-0.227$ & $-0.438$ & $-0.481$ & $-0.332$ & $12.753$ \\
		$44$ modules over $4$ colors & $0.987$ & $0.973$ & $-0.345$ & $-0.224$ & $-0.349$ & $-0.230$ & $-0.479$ & $-0.290$ & $15.695$ \\
		$44$ modules over $4$ colors, uniform guess & $0.999$ & $0.973$ & $^\dagger0.000$ & $-0.589$ & $0.000$ & $-0.606$ & $-0.353$ & $-0.298$ & $11.785$ \\
		$44$ modules over $4$ colors, $3$-color guess & $0.993$ & $0.968$ & $-0.210$ & $-0.585$ & $-0.212$ & $-0.604$ & $-0.561$ & $-0.405$ & $13.804$ \\
		Non-modular & $0.987$ & $0.982$ & $-0.370$ & $-0.669$ & $-0.375$ & $-0.682$ & $-0.772$ & $-0.528$ & $15.314$
	\end{tabular}
	\label{tab:reusable}
\end{table}

\subsection{\replace{}{Dependence on initial guess}}\label{sec:reusableguess}

\begin{figure}[!b]
	\centering
	\begin{subfigure}{0.35\linewidth}
		\includegraphics[width=\linewidth]{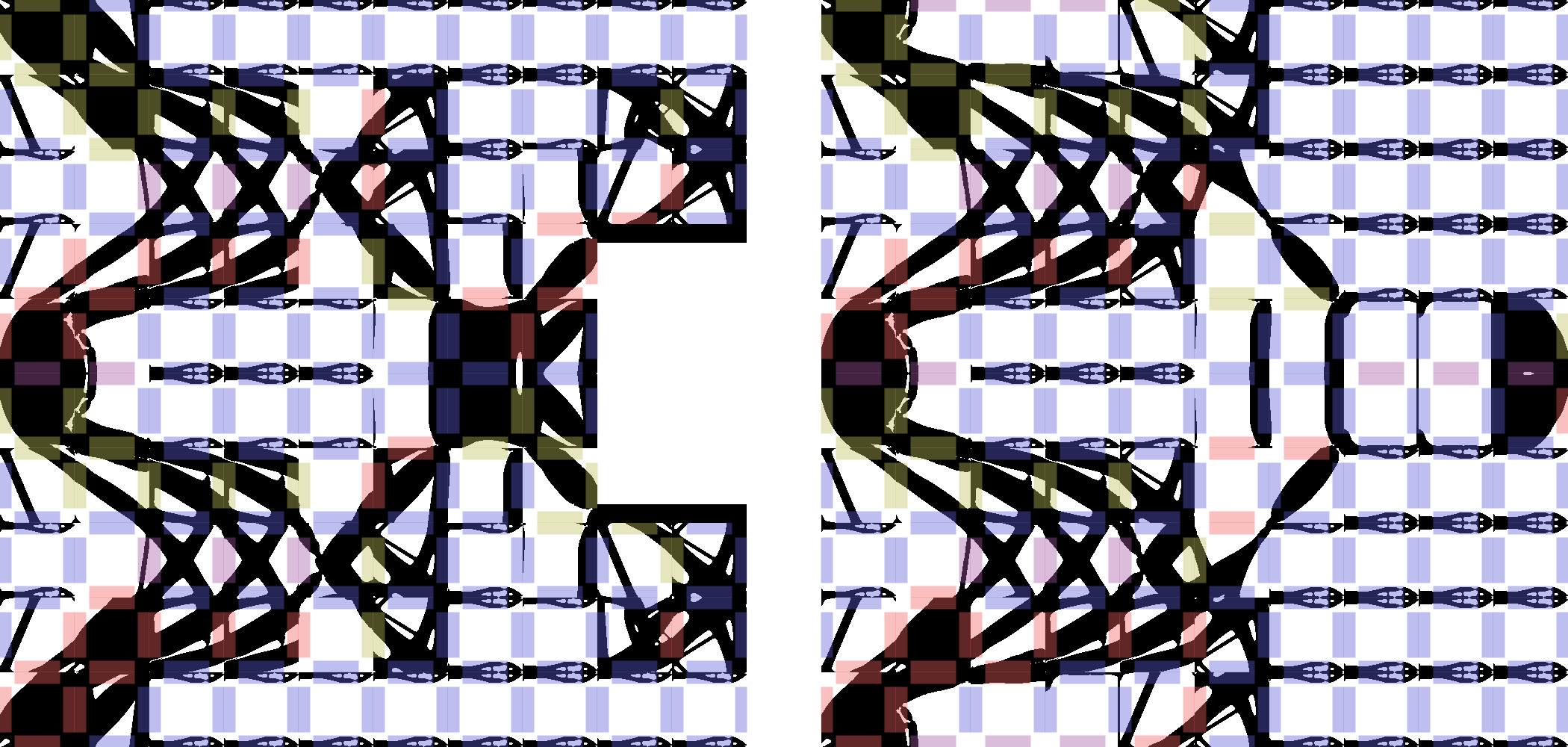}\\
		\vskip -1mm
		\includegraphics[width=\linewidth]{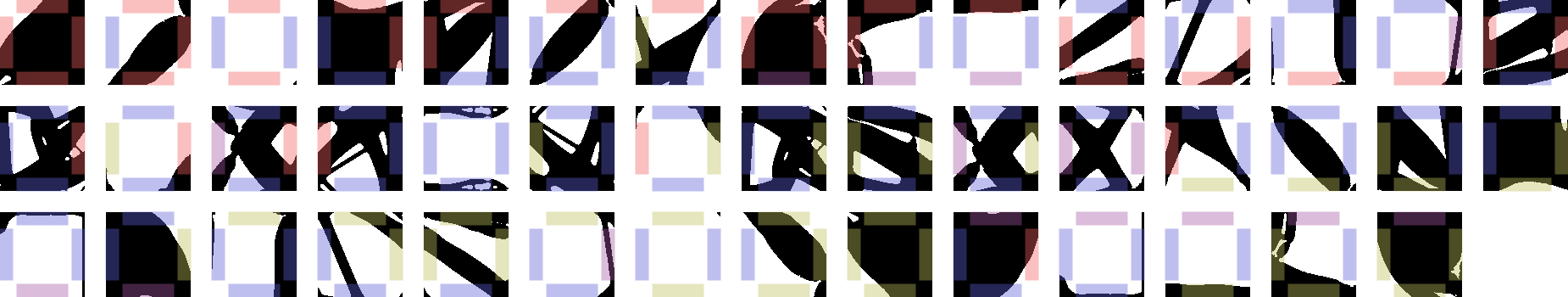}
		\caption{}
		\label{fig:guess_three}
	\end{subfigure}%
	\hspace{1cm}\begin{subfigure}{0.35\linewidth}
		\includegraphics[width=\linewidth]{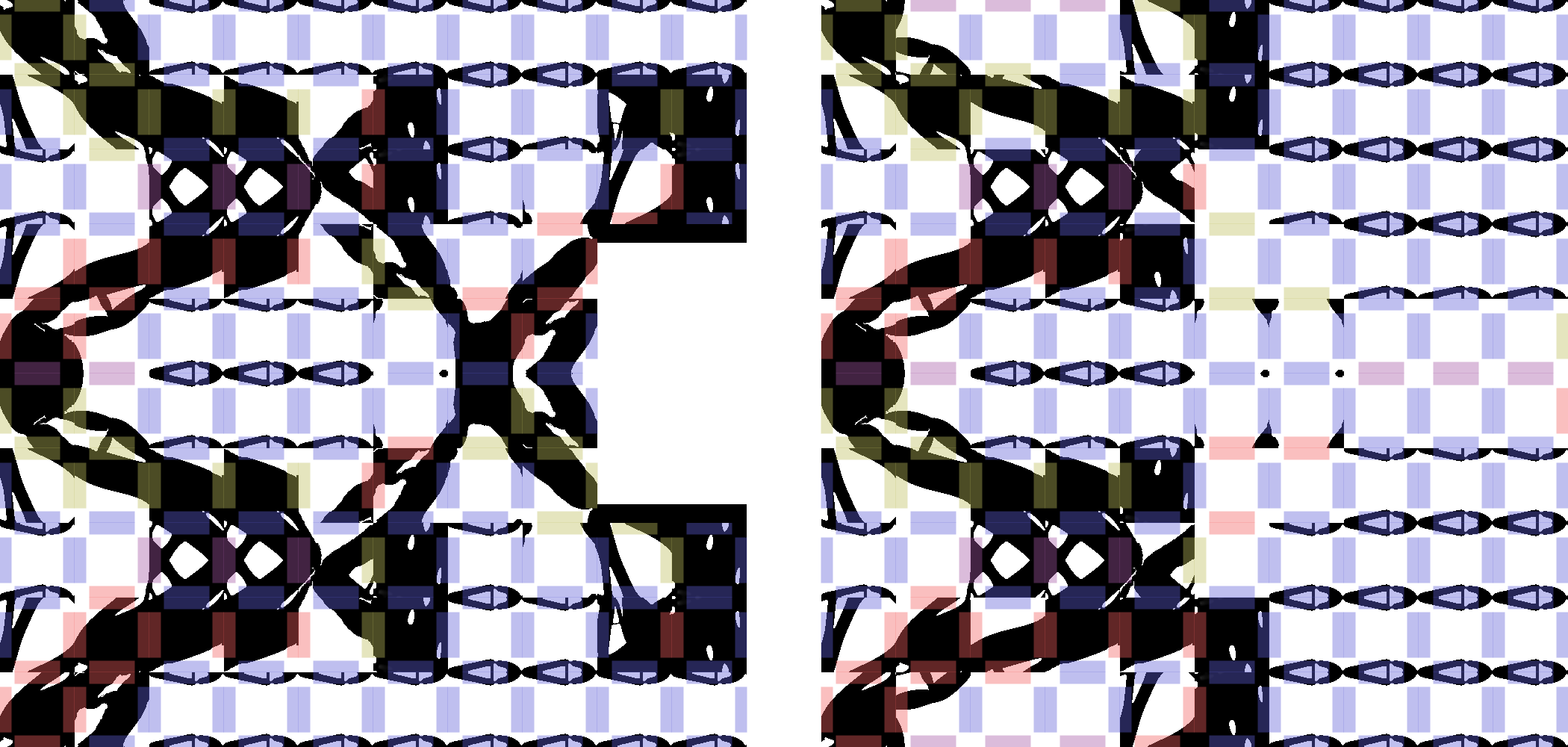}\\
		\vskip -1mm
		\includegraphics[width=\linewidth]{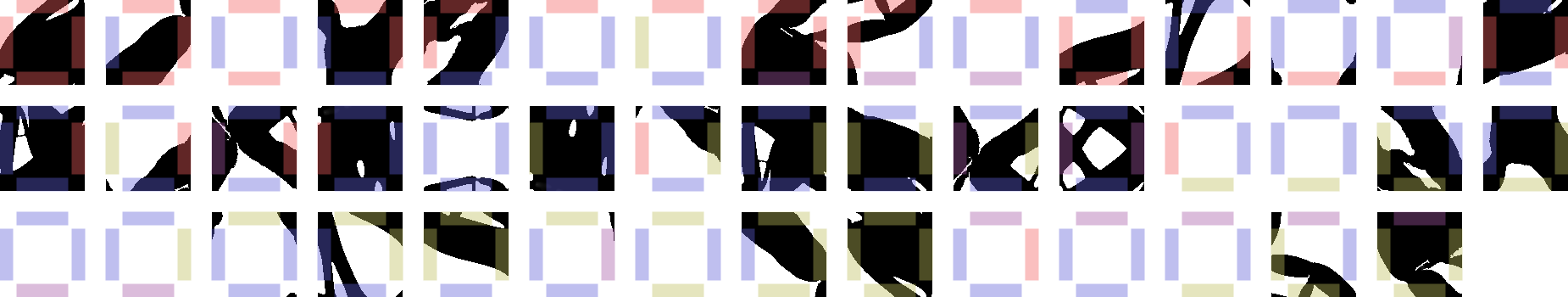}
		\caption{}
		\label{fig:guess_uniform}
	\end{subfigure}
	\caption{\replace{}{Designs of reusable modules in a gripper and inverter mechanisms obtained from (a) three-color-based initial guess, and (b) a uniform guess in TO.}}
\end{figure}

\replace{}{
	For the case of the $4$-color-based reusable modular mechanisms, we further illustrate the strong dependence of the TO part on supplied starting points. First, because the agglomerative clustering algorithm is deterministic, an increase in the number of module edge colors always generates assembly plans for which the starting points obtained for a lower number of edge colors are also feasible. As a simple example, we have initiated the $4$-color-based TO with the starting point obtained from the solution to the $3$-color-based modular FMO. The resulting design shown in Fig.~\ref{fig:guess_three} exhibits a slightly improved overall objective function $s_{\mathrm{TO},4} = -0.561$ compared to the results in Section \ref{sec:reusable}; with the gripper improved significantly ($s_{\mathrm{TO},4,\mathrm{grip}} = -0.585$), and the performance of the inverter mechanism degraded ($s_{\mathrm{TO},4,\mathrm{inv}} = -0.210$). Interestingly, the maximum von Mises stress is also reduced, recall Tab.~\ref{tab:reusable}.}

\replace{}{Another straightforward option is to adopt the common starting point of a uniform mass distribution, yielding $s_{\mathrm{TO},4} = -0.353$ and the designs in Fig.~\ref{fig:guess_uniform}. Here, the gripper exhibits $s_{\mathrm{TO},4,\mathrm{grip}} =-0.589$, but the inverter mechanism fails to work, showing that an appropriate choice of the starting point is very important and that it is difficult to rigorously compare the performance of different fixed assembly plans.}

\section{Conclusions}

In this contribution, we have developed a novel heuristic sequential framework for designing modular structures and mechanisms based on Wang tiling formalism. To the best of our knowledge, this manuscript presents the first study which allows modular compliant mechanisms with non-uniform modules to be designed, thus answering one of the challenges \replace{laid out}{outlined} in \citep{tyburec2020,wu2021b}.

Our strategy relies on a solution to the Free Material Optimization (FMO) problem \eqref{eq:fmo} on \replace{a coarse finite element mesh, smoothing the optimal material stiffness field to obtain physical results}{a module interface-related} finite element mesh, and clustering \replace{}{of} the resulting elasticity \replace{matrices}{tensors}. Topologies of individual modules in the assembly plans generated by the previous steps are eventually found by standard Topology Optimization (TO) extended to structural modularity, which we initialize using a well-defined yet problem-specific initial guess. Compared to our prior study on \replace{}{the} modular-topology optimization of truss structures \citep{tyburec2020}, this framework constitutes a computationally more attractive scheme that requires neither a repetitive solution to the module topology optimization problem nor its convexity. Consequently, the new approach also applies to non-convex problems arising, e.g., in continuum topology optimization.

As the first step of our method, we \replace{proposed solving}{solved} the FMO problem \eqref{eq:fmo} on a module-interface-based mesh discretization: \replace{we solve}{solving} for the (locally-)optimal elasticity \replace{matrices}{tensors} in the discretized domain using the \textsc{Pennon} \citep{kocvara2003} optimizer. Interestingly, the \replace{}{tensors} are often prone to the checkerboard phenomenon, at least when the most common bilinear quadrilateral finite elements are employed. Although the resulting degradation of structural responses is very significant and can be verified simply by projecting the optimal material properties onto a refined mesh, the checkerboard issue has surprisingly never been studied or even reported with connection to FMO problems\replace{, and thus it remains open to rigorous examination}{}. The reason behind this can be attributed to the fact that the checkerboard patterns in tensorial field are hard to identify in general---contrary to, e.g., the scalar density fields in TO. However, the checkerboard issue in FMO is often discernible when visualizing traces of the optimized material stiffnesses, recall Figs.~\ref{fig:bar} or cf. \citep[Figs. 4b and 4e]{hu2020}. \replace{Because we use the FMO problem \eqref{eq:fmo} only as a heuristic tool only, we eliminate the checkerboard using a single, computationally inexpensive post-processing step performed on the optimization outputs. The essence of post-processing adopts the idea of averaging discretized element stiffnesses onto nodal ones and a~backward projection onto element stiffnesses. Because these mappings eliminate the checkerboard-like distribution of stiffnesses, the post-processed designs are more robust with respect to mesh refinement. Furthermore, due to its convexity, the post-processing step preserves both volume and the convex eigenvalue-based constraints. Admittedly, however, the objective function usually increases with the increase magnitude depending on the level of refinement, approaching zero for infinitely fine discretizations.}{Here, we eliminate the checkerboard issue by adopting a refined design-element discretization, in a manner similar to \cite{Mukherjee2021}.}

The particular shape of the finite element discretization used in the problem \eqref{eq:fmo} was adopted to provide a simple interface-based definition of individual modules, i.e., each module being defined by four edges and the corresponding quadruple of elasticity matrices. This definition, which is equivalent to Wang tiles, allows us to control the number of interface types among the modules. In this manuscript, we have forbidden rotations of modules, so the clustering of the vertical and horizontal edges \replace{were}{was} performed independently. 

To this goal, we developed a deterministic agglomerative clustering algorithm tailored to satisfy structural symmetries when present. While in very specific instances of linear elasticity one may rely on partitioning simpler datasets than \replace{the}{with} stiffness tensors, recall Section \ref{sec:introclustering}, our clustering covers the most general class of problems, and therefore also applies to multiple-load case minimum-compliance optimization or compliant mechanism design problems.

The resulting clustered designs form modular assembly plans in the form of valid Wang tilings, and their performance can readily be assessed using FMO formulation extended to structural modularity \eqref{eq:mfmo}. However, the modular FMO problem also appears appealing from another viewpoint: its outputs in the form of the traces of module stiffnesses can be advantageously supplied to TO as initial guesses. Subsequently, TO---with the Optimality Criteria method with sensitivity filtering and a continuation---was used to arrive at optimized topologies.

We demonstrated the applicability of the developed framework on three classical problems: \replace{}{the} design of a minimum-compliant modular MBB beam, \replace{}{of} a force inverter, and \replace{}{of} a modular gripper mechanism. Ultimately, we introduced a new problem \replace{of}{for} designing topologies and assembly of modules that can be reused in both the gripper and inverter mechanisms. 

For all these problems, we have shown that with an increasing number of colors, the number of module types not only grows, but overall structural performance also improves. In the ultimate limit, we would reach the performance of a non-modular design. \replace{}{A second interesting observation emerged from the von Mises stress distributions. For the case of compliance minimization problems, modularity seems to increase the maximum stress value compared to non-modular designs, but the converse is true for compliant mechanisms---few strong singularities in compliant hinges in non-modular designs are usually spread into a larger number of weaker ones, reducing the overall maximum stress level.}

This work can be extended in several directions\replace{. In addition to the}{, starting with a} straightforward extension to three-dimensional problems\replace{, we plan to focus on the proper handling of the checkerboard phenomenon, e.g., via a filtering scheme inside FMO or by using higher-order finite elements}. \replace{With}{Assuming that} simpler manufacturability \replace{being}{is} one benefi\replace{t}{cial future outpu}t of modularity, we will try to develop constraints to secure a continuous material distribution inside individual modules \replace{}{and adopt a robust scheme to suppress the compliant hinges in mechanism designs~\cite{wu2021b}}. Another promising extension constitutes the design of multi-functional mechanisms under finite strains. Such extension, however, would require extending the clustering procedure in Section~\ref{sec:clustering} to deal with \replace{}{(pseudo)}time-dependent datasets. In addition, modularity can also be exploited \replace{inside}{within} optimization and in finite element analyses by efficient preconditioners or domain decomposition methods \cite{medricky2021}. Finally, \replace{}{to address the challenge of manufacturability,} we plan to manufacture the selected modular designs and validate \replace{}{experimentally} the structural response of the prototypes against \replace{}{numerical model's} predictions.

\section*{Declaration of competing interest}
The authors declare that they have no known competing financial interests or personal relationships that could have appeared to influence the work reported in this paper.

\section*{CRediT author statement}

\replace{}{
	\noindent\textbf{Marek Tyburec} Conceptualization, Methodology, Investigation, Software, Writing- Original draft preparation, \textbf{Martin Do{\v{s}}k{\'{a}}{\v{r}}} Conceptualization, Methodology, Software, Writing- Original draft preparation, \textbf{Jan Zeman} Supervision, Writing- Reviewing and Editing, Funding acquisition, Project administration, \textbf{Martin Kru{\v{z}}{\'{i}}k} Supervision, Writing- Reviewing and Editing.}

\section*{Acknowledgement}
\replace{}{We thank Stephanie Krueger for a proofreading of the initial versions of this manuscript and to the anonymous reviewers, whose comments allowed to improve the manuscript substantially.}

The authors acknowledge the support from the Czech Science Foundation, project No. 19-26143X.

\renewcommand{\thesubsection}{\Alph{subsection}}
\section*{Appendix}

\subsection{Material stiffness matrix in symmetric designs}\label{app:symmetry}

Let $\sevek{\sigma}^{(i)}, \sevek{\varepsilon}^{(i)}$, and $\sevek{\sigma}^{(j)}, \sevek{\varepsilon}^{(j)}$ be the stress and engineering strain column vectors in the element $i$ and $j$, respectively. Consider now a symmetric distribution of material stiffnesses and displacement fields in a design domain, and the axis of symmetry aligned with the $1$ or $2$ axis. Then, for the two elements $i$ and $j$ corresponding to each other in the symmetry, we have $\sigma^{(i)}_{11} = \sigma^{(j)}_{11}$, $\sigma^{(i)}_{22} = \sigma^{(j)}_{22}$, but $\sigma^{(i)}_{12} = -\sigma^{(j)}_{12}$. The strains possess similar symmetries, i.e., $\varepsilon^{(i)}_{11} = \varepsilon^{(j)}_{11}$, $\varepsilon^{(i)}_{22} = \varepsilon^{(j)}_{22}$, and $\gamma^{(i)}_{12} = -\gamma^{(j)}_{12}$.

Considering five test symmetric strain column vectors, the stress symmetry requires
\begin{equation}
\semtrx{E}^{(i)} \begin{pmatrix}
\varepsilon_{11}^{(i)} & 0 & 0 & \varepsilon_{11}^{(i)} & \varepsilon_{11}^{(i)}\\
0 & \varepsilon^{(i)}_{22} & 0 & \varepsilon_{22}^{(i)} & \varepsilon_{22}^{(i)}\\
0& 0 &\gamma^{(i)}_{12} & 0 &\gamma^{(i)}_{12}
\end{pmatrix} = 
\semtrx{E}^{(j)} \begin{pmatrix}
\varepsilon^{(i)}_{11} & 0 & 0 & \varepsilon_{11}^{(i)} & \varepsilon_{11}^{(i)}\\
0 & \varepsilon^{(i)}_{22} & 0 & \varepsilon_{22}^{(i)} & \varepsilon_{22}^{(i)}\\
0& 0 & -(-\gamma^{(i)}_{12}) & 0 & -(-\gamma^{(i)}_{12})
\end{pmatrix}.
\end{equation}
From the first three test strain fields it follows that the diagonal entries of the elastic material stiffness matrix must match, i.e., $e_{1111}^{(i)} = e_{1111}^{(j)}$, $e_{2222}^{(i)} = e_{2222}^{(j)}$, and $e_{1212}^{(i)} = e_{1212}^{(j)}$. Using this observation, the fourth test strain yields $e_{1122}^{(i)} = e_{1122}^{(j)}$. The relation of the two remaining components of the elastic tensor are determined from the stress balance induced by the fifth field. Then, $e_{1112}^{(i)} = -e_{1112}^{(j)}$ and $e_{2212}^{(i)} = -e_{2212}^{(j)}$.

\subsection{Derivation of sensitivities}\label{app:sensitivities}

Let us assume an unknown field denoted by $\sevek{y}$, which encompasses the elements of $\semtrx{E}$ in the case of \eqref{eq:fmo} and $\sevek{\rho}$ in \eqref{eq:mto}. In addition, we define 
\begin{equation}\label{eq:con}
a\left(\sevek{y}\right) \coloneqq \sevek{a}^\mathrm{T} \semtrx{K}(\sevek{y})^{-1} \sevek{f} -s,
\end{equation}
so that the inequalities \eqref{eq:fmo_mutual} or \eqref{eq:mto_mutual} are equivalent to
\begin{equation}
a(\sevek{y}) \le 0.
\end{equation}
In the following subsections, we derive analytical expressions for the first- and second-order derivatives of $a(\sevek{y})$ with respect to $\sevek{y}$.

\subsubsection{Gradient}
\label{sec:general_sensitivity}

From Eq.~\eqref{eq:con} it follows that
\begin{equation}\label{eq:dir}
\pd{a(\sevek{y})}{y_i} = \sevek{a}^\mathrm{T} \pd{\semtrx{K}(\sevek{y})^\mathrm{-1}}{y_i} \sevek{f}.
\end{equation}
After differentiating the identity
\begin{equation}
\semtrx{I} = \semtrx{K}(\sevek{y})^{-1} \semtrx{K}(\sevek{y}), 
\end{equation}
with respect to $y_i$ and re-arranging the terms, we obtain
\begin{equation}\label{eq:inv}
\pd{\semtrx{K}(\sevek{y})^{-1}}{y_i} = -\semtrx{K}(\sevek{y})^{-1} \pd{\semtrx{K}(\sevek{y})}{y_i} \semtrx{K}(\sevek{y})^{-1}.
\end{equation}
Inserting \eqref{eq:inv} into \eqref{eq:dir}, we obtain
\begin{equation}
\pd{a(\sevek{y})}{y_i}
= - \sevek{a}^\mathrm{T} \semtrx{K}(\sevek{y})^{-1} \pd{\semtrx{K}(\sevek{y})}{y_i} \semtrx{K}(\sevek{y})^{-1} \sevek{f} = -\sevek{\lambda}^\mathrm{T} \pd{\semtrx{K}(\sevek{y})}{y_i} \sevek{u},
\end{equation}
with $\sevek{\lambda}$ and $\sevek{u}$ being solutions to
\begin{subequations}
	\begin{align}
	\semtrx{K}(\sevek{y}) \sevek{u} &= \sevek{f},\\
	\semtrx{K}(\sevek{y}) \sevek{\lambda} &= \sevek{a}.
	\end{align}
\end{subequations}

\subsubsection{Hessian}\label{app:hessian}

In order to compute the Hessian matrix, which we apply only in solutions of the FMO problems \eqref{eq:fmo} and \eqref{eq:mfmo}, we first differentiate \eqref{eq:dir} with respect to $y_j$, i.e., 
\begin{equation}\label{eq:dir2}
\pd{^2 a(\sevek{y})}{y_i \partial y_j} = \sevek{a}^\mathrm{T} \pd{^2 \semtrx{K}(\sevek{y})^\mathrm{-1}}{y_i \partial y_j} \sevek{f}
\end{equation}
To compute $\pd{\semtrx{K}(\sevek{y})^\mathrm{-1}}{y_i \partial y_j}$, we differentiate \eqref{eq:inv} with respect to $y_j$ to obtain
\begin{equation}
\pd{^2 \semtrx{K}(\sevek{y})^{-1}}{y_i \partial y_j} =  -\pd{\semtrx{K}(\sevek{y})^{-1}}{y_j} \pd{\semtrx{K}(\sevek{y})}{y_i} \semtrx{K}(\sevek{y})^{-1} -  \semtrx{K}(\sevek{y})^{-1}\pd{\semtrx{K}(\sevek{y})}{y_i} \pd{\semtrx{K}(\sevek{y})^{-1}}{y_j} - \semtrx{K}(\sevek{y})^{-1} \pd{^2 \semtrx{K}(\sevek{y})}{y_i \partial y_j} \semtrx{K}(\sevek{y})^{-1}.
\end{equation}
Using the identity \eqref{eq:inv} and exploiting that the stiffness matrix is symmetric, we receive
\begin{equation}\label{eq:k2}
\pd{^2 \semtrx{K}(\sevek{y})^{-1}}{y_i \partial y_j} =  2\semtrx{K}(\sevek{y})^{-1} \pd{\semtrx{K}(\sevek{y})}{y_i} \semtrx{K}(\sevek{y})^{-1} \pd{\semtrx{K}(\sevek{y})}{y_j} \semtrx{K}(\sevek{y})^{-1} - \semtrx{K}(\sevek{y})^{-1} \pd{^2 \semtrx{K}(\sevek{y})}{y_i \partial y_j} \semtrx{K}(\sevek{y})^{-1}.
\end{equation}
In the FMO problem \eqref{eq:fmo}, the stiffness matrix is a linear function of $\sevek{y}$, so the second term vanishes. Therefore, equations \eqref{eq:inv}, \eqref{eq:k2} combined with \eqref{eq:dir2} yield
\begin{equation}
\pd{^2 a(\sevek{y})}{y_i \partial y_j} = 2 \sevek{a}^\mathrm{T} \semtrx{K}(\sevek{y})^{-1} \pd{\semtrx{K}(\sevek{y})}{y_i} \semtrx{K}(\sevek{y})^{-1} \pd{\semtrx{K}(\sevek{y})}{y_j} \semtrx{K}(\sevek{y})^{-1} \sevek{f} = 2 \sevek{\lambda}^\mathrm{T} \pd{\semtrx{K}(\sevek{y})}{y_i} \semtrx{K}(\sevek{y})^{-1} \pd{\semtrx{K}(\sevek{y})}{y_j} \sevek{u}.
\end{equation}
Let us now define matrices $\semtrx{U}, \sevek{\Lambda}$ with the $i$-th columns computed as
\begin{subequations}
	\begin{align}
	\semtrx{U}_i (\sevek{y})&=\pd{\semtrx{K}(\sevek{y})}{y_i} \sevek{u},\\
	\semtrx{\Lambda}_i (\sevek{y})&=\pd{\semtrx{K}(\sevek{y})}{y_i} \sevek{\lambda}.
	\end{align}
\end{subequations}
Then, the Hessian matrix $\semtrx{H} (\sevek{y})$ reads as
\begin{equation}\label{eq:hessian}
\semtrx{H}(\sevek{y}) = 2 \left[\semtrx{\Lambda}(\sevek{y})\right]^\mathrm{T} \semtrx{K}(\sevek{y})^{-1} \semtrx{U}(\sevek{y}).
\end{equation}
When \eqref{eq:fmo_mutual} or \eqref{eq:mto_mutual} is a compliance constraint, $\semtrx{\Lambda}(\sevek{y}) = \semtrx{U}(\sevek{y})$. Because $\semtrx{K}(\sevek{y})^{-1} \succ 0$, we also have $\semtrx{H}(\sevek{y}) \succ 0$ for all $\sevek{y}$. Consequently, the constraint \eqref{eq:fmo_mutual} is convex.

\subsection{Scaling of the optimization problem}\label{app:scaling}

In the case of $\sevek{f} \neq \sevek{a}$, numerical solution to \eqref{eq:fmo} appears to be sensitive to scaling of the design variables. To make the solution process more robust, we substitute $\semtrx{E}^{(\ell)}$ with $\overline{\epsilon} \tilde{\semtrx{E}}^{(\ell)}$, which implies that $\mathrm{Tr}(\tilde{\semtrx{E}}^{(\ell)}) \in \langle 3 \underline{\epsilon}/\overline{\epsilon}, 1 \rangle$. Consequently, we arrive at the modified formulation
\begin{subequations}
	\begin{alignat}{4}
	\min_{s_\mathrm{sc}, \tilde{\semtrx{E}}^{(1)}, \dots, \tilde{\semtrx{E}}^{(n_\mathrm{\ell})}}\; && s_\mathrm{sc}\hspace{41.5mm}\\
	\text{s.t.}\; && s_\mathrm{sc}-\sevek{a}^\mathrm{T} \left[\frac{\semtrx{K}_0}{\overline{\epsilon}} + \sum_{\ell=1}^{n_\mathrm{\ell}} \semtrx{K}^{(\ell)}\left(\tilde{\semtrx{E}}^{(\ell)}\right)\right]^{-1} \sevek{f} &&\;\ge\;& 0,\\
	&& \frac{\sum_{\ell=1}^{n_\mathrm{\ell}} v^{(\ell)}}{\overline{V}} - \sum_{\ell=1}^{n_\mathrm{\ell}} \frac{v^{(\ell)}\mathrm{Tr} \left(\tilde{\semtrx{E}}^{(\ell)}\right)}{\overline{V}} &&\;\ge\;& 0,\\
	&& 1 - \mathrm{Tr}\left(\tilde{\semtrx{E}}^{(\ell)}\right) &&\;\ge\;& 0, &\;\;\forall e \in \{1,\dots, n_\mathrm{\ell}\},\\
	&& \tilde{\semtrx{E}}^{(\ell)} &&\;\succeq\;& \underline{\epsilon}/\overline{\epsilon}, &\;\;\forall e \in \{1,\dots, n_\mathrm{\ell}\},
	\end{alignat}
\end{subequations}
from which the (locally) optimal solution is recovered as
\begin{subequations}
	\begin{align}
	s &= s_\mathrm{sc}/\overline{\epsilon}\\
	\semtrx{E}^{(\ell)} &= \overline{\epsilon} \tilde{\semtrx{E}}^{(\ell)}, \quad \forall e \in \{1,\dots,n_\mathrm{\ell}\}.
	\end{align}
\end{subequations}

\bibliography{liter}

\end{document}

%% file: vM_colorbar.pdf_tex
\begingroup%
  \makeatletter%
  \providecommand\color[2][]{%
    \errmessage{(Inkscape) Color is used for the text in Inkscape, but the package 'color.sty' is not loaded}%
    \renewcommand\color[2][]{}%
  }%
  \providecommand\transparent[1]{%
    \errmessage{(Inkscape) Transparency is used (non-zero) for the text in Inkscape, but the package 'transparent.sty' is not loaded}%
    \renewcommand\transparent[1]{}%
  }%
  \providecommand\rotatebox[2]{#2}%
  \newcommand*\fsize{\dimexpr\f@size pt\relax}%
  \newcommand*\lineheight[1]{\fontsize{\fsize}{#1\fsize}\selectfont}%
  \ifx\svgwidth\undefined%
    \setlength{\unitlength}{64.46869798bp}%
    \ifx\svgscale\undefined%
      \relax%
    \else%
      \setlength{\unitlength}{\unitlength * \real{\svgscale}}%
    \fi%
  \else%
    \setlength{\unitlength}{\svgwidth}%
  \fi%
  \global\let\svgwidth\undefined%
  \global\let\svgscale\undefined%
  \makeatother%
  \begin{picture}(1,4.05810693)%
    \lineheight{1}%
    \setlength\tabcolsep{0pt}%
    \put(0.43822454,0.09086233){\makebox(0,0)[lt]{\lineheight{1.25}\smash{\begin{tabular}[t]{l}\Large{0}\end{tabular}}}}%
    \put(0.96155474,2.03273433){\rotatebox{90}{\makebox(0,0)[t]{\lineheight{1.25}\smash{\begin{tabular}[t]{c}\LARGE{von Mises stress $\sigma_{ \mathrm{v}}$}\end{tabular}}}}}%
    \put(0,0){\includegraphics[width=\unitlength,page=1]{vM_colorbar.pdf}}%
    \put(0.43822454,0.5641546){\makebox(0,0)[lt]{\lineheight{1.25}\smash{\begin{tabular}[t]{l}\Large{2}\end{tabular}}}}%
    \put(0.43822454,1.03744688){\makebox(0,0)[lt]{\lineheight{1.25}\smash{\begin{tabular}[t]{l}\Large{4}\end{tabular}}}}%
    \put(0.43822454,1.51073915){\makebox(0,0)[lt]{\lineheight{1.25}\smash{\begin{tabular}[t]{l}\Large{6}\end{tabular}}}}%
    \put(0.43822454,1.98403142){\makebox(0,0)[lt]{\lineheight{1.25}\smash{\begin{tabular}[t]{l}\Large{8}\end{tabular}}}}%
    \put(0.43822454,2.45732353){\makebox(0,0)[lt]{\lineheight{1.25}\smash{\begin{tabular}[t]{l}\Large{10}\end{tabular}}}}%
    \put(0.43822454,2.93061581){\makebox(0,0)[lt]{\lineheight{1.25}\smash{\begin{tabular}[t]{l}\Large{12}\end{tabular}}}}%
    \put(0.43822454,3.40390817){\makebox(0,0)[lt]{\lineheight{1.25}\smash{\begin{tabular}[t]{l}\Large{14}\end{tabular}}}}%
    \put(0.43822454,3.87720034){\makebox(0,0)[lt]{\lineheight{1.25}\smash{\begin{tabular}[t]{l}\Large{16}\end{tabular}}}}%
  \end{picture}%
\endgroup%

%% file: paper.bbl
\begin{thebibliography}{86}
\providecommand{\natexlab}[1]{#1}
\providecommand{\url}[1]{\texttt{#1}}
\providecommand{\urlprefix}{URL }
\expandafter\ifx\csname urlstyle\endcsname\relax
  \providecommand{\doi}[1]{doi:\discretionary{}{}{}#1}\else
  \providecommand{\doi}[1]{doi:\discretionary{}{}{}\begingroup
  \urlstyle{rm}\url{#1}\endgroup}\fi
\providecommand{\bibinfo}[2]{#2}

\bibitem[{Zhu et~al.(2021)Zhu, Zhou, Wang, Zhou, Yuan, and Zhang}]{zhu2021}
\bibinfo{author}{J.~Zhu}, \bibinfo{author}{H.~Zhou}, \bibinfo{author}{C.~Wang},
  \bibinfo{author}{L.~Zhou}, \bibinfo{author}{S.~Yuan},
  \bibinfo{author}{W.~Zhang}, \bibinfo{title}{A review of topology optimization
  for additive manufacturing: Status and challenges}, \bibinfo{journal}{Chinese
  Journal of Aeronautics} \bibinfo{volume}{34}~(\bibinfo{number}{1})
  (\bibinfo{year}{2021}) \bibinfo{pages}{91--110},
  \doi{\bibinfo{doi}{10.1016/j.cja.2020.09.020}}.

\bibitem[{Csal{\'{o}}di et~al.(2021)Csal{\'{o}}di, S\"{u}le, Jask{\'{o}},
  Holczinger, and Abonyi}]{csaldi2021}
\bibinfo{author}{R.~Csal{\'{o}}di}, \bibinfo{author}{Z.~S\"{u}le},
  \bibinfo{author}{S.~Jask{\'{o}}}, \bibinfo{author}{T.~Holczinger},
  \bibinfo{author}{J.~Abonyi}, \bibinfo{title}{Industry 4.0-driven development
  of optimization algorithms: A systematic overview},
  \bibinfo{journal}{Complexity} \bibinfo{volume}{2021} (\bibinfo{year}{2021})
  \bibinfo{pages}{1--22}, \doi{\bibinfo{doi}{10.1155/2021/6621235}}.

\bibitem[{Liu and Ma(2016)}]{liu2016}
\bibinfo{author}{J.~Liu}, \bibinfo{author}{Y.~Ma}, \bibinfo{title}{Sustainable
  design-oriented level set topology optimization}, \bibinfo{journal}{Journal
  of Mechanical Design} \bibinfo{volume}{139}~(\bibinfo{number}{1}),
  \doi{\bibinfo{doi}{10.1115/1.4035052}}.

\bibitem[{Tajs-Zieli{\'{n}}ska and Bochenek(2021)}]{tajsZieliska2021}
\bibinfo{author}{K.~Tajs-Zieli{\'{n}}ska}, \bibinfo{author}{B.~Bochenek},
  \bibinfo{title}{Multi-domain and multi-material topology optimization in
  design and strengthening of innovative sustainable structures},
  \bibinfo{journal}{Sustainability} \bibinfo{volume}{13}~(\bibinfo{number}{6})
  (\bibinfo{year}{2021}) \bibinfo{pages}{3435},
  \doi{\bibinfo{doi}{10.3390/su13063435}}.

\bibitem[{J\'ilek et~al.(2021)J\'ilek, Somr, Kulich, Zeman, and
  P\v{r}eu\v{c}il}]{jilek2021}
\bibinfo{author}{M.~J\'ilek}, \bibinfo{author}{M.~Somr},
  \bibinfo{author}{M.~Kulich}, \bibinfo{author}{J.~Zeman},
  \bibinfo{author}{L.~P\v{r}eu\v{c}il}, \bibinfo{title}{Towards a passive
  self-assembling macroscale multi-robot system}, \bibinfo{journal}{IEEE
  Robotics and Automation Letters} \bibinfo{volume}{6}~(\bibinfo{number}{4})
  (\bibinfo{year}{2021}) \bibinfo{pages}{7293--7300},
  \doi{\bibinfo{doi}{10.1109/LRA.2021.3096748}}.

\bibitem[{Yasuda et~al.(2021)Yasuda, Buskohl, Gillman, Murphey, Stepney, Vaia,
  and Raney}]{yasuda2021}
\bibinfo{author}{H.~Yasuda}, \bibinfo{author}{P.~R. Buskohl},
  \bibinfo{author}{A.~Gillman}, \bibinfo{author}{T.~D. Murphey},
  \bibinfo{author}{S.~Stepney}, \bibinfo{author}{R.~A. Vaia},
  \bibinfo{author}{J.~R. Raney}, \bibinfo{title}{Mechanical computing},
  \bibinfo{journal}{Nature} \bibinfo{volume}{598}~(\bibinfo{number}{7879})
  (\bibinfo{year}{2021}) \bibinfo{pages}{39--48},
  \doi{\bibinfo{doi}{10.1038/s41586-021-03623-y}}.

\bibitem[{Ion et~al.(2017)Ion, Wall, Kovacs, and Baudisch}]{ion2017}
\bibinfo{author}{A.~Ion}, \bibinfo{author}{L.~Wall},
  \bibinfo{author}{R.~Kovacs}, \bibinfo{author}{P.~Baudisch},
  \bibinfo{title}{Digital mechanical metamaterials}, in:
  \bibinfo{booktitle}{Proceedings of the 2017 {CHI} Conference on Human Factors
  in Computing Systems}, \bibinfo{publisher}{{ACM}}, \bibinfo{pages}{977--988},
  \doi{\bibinfo{doi}{10.1145/3025453.3025624}}, \bibinfo{year}{2017}.

\bibitem[{Ne{\v{z}}erka et~al.(2018)Ne{\v{z}}erka, Somr, Janda, Vorel,
  Do{\v{s}}k{\'{a}}{\v{r}}, Anto{\v{s}}, Zeman, and Nov{\'{a}}k}]{nezerka2018}
\bibinfo{author}{V.~Ne{\v{z}}erka}, \bibinfo{author}{M.~Somr},
  \bibinfo{author}{T.~Janda}, \bibinfo{author}{J.~Vorel},
  \bibinfo{author}{M.~Do{\v{s}}k{\'{a}}{\v{r}}},
  \bibinfo{author}{J.~Anto{\v{s}}}, \bibinfo{author}{J.~Zeman},
  \bibinfo{author}{J.~Nov{\'{a}}k}, \bibinfo{title}{A jigsaw puzzle
  metamaterial concept}, \bibinfo{journal}{Composite Structures}
  \bibinfo{volume}{202} (\bibinfo{year}{2018}) \bibinfo{pages}{1275--1279},
  \doi{\bibinfo{doi}{10.1016/j.compstruct.2018.06.015}}.

\bibitem[{Tugilimana et~al.(2019)Tugilimana, Coelho, and
  Thrall}]{tugilimana2019}
\bibinfo{author}{A.~Tugilimana}, \bibinfo{author}{R.~F. Coelho},
  \bibinfo{author}{A.~P. Thrall}, \bibinfo{title}{An integrated design
  methodology for modular trusses including dynamic grouping, module spatial
  orientation, and topology optimization}, \bibinfo{journal}{Structural and
  Multidisciplinary Optimization} \bibinfo{volume}{60}~(\bibinfo{number}{2})
  (\bibinfo{year}{2019}) \bibinfo{pages}{613--638},
  \doi{\bibinfo{doi}{10.1007/s00158-019-02230-w}}.

\bibitem[{Wu et~al.(2021{\natexlab{a}})Wu, Sigmund, and Du}]{wu2021b}
\bibinfo{author}{K.~Wu}, \bibinfo{author}{O.~Sigmund}, \bibinfo{author}{J.~Du},
  \bibinfo{title}{Design of metamaterial mechanisms using robust topology
  optimization and variable linking scheme}, \bibinfo{journal}{Structural and
  Multidisciplinary Optimization} \bibinfo{volume}{63}~(\bibinfo{number}{4})
  (\bibinfo{year}{2021}{\natexlab{a}}) \bibinfo{pages}{1975--1988},
  \doi{\bibinfo{doi}{10.1007/s00158-020-02791-1}}.

\bibitem[{Franchetti and Kress(2016)}]{franchetti2016}
\bibinfo{author}{M.~Franchetti}, \bibinfo{author}{C.~Kress}, \bibinfo{title}{An
  economic analysis comparing the cost feasibility of replacing injection
  molding processes with emerging additive manufacturing techniques},
  \bibinfo{journal}{Internatial Journal of Advanced Manufacturing}
  \bibinfo{volume}{88}~(\bibinfo{number}{9-12}) (\bibinfo{year}{2016})
  \bibinfo{pages}{2573--2579}, \doi{\bibinfo{doi}{10.1007/s00170-016-8968-7}}.

\bibitem[{Jenett et~al.(2020)Jenett, Cameron, Tourlomousis, Rubio, Ochalek, and
  Gershenfeld}]{Jenett2020}
\bibinfo{author}{B.~Jenett}, \bibinfo{author}{C.~Cameron},
  \bibinfo{author}{F.~Tourlomousis}, \bibinfo{author}{A.~P. Rubio},
  \bibinfo{author}{M.~Ochalek}, \bibinfo{author}{N.~Gershenfeld},
  \bibinfo{title}{Discretely assembled mechanical metamaterials},
  \bibinfo{journal}{Science Advances}
  \bibinfo{volume}{6}~(\bibinfo{number}{47}),
  \doi{\bibinfo{doi}{10.1126/sciadv.abc9943}}.

\bibitem[{Tyburec et~al.(2020)Tyburec, Zeman, Do{\v{s}}k{\'{a}}{\v{r}},
  Kru{\v{z}}{\'{\i}}k, and Lep{\v{s}}}]{tyburec2020}
\bibinfo{author}{M.~Tyburec}, \bibinfo{author}{J.~Zeman},
  \bibinfo{author}{M.~Do{\v{s}}k{\'{a}}{\v{r}}},
  \bibinfo{author}{M.~Kru{\v{z}}{\'{\i}}k}, \bibinfo{author}{M.~Lep{\v{s}}},
  \bibinfo{title}{Modular-topology optimization with {Wang} tilings: {An}
  application to truss structures}, \bibinfo{journal}{Structural and
  Multidisciplinary Optimization}
  \doi{\bibinfo{doi}{10.1007/s00158-020-02744-8}}.

\bibitem[{Wu et~al.(2021{\natexlab{b}})Wu, Sigmund, and Groen}]{wu2021}
\bibinfo{author}{J.~Wu}, \bibinfo{author}{O.~Sigmund}, \bibinfo{author}{J.~P.
  Groen}, \bibinfo{title}{Topology optimization of multi-scale structures: {A}
  review}, \bibinfo{journal}{Structural and Multidisciplinary Optimization}
  \bibinfo{volume}{63}~(\bibinfo{number}{3})
  (\bibinfo{year}{2021}{\natexlab{b}}) \bibinfo{pages}{1455--1480},
  \doi{\bibinfo{doi}{10.1007/s00158-021-02881-8}}.

\bibitem[{Bourgat(1979)}]{bourgat1979}
\bibinfo{author}{J.~F. Bourgat}, \bibinfo{title}{Numerical experiments of the
  homogenization method}, in: \bibinfo{booktitle}{Lecture Notes in
  Mathematics}, \bibinfo{publisher}{Springer Berlin Heidelberg},
  \bibinfo{pages}{330--356}, \doi{\bibinfo{doi}{10.1007/bfb0063630}},
  \bibinfo{year}{1979}.

\bibitem[{Sigmund(1994)}]{sigmund1994}
\bibinfo{author}{O.~Sigmund}, \bibinfo{title}{Materials with prescribed
  constitutive parameters: {An} inverse homogenization problem},
  \bibinfo{journal}{International Journal of Solids and Structures}
  \bibinfo{volume}{31}~(\bibinfo{number}{17}) (\bibinfo{year}{1994})
  \bibinfo{pages}{2313--2329},
  \doi{\bibinfo{doi}{10.1016/0020-7683(94)90154-6}}.

\bibitem[{Rodrigues et~al.(2002)Rodrigues, Guedes, and Bendsoe}]{rodrigues2002}
\bibinfo{author}{H.~Rodrigues}, \bibinfo{author}{J.~Guedes},
  \bibinfo{author}{M.~Bendsoe}, \bibinfo{title}{Hierarchical optimization of
  material and structure}, \bibinfo{journal}{Structural and Multidisciplinary
  Optimization} \bibinfo{volume}{24}~(\bibinfo{number}{1})
  (\bibinfo{year}{2002}) \bibinfo{pages}{1--10},
  \doi{\bibinfo{doi}{10.1007/s00158-002-0209-z}}.

\bibitem[{Schury et~al.(2012)Schury, Stingl, and Wein}]{schury2012}
\bibinfo{author}{F.~Schury}, \bibinfo{author}{M.~Stingl},
  \bibinfo{author}{F.~Wein}, \bibinfo{title}{Efficient two-scale optimization
  of manufacturable graded structures}, \bibinfo{journal}{{SIAM} Journal on
  Scientific Computing} \bibinfo{volume}{34}~(\bibinfo{number}{6})
  (\bibinfo{year}{2012}) \bibinfo{pages}{B711--B733},
  \doi{\bibinfo{doi}{10.1137/110850335}}.

\bibitem[{Garner et~al.(2019)Garner, Kolken, Wang, Zadpoor, and
  Wu}]{garner2019}
\bibinfo{author}{E.~Garner}, \bibinfo{author}{H.~M. Kolken},
  \bibinfo{author}{C.~C. Wang}, \bibinfo{author}{A.~A. Zadpoor},
  \bibinfo{author}{J.~Wu}, \bibinfo{title}{Compatibility in microstructural
  optimization for additive manufacturing}, \bibinfo{journal}{Additive
  Manufacturing} \bibinfo{volume}{26} (\bibinfo{year}{2019})
  \bibinfo{pages}{65--75}, \doi{\bibinfo{doi}{10.1016/j.addma.2018.12.007}}.

\bibitem[{Liu et~al.(2008)Liu, Yan, and Cheng}]{liu2008}
\bibinfo{author}{L.~Liu}, \bibinfo{author}{J.~Yan}, \bibinfo{author}{G.~Cheng},
  \bibinfo{title}{Optimum structure with homogeneous optimum truss-like
  material}, \bibinfo{journal}{Computers {\&} Structures}
  \bibinfo{volume}{86}~(\bibinfo{number}{13-14}) (\bibinfo{year}{2008})
  \bibinfo{pages}{1417--1425},
  \doi{\bibinfo{doi}{10.1016/j.compstruc.2007.04.030}}.

\bibitem[{Tugilimana et~al.(2016)Tugilimana, Thrall, Descamps, and
  Coelho}]{tugilimana2016}
\bibinfo{author}{A.~Tugilimana}, \bibinfo{author}{A.~P. Thrall},
  \bibinfo{author}{B.~Descamps}, \bibinfo{author}{R.~F. Coelho},
  \bibinfo{title}{Spatial orientation and topology optimization of modular
  trusses}, \bibinfo{journal}{Structural and Multidisciplinary Optimization}
  \bibinfo{volume}{55}~(\bibinfo{number}{2}) (\bibinfo{year}{2016})
  \bibinfo{pages}{459--476}, \doi{\bibinfo{doi}{10.1007/s00158-016-1501-7}}.

\bibitem[{Zhang et~al.(2018)Zhang, Xiao, Li, Gao, and Chu}]{zhang2018a}
\bibinfo{author}{Y.~Zhang}, \bibinfo{author}{M.~Xiao}, \bibinfo{author}{H.~Li},
  \bibinfo{author}{L.~Gao}, \bibinfo{author}{S.~Chu},
  \bibinfo{title}{Multiscale concurrent topology optimization for cellular
  structures with multiple microstructures based on ordered {SIMP}
  interpolation}, \bibinfo{journal}{Computational Materials Science}
  \bibinfo{volume}{155} (\bibinfo{year}{2018}) \bibinfo{pages}{74--91},
  \doi{\bibinfo{doi}{10.1016/j.commatsci.2018.08.030}}.

\bibitem[{Groen and Sigmund(2017)}]{groen2017}
\bibinfo{author}{J.~P. Groen}, \bibinfo{author}{O.~Sigmund},
  \bibinfo{title}{Homogenization-based topology optimization for
  high-resolution manufacturable microstructures},
  \bibinfo{journal}{International Journal for Numerical Methods in Engineering}
  \bibinfo{volume}{113}~(\bibinfo{number}{8}) (\bibinfo{year}{2017})
  \bibinfo{pages}{1148--1163}, \doi{\bibinfo{doi}{10.1002/nme.5575}}.

\bibitem[{Allaire et~al.(2019)Allaire, Geoffroy-Donders, and
  Pantz}]{allaire2019}
\bibinfo{author}{G.~Allaire}, \bibinfo{author}{P.~Geoffroy-Donders},
  \bibinfo{author}{O.~Pantz}, \bibinfo{title}{Topology optimization of
  modulated and oriented periodic microstructures by the homogenization
  method}, \bibinfo{journal}{Computers {\&} Mathematics with Applications}
  \bibinfo{volume}{78}~(\bibinfo{number}{7}) (\bibinfo{year}{2019})
  \bibinfo{pages}{2197--2229},
  \doi{\bibinfo{doi}{10.1016/j.camwa.2018.08.007}}.

\bibitem[{Pantz and Trabelsi(2008)}]{pantz2008}
\bibinfo{author}{O.~Pantz}, \bibinfo{author}{K.~Trabelsi}, \bibinfo{title}{A
  post-treatment of the homogenization method for shape optimization},
  \bibinfo{journal}{SIAM Journal on Control and Optimization}
  \bibinfo{volume}{47}~(\bibinfo{number}{3}) (\bibinfo{year}{2008})
  \bibinfo{pages}{1380--1398}, \doi{\bibinfo{doi}{10.1137/070688900}}.

\bibitem[{Groen et~al.(2020)Groen, Stutz, Aage, B{\ae}rentzen, and
  Sigmund}]{groen2020}
\bibinfo{author}{J.~P. Groen}, \bibinfo{author}{F.~C. Stutz},
  \bibinfo{author}{N.~Aage}, \bibinfo{author}{J.~A. B{\ae}rentzen},
  \bibinfo{author}{O.~Sigmund}, \bibinfo{title}{De-homogenization of optimal
  multi-scale {3D} topologies}, \bibinfo{journal}{Computer Methods in Applied
  Mechanics and Engineering} \bibinfo{volume}{364} (\bibinfo{year}{2020})
  \bibinfo{pages}{112979}, \doi{\bibinfo{doi}{10.1016/j.cma.2020.112979}}.

\bibitem[{Geoffroy-Donders et~al.(2020)Geoffroy-Donders, Allaire, and
  Pantz}]{geoffroydonders2020}
\bibinfo{author}{P.~Geoffroy-Donders}, \bibinfo{author}{G.~Allaire},
  \bibinfo{author}{O.~Pantz}, \bibinfo{title}{3-D topology optimization of
  modulated and oriented periodic microstructures by the homogenization
  method}, \bibinfo{journal}{Journal of Computational Physics}
  \bibinfo{volume}{401} (\bibinfo{year}{2020}) \bibinfo{pages}{108994},
  \doi{\bibinfo{doi}{10.1016/j.jcp.2019.108994}}.

\bibitem[{Jung et~al.(2022)Jung, Lee, Nomura, and Dede}]{Jung2022}
\bibinfo{author}{T.~Jung}, \bibinfo{author}{J.~Lee},
  \bibinfo{author}{T.~Nomura}, \bibinfo{author}{E.~M. Dede},
  \bibinfo{title}{Inverse design of three-dimensional fiber reinforced
  composites with spatially-varying fiber size and orientation using multiscale
  topology optimization}, \bibinfo{journal}{Composite Structures}
  \bibinfo{volume}{279} (\bibinfo{year}{2022}) \bibinfo{pages}{114768},
  \doi{\bibinfo{doi}{10.1016/j.compstruct.2021.114768}}.

\bibitem[{Sivapuram et~al.(2016)Sivapuram, Dunning, and Kim}]{sivapuram2016}
\bibinfo{author}{R.~Sivapuram}, \bibinfo{author}{P.~D. Dunning},
  \bibinfo{author}{H.~A. Kim}, \bibinfo{title}{Simultaneous material and
  structural optimization by multiscale topology optimization},
  \bibinfo{journal}{Structural and Multidisciplinary Optimization}
  \bibinfo{volume}{54}~(\bibinfo{number}{5}) (\bibinfo{year}{2016})
  \bibinfo{pages}{1267--1281}, \doi{\bibinfo{doi}{10.1007/s00158-016-1519-x}}.

\bibitem[{Ferrer et~al.(2018)Ferrer, Cante, Hern{\'{a}}ndez, and
  Oliver}]{ferrer2018}
\bibinfo{author}{A.~Ferrer}, \bibinfo{author}{J.~Cante},
  \bibinfo{author}{J.~Hern{\'{a}}ndez}, \bibinfo{author}{J.~Oliver},
  \bibinfo{title}{Two-scale topology optimization in computational material
  design: {An} integrated approach}, \bibinfo{journal}{Internation Journal of
  Numerical Methods in Engineering} \bibinfo{volume}{114}~(\bibinfo{number}{3})
  (\bibinfo{year}{2018}) \bibinfo{pages}{232--254},
  \doi{\bibinfo{doi}{10.1002/nme.5742}}.

\bibitem[{Alexandersen and Lazarov(2015{\natexlab{a}})}]{alexandersen2015}
\bibinfo{author}{J.~Alexandersen}, \bibinfo{author}{B.~S. Lazarov},
  \bibinfo{title}{Topology optimisation of manufacturable microstructural
  details without length scale separation using a spectral coarse basis
  preconditioner}, \bibinfo{journal}{Computer Methods in Applied Mechanics and
  Engineering} \bibinfo{volume}{290} (\bibinfo{year}{2015}{\natexlab{a}})
  \bibinfo{pages}{156--182}, \doi{\bibinfo{doi}{10.1016/j.cma.2015.02.028}}.

\bibitem[{Alexandersen and Lazarov(2015{\natexlab{b}})}]{alexandersen2015a}
\bibinfo{author}{J.~Alexandersen}, \bibinfo{author}{B.~S. Lazarov},
  \bibinfo{title}{Tailoring macroscale response of mechanical and heat transfer
  systems by topology optimization of microstructural details}, in:
  \bibinfo{booktitle}{Computational Methods in Applied Sciences},
  \bibinfo{publisher}{Springer International Publishing},
  \bibinfo{pages}{267--288}, \doi{\bibinfo{doi}{10.1007/978-3-319-18320-6_15}},
  \bibinfo{year}{2015}{\natexlab{b}}.

\bibitem[{Coulais et~al.(2016)Coulais, Teomy, de~Reus, Shokef, and van
  Hecke}]{coulais2016}
\bibinfo{author}{C.~Coulais}, \bibinfo{author}{E.~Teomy},
  \bibinfo{author}{K.~de~Reus}, \bibinfo{author}{Y.~Shokef},
  \bibinfo{author}{M.~van Hecke}, \bibinfo{title}{Combinatorial design of
  textured mechanical metamaterials}, \bibinfo{journal}{Nature}
  \bibinfo{volume}{535}~(\bibinfo{number}{7613}) (\bibinfo{year}{2016})
  \bibinfo{pages}{529--532}, \doi{\bibinfo{doi}{10.1038/nature18960}}.

\bibitem[{Wu et~al.(2020)Wu, Liu, Wang, Zhai, Zhuang, Krishnaraju, Wang, and
  Jiang}]{wu2020}
\bibinfo{author}{L.~Wu}, \bibinfo{author}{L.~Liu}, \bibinfo{author}{Y.~Wang},
  \bibinfo{author}{Z.~Zhai}, \bibinfo{author}{H.~Zhuang},
  \bibinfo{author}{D.~Krishnaraju}, \bibinfo{author}{Q.~Wang},
  \bibinfo{author}{H.~Jiang}, \bibinfo{title}{A machine learning-based method
  to design modular metamaterials}, \bibinfo{journal}{Extreme Mechanics
  Letters} \bibinfo{volume}{36} (\bibinfo{year}{2020}) \bibinfo{pages}{100657},
  \doi{\bibinfo{doi}{10.1016/j.eml.2020.100657}}.

\bibitem[{Yang et~al.(2020)Yang, Chen, Yang, and Silverberg}]{yang2020}
\bibinfo{author}{N.~Yang}, \bibinfo{author}{C.-W. Chen},
  \bibinfo{author}{J.~Yang}, \bibinfo{author}{J.~L. Silverberg},
  \bibinfo{title}{Emergent reconfigurable mechanical metamaterial tessellations
  with an exponentially large number of discrete configurations},
  \bibinfo{journal}{Materials {\&} Design} \bibinfo{volume}{196}
  (\bibinfo{year}{2020}) \bibinfo{pages}{109143},
  \doi{\bibinfo{doi}{10.1016/j.matdes.2020.109143}}.

\bibitem[{Jenett et~al.(2017)Jenett, Calisch, Cellucci, Cramer, Gershenfeld,
  Swei, and Cheung}]{Jenett2017}
\bibinfo{author}{B.~Jenett}, \bibinfo{author}{S.~Calisch},
  \bibinfo{author}{D.~Cellucci}, \bibinfo{author}{N.~Cramer},
  \bibinfo{author}{N.~Gershenfeld}, \bibinfo{author}{S.~Swei},
  \bibinfo{author}{K.~C. Cheung}, \bibinfo{title}{Digital morphing wing: Active
  wing shaping concept using composite lattice-based cellular structures},
  \bibinfo{journal}{Soft Robotics} \bibinfo{volume}{4}~(\bibinfo{number}{1})
  (\bibinfo{year}{2017}) \bibinfo{pages}{33--48},
  \doi{\bibinfo{doi}{10.1089/soro.2016.0032}}.

\bibitem[{Cheung and Gershenfeld(2013)}]{cheung2013}
\bibinfo{author}{K.~C. Cheung}, \bibinfo{author}{N.~Gershenfeld},
  \bibinfo{title}{Reversibly assembled cellular composite materials},
  \bibinfo{journal}{Science} \bibinfo{volume}{341}~(\bibinfo{number}{6151})
  (\bibinfo{year}{2013}) \bibinfo{pages}{1219--1221},
  \doi{\bibinfo{doi}{10.1126/science.1240889}}.

\bibitem[{Gregg et~al.(2018)Gregg, Kim, and Cheung}]{gregg2018}
\bibinfo{author}{C.~E. Gregg}, \bibinfo{author}{J.~H. Kim},
  \bibinfo{author}{K.~C. Cheung}, \bibinfo{title}{Ultra-light and scalable
  composite lattice materials}, \bibinfo{journal}{Advanced Engineering
  Materials} \bibinfo{volume}{20}~(\bibinfo{number}{9}) (\bibinfo{year}{2018})
  \bibinfo{pages}{1800213}, \doi{\bibinfo{doi}{10.1002/adem.201800213}}.

\bibitem[{Yang et~al.(2018)Yang, Liu, Gao, Yue, and Xu}]{yang2018}
\bibinfo{author}{W.~Yang}, \bibinfo{author}{Q.~Liu}, \bibinfo{author}{Z.~Gao},
  \bibinfo{author}{Z.~Yue}, \bibinfo{author}{B.~Xu},
  \bibinfo{title}{Theoretical search for heterogeneously architected 2D
  structures}, \bibinfo{journal}{Proceedings of the National Academy of
  Sciences} \bibinfo{volume}{115}~(\bibinfo{number}{31}) (\bibinfo{year}{2018})
  \bibinfo{pages}{E7245--E7254}, \doi{\bibinfo{doi}{10.1073/pnas.1806769115}}.

\bibitem[{Silverberg et~al.(2014)Silverberg, Evans, McLeod, Hayward, Hull,
  Santangelo, and Cohen}]{silverberg2014}
\bibinfo{author}{J.~L. Silverberg}, \bibinfo{author}{A.~A. Evans},
  \bibinfo{author}{L.~McLeod}, \bibinfo{author}{R.~C. Hayward},
  \bibinfo{author}{T.~Hull}, \bibinfo{author}{C.~D. Santangelo},
  \bibinfo{author}{I.~Cohen}, \bibinfo{title}{Using origami design principles
  to fold reprogrammable mechanical metamaterials}, \bibinfo{journal}{Science}
  \bibinfo{volume}{345}~(\bibinfo{number}{6197}) (\bibinfo{year}{2014})
  \bibinfo{pages}{647--650}, \doi{\bibinfo{doi}{10.1126/science.1252876}}.

\bibitem[{Yang and Silverberg(2017)}]{yang2017}
\bibinfo{author}{N.~Yang}, \bibinfo{author}{J.~L. Silverberg},
  \bibinfo{title}{Decoupling local mechanics from large-scale structure in
  modular metamaterials}, \bibinfo{journal}{Proceedings of the National Academy
  of Sciences} \bibinfo{volume}{114}~(\bibinfo{number}{14})
  (\bibinfo{year}{2017}) \bibinfo{pages}{3590--3595},
  \doi{\bibinfo{doi}{10.1073/pnas.1620714114}}.

\bibitem[{Ma et~al.(2022)Ma, Jiang, and Chen}]{ma2022}
\bibinfo{author}{J.~Ma}, \bibinfo{author}{X.~Jiang}, \bibinfo{author}{Y.~Chen},
  \bibinfo{title}{A {3D} modular meta-structure with continuous mechanism
  motion and bistability}, \bibinfo{journal}{Extreme Mechanics Letters}
  \bibinfo{volume}{51} (\bibinfo{year}{2022}) \bibinfo{pages}{101584},
  \doi{\bibinfo{doi}{10.1016/j.eml.2021.101584}}.

\bibitem[{Lu and Tong(2021)}]{lu2021}
\bibinfo{author}{Y.~Lu}, \bibinfo{author}{L.~Tong}, \bibinfo{title}{Concurrent
  topology optimization of cellular structures and anisotropic materials},
  \bibinfo{journal}{Computers {\&} Structures} \bibinfo{volume}{255}
  (\bibinfo{year}{2021}) \bibinfo{pages}{106624},
  \doi{\bibinfo{doi}{10.1016/j.compstruc.2021.106624}}.

\bibitem[{Zhang and Sun(2006)}]{zhang2006}
\bibinfo{author}{W.~Zhang}, \bibinfo{author}{S.~Sun},
  \bibinfo{title}{Scale-related topology optimization of cellular materials and
  structures} \bibinfo{volume}{68}~(\bibinfo{number}{9}) (\bibinfo{year}{2006})
  \bibinfo{pages}{993--1011}, \doi{\bibinfo{doi}{10.1002/nme.1743}}.

\bibitem[{Li et~al.(2018)Li, Luo, Gao, and Qin}]{li2018}
\bibinfo{author}{H.~Li}, \bibinfo{author}{Z.~Luo}, \bibinfo{author}{L.~Gao},
  \bibinfo{author}{Q.~Qin}, \bibinfo{title}{Topology optimization for
  concurrent design of structures with multi-patch microstructures by level
  sets}, \bibinfo{journal}{Computer Methods in Applied Mechanics and
  Engineering} \bibinfo{volume}{331} (\bibinfo{year}{2018})
  \bibinfo{pages}{536--561}, \doi{\bibinfo{doi}{10.1016/j.cma.2017.11.033}}.

\bibitem[{Gao et~al.(2019{\natexlab{a}})Gao, Luo, Li, and Gao}]{gao2019a}
\bibinfo{author}{J.~Gao}, \bibinfo{author}{Z.~Luo}, \bibinfo{author}{H.~Li},
  \bibinfo{author}{L.~Gao}, \bibinfo{title}{Topology optimization for
  multiscale design of porous composites with multi-domain microstructures},
  \bibinfo{journal}{Computer Methods in Applied Mechanics and Engineering}
  \bibinfo{volume}{344} (\bibinfo{year}{2019}{\natexlab{a}})
  \bibinfo{pages}{451--476}, \doi{\bibinfo{doi}{10.1016/j.cma.2018.10.017}}.

\bibitem[{Gao et~al.(2019{\natexlab{b}})Gao, Luo, Li, Li, and Gao}]{gao2019b}
\bibinfo{author}{J.~Gao}, \bibinfo{author}{Z.~Luo}, \bibinfo{author}{H.~Li},
  \bibinfo{author}{P.~Li}, \bibinfo{author}{L.~Gao}, \bibinfo{title}{Dynamic
  multiscale topology optimization for multi-regional micro-structured cellular
  composites}, \bibinfo{journal}{Composite Structures} \bibinfo{volume}{211}
  (\bibinfo{year}{2019}{\natexlab{b}}) \bibinfo{pages}{401--417},
  \doi{\bibinfo{doi}{10.1016/j.compstruct.2018.12.031}}.

\bibitem[{Yan et~al.(2020)Yan, Sui, Fan, Duan, and Yu}]{yan2020}
\bibinfo{author}{J.~Yan}, \bibinfo{author}{Q.~Sui}, \bibinfo{author}{Z.~Fan},
  \bibinfo{author}{Z.~Duan}, \bibinfo{author}{T.~Yu},
  \bibinfo{title}{Clustering-based multiscale topology optimization of
  thermo-elastic lattice structures}, \bibinfo{journal}{Computational
  Mechanics} \bibinfo{volume}{66}~(\bibinfo{number}{4}) (\bibinfo{year}{2020})
  \bibinfo{pages}{979--1002}, \doi{\bibinfo{doi}{10.1007/s00466-020-01892-4}}.

\bibitem[{Nikravesh et~al.(2022)Nikravesh, Zhang, Liu, and
  Frantziskonis}]{Nikravesh2022}
\bibinfo{author}{Y.~Nikravesh}, \bibinfo{author}{Y.~Zhang},
  \bibinfo{author}{J.~Liu}, \bibinfo{author}{G.~N. Frantziskonis},
  \bibinfo{title}{A partition and microstructure based method applicable to
  large-scale topology optimization}, \bibinfo{journal}{Mechanics of Materials}
  \bibinfo{volume}{166} (\bibinfo{year}{2022}) \bibinfo{pages}{104234},
  \doi{\bibinfo{doi}{10.1016/j.mechmat.2022.104234}}.

\bibitem[{Xu and Cheng(2018)}]{xu2018}
\bibinfo{author}{L.~Xu}, \bibinfo{author}{G.~Cheng}, \bibinfo{title}{Two-scale
  concurrent topology optimization with multiple micro materials based on
  principal stress orientation}, \bibinfo{journal}{Structural and
  Multidisciplinary Optimization} \bibinfo{volume}{57}~(\bibinfo{number}{5})
  (\bibinfo{year}{2018}) \bibinfo{pages}{2093--2107},
  \doi{\bibinfo{doi}{10.1007/s00158-018-1916-4}}.

\bibitem[{Qiu et~al.(2020)Qiu, Li, Liu, and Xu}]{qiu2020}
\bibinfo{author}{Z.~Qiu}, \bibinfo{author}{Q.~Li}, \bibinfo{author}{S.~Liu},
  \bibinfo{author}{R.~Xu}, \bibinfo{title}{Clustering-based concurrent topology
  optimization with macrostructure, components, and materials},
  \bibinfo{journal}{Structural and Multidisciplinary Optimization}
  \doi{\bibinfo{doi}{10.1007/s00158-020-02755-5}}.

\bibitem[{Kumar and Suresh(2019)}]{kumar2019}
\bibinfo{author}{T.~Kumar}, \bibinfo{author}{K.~Suresh}, \bibinfo{title}{A
  density-and-strain-based K-clustering approach to microstructural topology
  optimization}, \bibinfo{journal}{Structural and Multidisciplinary
  Optimization} \bibinfo{volume}{61}~(\bibinfo{number}{4})
  (\bibinfo{year}{2019}) \bibinfo{pages}{1399--1415},
  \doi{\bibinfo{doi}{10.1007/s00158-019-02422-4}}.

\bibitem[{Yang and Li(2019)}]{yang2019}
\bibinfo{author}{X.~Yang}, \bibinfo{author}{M.~Li}, \bibinfo{title}{Free
  isotropic material optimization via second order cone programming},
  \bibinfo{journal}{Computer-Aided Design} \bibinfo{volume}{115}
  (\bibinfo{year}{2019}) \bibinfo{pages}{52--63},
  \doi{\bibinfo{doi}{10.1016/j.cad.2019.05.002}}.

\bibitem[{Hu et~al.(2020)Hu, Li, Yang, and Gao}]{hu2020}
\bibinfo{author}{J.~Hu}, \bibinfo{author}{M.~Li}, \bibinfo{author}{X.~Yang},
  \bibinfo{author}{S.~Gao}, \bibinfo{title}{Cellular structure design based on
  free material optimization under connectivity control},
  \bibinfo{journal}{Computer-Aided Design} \bibinfo{volume}{127}
  (\bibinfo{year}{2020}) \bibinfo{pages}{102854},
  \doi{\bibinfo{doi}{10.1016/j.cad.2020.102854}}.

\bibitem[{Zowe et~al.(1997)Zowe, Ko{\v{c}}vara, and Bends{\o}e}]{zowe1997}
\bibinfo{author}{J.~Zowe}, \bibinfo{author}{M.~Ko{\v{c}}vara},
  \bibinfo{author}{M.~P. Bends{\o}e}, \bibinfo{title}{Free material
  optimization via mathematical programming}, \bibinfo{journal}{Mathematical
  Programming} \bibinfo{volume}{79}~(\bibinfo{number}{1-3})
  (\bibinfo{year}{1997}) \bibinfo{pages}{445--466},
  \doi{\bibinfo{doi}{10.1007/bf02614328}}.

\bibitem[{Wang(1961)}]{wang1961}
\bibinfo{author}{H.~Wang}, \bibinfo{title}{Proving theorems by pattern
  recognition--{II}}, \bibinfo{journal}{Bell System Technical Journal}
  \bibinfo{volume}{40}~(\bibinfo{number}{1}) (\bibinfo{year}{1961})
  \bibinfo{pages}{1--41},
  \doi{\bibinfo{doi}{10.1002/j.1538-7305.1961.tb03975.x}}.

\bibitem[{Berger(1966)}]{berger1966}
\bibinfo{author}{R.~Berger}, \bibinfo{title}{The undecidability of the domino
  problem}, \bibinfo{journal}{Memoirs of the American Mathematical Society}
  \bibinfo{volume}{0}~(\bibinfo{number}{66}), ISSN \bibinfo{issn}{0065-9266},
  \doi{\bibinfo{doi}{10.1090/memo/0066}}.

\bibitem[{Gr\"{u}nbaum and Shephard(2016)}]{grunbaum1987}
\bibinfo{author}{B.~Gr\"{u}nbaum}, \bibinfo{author}{G.~C. Shephard},
  \bibinfo{title}{Tilings and patterns}, \bibinfo{publisher}{Dover
  Publications, Inc}, \bibinfo{address}{Mineola, New York}, ISBN
  \bibinfo{isbn}{9780486469812}, \bibinfo{year}{2016}.

\bibitem[{Winfree et~al.(1998)Winfree, Liu, Wenzler, and Seeman}]{winfree1998}
\bibinfo{author}{E.~Winfree}, \bibinfo{author}{F.~Liu}, \bibinfo{author}{L.~A.
  Wenzler}, \bibinfo{author}{N.~C. Seeman}, \bibinfo{title}{Design and
  self-assembly of two-dimensional {DNA} crystals}, \bibinfo{journal}{Nature}
  \bibinfo{volume}{394}~(\bibinfo{number}{6693}) (\bibinfo{year}{1998})
  \bibinfo{pages}{539--544}, \doi{\bibinfo{doi}{10.1038/28998}}.

\bibitem[{Winfree(2000)}]{winfree2000}
\bibinfo{author}{E.~Winfree}, \bibinfo{title}{Algorithmic self-assembly of
  {DNA}: Theoretical motivations and 2{D} assembly experiments},
  \bibinfo{journal}{Journal of Biomolecular Structure and Dynamics}
  \bibinfo{volume}{17}~(\bibinfo{number}{sup1}) (\bibinfo{year}{2000})
  \bibinfo{pages}{263--270},
  \doi{\bibinfo{doi}{10.1080/07391102.2000.10506630}}.

\bibitem[{Cohen et~al.(2003)Cohen, Shade, Hiller, and Deussen}]{cohen2003}
\bibinfo{author}{M.~F. Cohen}, \bibinfo{author}{J.~Shade},
  \bibinfo{author}{S.~Hiller}, \bibinfo{author}{O.~Deussen},
  \bibinfo{title}{Wang tiles for image and texture generation},
  \bibinfo{journal}{ACM Transactions on Graphics}
  \bibinfo{volume}{22}~(\bibinfo{number}{3}) (\bibinfo{year}{2003})
  \bibinfo{pages}{287--294}, \doi{\bibinfo{doi}{10.1145/882262.882265}}.

\bibitem[{Zhang and Kim(2008)}]{zhang2008}
\bibinfo{author}{X.~Zhang}, \bibinfo{author}{Y.~J. Kim},
  \bibinfo{title}{Efficient texture synthesis using strict {Wang} {tiles}},
  \bibinfo{journal}{Graphical Models}
  \bibinfo{volume}{70}~(\bibinfo{number}{3}) (\bibinfo{year}{2008})
  \bibinfo{pages}{43--56}, ISSN \bibinfo{issn}{15240703},
  \doi{\bibinfo{doi}{10.1016/j.gmod.2007.10.002}}.

\bibitem[{Hiller et~al.(2001)Hiller, Deussen, and Keller}]{hiller2001}
\bibinfo{author}{S.~Hiller}, \bibinfo{author}{O.~Deussen},
  \bibinfo{author}{A.~Keller}, \bibinfo{title}{Tiled {blue} {noise} {samples}},
  in: \bibinfo{booktitle}{Proceedings of the {Vision} {Modeling} and
  {Visualization} {Conference} 2001}, {VMV} '01, \bibinfo{publisher}{Aka GmbH},
  \bibinfo{address}{Stuttgart, Germany}, ISBN \bibinfo{isbn}{3-89838-028-9},
  \bibinfo{pages}{265--272}, \bibinfo{year}{2001}.

\bibitem[{Nov\'ak et~al.(2012)Nov\'ak, Ku\v{c}erov\'a, and Zeman}]{novak2012}
\bibinfo{author}{J.~Nov\'ak}, \bibinfo{author}{A.~Ku\v{c}erov\'a},
  \bibinfo{author}{J.~Zeman}, \bibinfo{title}{Compressing random
  microstructures via stochastic {Wang} tilings}, \bibinfo{journal}{Physical
  Review E} \bibinfo{volume}{86}~(\bibinfo{number}{4}) (\bibinfo{year}{2012})
  \bibinfo{pages}{040104(R)}, ISSN \bibinfo{issn}{1539-3755, 1550-2376},
  \doi{\bibinfo{doi}{10.1103/PhysRevE.86.040104}}.

\bibitem[{Do{\v{s}}k{\'{a}}{\v{r}} et~al.(2014)Do{\v{s}}k{\'{a}}{\v{r}},
  Nov{\'{a}}k, and Zeman}]{doskar2014}
\bibinfo{author}{M.~Do{\v{s}}k{\'{a}}{\v{r}}},
  \bibinfo{author}{J.~Nov{\'{a}}k}, \bibinfo{author}{J.~Zeman},
  \bibinfo{title}{Aperiodic compression and reconstruction of real-world
  material systems based on {Wang} tiles}, \bibinfo{journal}{Physical Review E}
  \bibinfo{volume}{90}~(\bibinfo{number}{6}),
  \doi{\bibinfo{doi}{10.1103/physreve.90.062118}}.

\bibitem[{Do\v{s}k\'a\v{r} et~al.(2020)Do\v{s}k\'a\v{r}, Zeman, Rypl, and
  Nov\'ak}]{doskar2020}
\bibinfo{author}{M.~Do\v{s}k\'a\v{r}}, \bibinfo{author}{J.~Zeman},
  \bibinfo{author}{D.~Rypl}, \bibinfo{author}{J.~Nov\'ak},
  \bibinfo{title}{Level-set based design of {Wang} tiles for modelling complex
  microstructures}, \bibinfo{journal}{Computer-Aided Design}
  \bibinfo{volume}{123} (\bibinfo{year}{2020}) \bibinfo{pages}{102827}, ISSN
  \bibinfo{issn}{00104485}, \doi{\bibinfo{doi}{10.1016/j.cad.2020.102827}}.

\bibitem[{Nov\'ak et~al.(2013)Nov\'ak, Ku\v{c}erov\'a, and Zeman}]{novak2013}
\bibinfo{author}{J.~Nov\'ak}, \bibinfo{author}{A.~Ku\v{c}erov\'a},
  \bibinfo{author}{J.~Zeman}, \bibinfo{title}{Microstructural enrichment
  functions based on stochastic {Wang} tilings}, \bibinfo{journal}{Modelling
  and Simulation in Materials Science and Engineering}
  \bibinfo{volume}{21}~(\bibinfo{number}{2}) (\bibinfo{year}{2013})
  \bibinfo{pages}{025014}, \doi{\bibinfo{doi}{10.1088/0965-0393/21/2/025014}}.

\bibitem[{Do{\v{s}}k{\'{a}}{\v{r}} and Nov{\'{a}}k(2016)}]{doskar2016}
\bibinfo{author}{M.~Do{\v{s}}k{\'{a}}{\v{r}}},
  \bibinfo{author}{J.~Nov{\'{a}}k}, \bibinfo{title}{A jigsaw puzzle framework
  for homogenization of high porosity foams}, \bibinfo{journal}{Computers {\&}
  Structures} \bibinfo{volume}{166} (\bibinfo{year}{2016})
  \bibinfo{pages}{33--41},
  \doi{\bibinfo{doi}{10.1016/j.compstruc.2016.01.003}}.

\bibitem[{Do{\v{s}}k{\'{a}}{\v{r}} et~al.(2018)Do{\v{s}}k{\'{a}}{\v{r}}, Zeman,
  Jaru\v{s}kov\'a, and Nov\'ak}]{doskar2018}
\bibinfo{author}{M.~Do{\v{s}}k{\'{a}}{\v{r}}}, \bibinfo{author}{J.~Zeman},
  \bibinfo{author}{D.~Jaru\v{s}kov\'a}, \bibinfo{author}{J.~Nov\'ak},
  \bibinfo{title}{Wang tiling aided statistical determination of the
  {Representative} {Volume} {Element} size of random heterogeneous materials},
  \bibinfo{journal}{European Journal of Mechanics - A/Solids}
  \bibinfo{volume}{70} (\bibinfo{year}{2018}) \bibinfo{pages}{280--295}, ISSN
  \bibinfo{issn}{09977538},
  \doi{\bibinfo{doi}{10.1016/j.euromechsol.2017.12.002}}.

\bibitem[{Do{\v{s}}k{\'{a}}{\v{r}} et~al.(2021)Do{\v{s}}k{\'{a}}{\v{r}}, Zeman,
  Krysl, and Nov{\'{a}}k}]{doskar2021}
\bibinfo{author}{M.~Do{\v{s}}k{\'{a}}{\v{r}}}, \bibinfo{author}{J.~Zeman},
  \bibinfo{author}{P.~Krysl}, \bibinfo{author}{J.~Nov{\'{a}}k},
  \bibinfo{title}{Microstructure-informed reduced modes synthesized with {Wang}
  tiles and the Generalized Finite Element Method},
  \bibinfo{journal}{Computational Mechanics}
  \bibinfo{volume}{68}~(\bibinfo{number}{2}) (\bibinfo{year}{2021})
  \bibinfo{pages}{233--253}, \doi{\bibinfo{doi}{10.1007/s00466-021-02028-y}}.

\bibitem[{Bends{\o}e and Sigmund(2004)}]{bendsoe2003}
\bibinfo{author}{M.~P. Bends{\o}e}, \bibinfo{author}{O.~Sigmund},
  \bibinfo{title}{Topology optimization: Theory, methods, and applications},
  \bibinfo{publisher}{Springer Berlin Heidelberg}, ISBN
  \bibinfo{isbn}{978-3-662-05086-6},
  \doi{\bibinfo{doi}{10.1007/978-3-662-05086-6}}, \bibinfo{year}{2004}.

\bibitem[{Ko{\v{c}}vara et~al.(2008)Ko{\v{c}}vara, Stingl, and
  Zowe}]{kocvara2008}
\bibinfo{author}{M.~Ko{\v{c}}vara}, \bibinfo{author}{M.~Stingl},
  \bibinfo{author}{J.~Zowe}, \bibinfo{title}{Free material optimization: Recent
  progress}, \bibinfo{journal}{A Journal of Mathematical Programming and
  Operations Research} \bibinfo{volume}{57}~(\bibinfo{number}{1})
  (\bibinfo{year}{2008}) \bibinfo{pages}{79--100},
  \doi{\bibinfo{doi}{10.1080/02331930701778908}}.

\bibitem[{Fiala et~al.(2013)Fiala, Ko\v{c}vara, and Stingl}]{fiala2013}
\bibinfo{author}{J.~Fiala}, \bibinfo{author}{M.~Ko\v{c}vara},
  \bibinfo{author}{M.~Stingl}, \bibinfo{title}{{PENLAB: A MATLAB solver for
  nonlinear semidefinite optimization}}, \bibinfo{note}{arXiv: 1311.5240},
  \bibinfo{year}{2013}.

\bibitem[{Ko{\v{c}}vara and Stingl(2003)}]{kocvara2003}
\bibinfo{author}{M.~Ko{\v{c}}vara}, \bibinfo{author}{M.~Stingl},
  \bibinfo{title}{{PENNON}: A code for convex nonlinear and semidefinite
  programming}, \bibinfo{journal}{Optimization Methods and Software}
  \bibinfo{volume}{18}~(\bibinfo{number}{3}) (\bibinfo{year}{2003})
  \bibinfo{pages}{317--333}, \doi{\bibinfo{doi}{10.1080/1055678031000098773}}.

\bibitem[{D{\'{\i}}az and Sigmund(1995)}]{daz1995}
\bibinfo{author}{A.~D{\'{\i}}az}, \bibinfo{author}{O.~Sigmund},
  \bibinfo{title}{Checkerboard patterns in layout optimization},
  \bibinfo{journal}{Structural Optimization}
  \bibinfo{volume}{10}~(\bibinfo{number}{1}) (\bibinfo{year}{1995})
  \bibinfo{pages}{40--45}, \doi{\bibinfo{doi}{10.1007/bf01743693}}.

\bibitem[{Andreassen et~al.(2010)Andreassen, Clausen, Schevenels, Lazarov, and
  Sigmund}]{andreassen2010}
\bibinfo{author}{E.~Andreassen}, \bibinfo{author}{A.~Clausen},
  \bibinfo{author}{M.~Schevenels}, \bibinfo{author}{B.~S. Lazarov},
  \bibinfo{author}{O.~Sigmund}, \bibinfo{title}{Efficient topology optimization
  in {MATLAB} using 88 lines of code}, \bibinfo{journal}{Structural and
  Multidisciplinary Optimization} \bibinfo{volume}{43}~(\bibinfo{number}{1})
  (\bibinfo{year}{2010}) \bibinfo{pages}{1--16},
  \doi{\bibinfo{doi}{10.1007/s00158-010-0594-7}}.

\bibitem[{Prathap(1985)}]{prathap1985}
\bibinfo{author}{G.~Prathap}, \bibinfo{title}{The poor bending response of the
  four-node plane stress quadrilateral}, \bibinfo{journal}{International
  Journal for Numerical Methods in Engineering}
  \bibinfo{volume}{21}~(\bibinfo{number}{5}) (\bibinfo{year}{1985})
  \bibinfo{pages}{825--835}, \doi{\bibinfo{doi}{10.1002/nme.1620210505}}.

\bibitem[{Zienkiewicz et~al.(2013)Zienkiewicz, Taylor, and
  Zhu}]{zienkiewicz2013}
\bibinfo{editor}{O.~Zienkiewicz}, \bibinfo{editor}{R.~Taylor},
  \bibinfo{editor}{J.~Zhu} (Eds.), \bibinfo{title}{The finite element method:
  Its basis and fundamentals}, \bibinfo{publisher}{Butterworth-Heinemann},
  \bibinfo{address}{Oxford}, \bibinfo{edition}{seventh edition} edn., ISBN
  \bibinfo{isbn}{978-1-85617-633-0},
  \doi{\bibinfo{doi}{10.1016/c2009-0-24909-9}}, \bibinfo{year}{2013}.

\bibitem[{Mukherjee et~al.(2021)Mukherjee, Lu, Dutta, Raghavan, Breitkopf, and
  Xiao}]{Mukherjee2021}
\bibinfo{author}{S.~Mukherjee}, \bibinfo{author}{D.~Lu},
  \bibinfo{author}{S.~Dutta}, \bibinfo{author}{B.~Raghavan},
  \bibinfo{author}{P.~Breitkopf}, \bibinfo{author}{M.~Xiao},
  \bibinfo{title}{Topology optimization of structures using higher order finite
  elements in analysis}, in: \bibinfo{editor}{B.~Das},
  \bibinfo{editor}{R.~Patgiri}, \bibinfo{editor}{S.~Bandyopadhyay},
  \bibinfo{editor}{V.~E. Balas} (Eds.), \bibinfo{booktitle}{Modeling,
  Simulation and Optimization}, \bibinfo{publisher}{Springer Singapore},
  \bibinfo{address}{Singapore}, ISBN \bibinfo{isbn}{978-981-15-9829-6},
  \bibinfo{pages}{791--800},
  \doi{\bibinfo{doi}{10.1007/978-981-15-9829-6\_61}}, \bibinfo{year}{2021}.

\bibitem[{Gan et~al.(2020)Gan, Ma, and Wu}]{gan2020}
\bibinfo{author}{G.~Gan}, \bibinfo{author}{C.~Ma}, \bibinfo{author}{J.~Wu},
  \bibinfo{title}{Data Clustering: Theory, Algorithms, and Applications},
  \bibinfo{publisher}{Society for Industrial and Applied Mathematics},
  \bibinfo{edition}{second} edn., \doi{\bibinfo{doi}{10.1137/1.9781611976335}},
  \bibinfo{year}{2020}.

\bibitem[{Bends{\o}e and Sigmund(1999)}]{bendsoe1999}
\bibinfo{author}{M.~P. Bends{\o}e}, \bibinfo{author}{O.~Sigmund},
  \bibinfo{title}{Material interpolation schemes in topology optimization},
  \bibinfo{journal}{Archive of Applied Mechanics (Ingenieur Archiv)}
  \bibinfo{volume}{69}~(\bibinfo{number}{9-10}) (\bibinfo{year}{1999})
  \bibinfo{pages}{635--654}, \doi{\bibinfo{doi}{10.1007/s004190050248}}.

\bibitem[{Groenwold and Etman(2009)}]{groenwold2009}
\bibinfo{author}{A.~A. Groenwold}, \bibinfo{author}{L.~F.~P. Etman},
  \bibinfo{title}{A simple heuristic for gray-scale suppression in optimality
  criterion-based topology optimization}, \bibinfo{journal}{Structural and
  Multidisciplinary Optimization} \bibinfo{volume}{39}~(\bibinfo{number}{2})
  (\bibinfo{year}{2009}) \bibinfo{pages}{217--225},
  \doi{\bibinfo{doi}{10.1007/s00158-008-0337-1}}.

\bibitem[{Sigmund(2007)}]{sigmund2007}
\bibinfo{author}{O.~Sigmund}, \bibinfo{title}{Morphology-based black and white
  filters for topology optimization}, \bibinfo{journal}{Structural and
  Multidisciplinary Optimization} \bibinfo{volume}{33}~(\bibinfo{number}{4-5})
  (\bibinfo{year}{2007}) \bibinfo{pages}{401--424}, ISSN
  \bibinfo{issn}{1615-147X, 1615-1488},
  \doi{\bibinfo{doi}{10.1007/s00158-006-0087-x}}.

\bibitem[{Tyburec et~al.(2022)Tyburec, Do\v{s}k\'a\v{r}, Zeman, and
  Kru\v{z}\'ik}]{tyburec2021}
\bibinfo{author}{M.~Tyburec}, \bibinfo{author}{M.~Do\v{s}k\'a\v{r}},
  \bibinfo{author}{J.~Zeman}, \bibinfo{author}{M.~Kru\v{z}\'ik},
  \bibinfo{title}{{Supplementary codes and datasets for ``Modular-topology
  optimization of structures and mechanismswith free material design and
  clustering''}}, \doi{\bibinfo{doi}{10.5281/zenodo.6127587}},
  \bibinfo{year}{2022}.

\bibitem[{Sigmund(1997)}]{sigmund1997}
\bibinfo{author}{O.~Sigmund}, \bibinfo{title}{On the design of compliant
  mechanisms using topology optimization}, \bibinfo{journal}{Mechanics of
  Structures and Machines} \bibinfo{volume}{25}~(\bibinfo{number}{4})
  (\bibinfo{year}{1997}) \bibinfo{pages}{493--524},
  \doi{\bibinfo{doi}{10.1080/08905459708945415}}.

\bibitem[{Med{\v{r}}ick{\'{y}} et~al.(2021)Med{\v{r}}ick{\'{y}},
  Do{\v{s}}k{\'{a}}{\v{r}}, Pultarov{\'{a}}, and Zeman}]{medricky2021}
\bibinfo{author}{T.~Med{\v{r}}ick{\'{y}}},
  \bibinfo{author}{M.~Do{\v{s}}k{\'{a}}{\v{r}}},
  \bibinfo{author}{I.~Pultarov{\'{a}}}, \bibinfo{author}{J.~Zeman},
  \bibinfo{title}{Comparison of {FETI}-based domain decomposition methods for
  topology optimization problems}, \bibinfo{note}{arXiv: 2111.11728},
  \bibinfo{year}{2021}.

\end{thebibliography}
